\documentclass{aa} 

\usepackage{graphicx}
\usepackage{txfonts}
\usepackage{array}
\usepackage{longtable}
\usepackage{natbib}
\usepackage{multirow}
\usepackage{textcomp}
\usepackage{amsmath}
\usepackage{afterpage}
\usepackage{amssymb}

\newcommand*{\thead}[1]{\multicolumn{1}{c}{#1}}

\begin{document}

   \title{VLT/FLAMES spectroscopy of red giant branch stars\\ in the Fornax dwarf spheroidal galaxy \thanks{Based on FLAMES observations collected at the European Southern Observatory, proposal number 080.B-0784.}}


   \author{B. Lemasle\inst{1}
          \and
          T.J.L. de Boer\inst{2} 
          \and
          V. Hill\inst{3}
	  \and
	  E. Tolstoy\inst{4}
	  \and
	  M. J. Irwin\inst{2}
	  \and 
	  P. Jablonka\inst{5,6}
	  \and 
	  K. Venn\inst{7}
	  \and 
	  G. Battaglia\inst{8,9}
	  \and 
	  E. Starkenburg\inst{7,10}
	  \and
	  M. Shetrone\inst{11}
	  \and 
	  B. Letarte\inst{12}
	  \and 
	  P. Fran\c cois\inst{6,13}
	  \and 
	  A. Helmi\inst{4}
	  \and 
	  F. Primas\inst{14}
	  \and 
	  A. Kaufer\inst{15}
	  \and 
	  T. Szeifert\inst{15}
}

   \institute{Anton Pannekoek Institute for Astronomy, University of Amsterdam, Science Park 904, PO Box 94249, 1090 GE, Amsterdam, The Netherlands\\
              \email{B.J.P.Lemasle@uva.nl}
         \and
		Institute of Astronomy, University of Cambridge, Madingley Road, Cambridge, CB3 0HA, UK
	 \and
		Observatoire de la C\^ote d'Azur, CNRS UMR 7293, BP4229, 06304, Nice Cedex 4, France
	 \and
		Kapteyn Astronomical Institute, University of Groningen, Postbus 800, 9700 AV, Groningen, The Netherlands
	 \and
		Laboratoire d’astrophysique, \'Ecole Polytechnique F\'ed\'erale de Lausanne (EPFL), Observatoire de Sauverny, 1290 Versoix, Switzerland 
	 \and
		GEPI, Observatoire de Paris, CNRS, Universit\'e Paris Diderot, Place Jules Janssen, 92190 Meudon, France 
	 \and
		Department of Physics and Astronomy, University of Victoria, 3800 Finnerty Road, Victoria, BC, V8P 1A1, Canada 
	 \and
		Istituto de Astrofisica de Canarias, calle via Lactea s/n, 38205 San Cristobal de La Laguna (Tenerife), Spain
	 \and
		Universidad de La Laguna, Dpto. Astrofísica, E-38206 La Laguna, Tenerife, Spain
	 \and
		CIFAR Global Scholar		
	 \and
		McDonald Observatory, The University of Texas at Austin, 1 University Station, C1400, Austin, TX 78712-0259, USA
	 \and
		Dept of Physics, North-West University, Mafikeng Campus, Private Bag X2046, Mmabatho 2745, South Africa
	 \and
		UPJV, Universit\'e de Picardie Jules Verne, 33 rue St. Leu, 80080 Amiens, France 
	 \and
		European Southern Observatory, Karl Schwarzschild Str. 2, 85748, Garching b. München, Germany
	 \and
		European Southern Observatory, Alonso de Cordova 3107, Santiago, Chile
}

   \date{Received September 15, 1996; accepted March 16, 1997}

 
  \abstract
   {Fornax is one of the most massive dwarf spheroidal galaxies in the Local Group. The Fornax field star population is dominated by intermediate age stars but star formation was going on over almost its entire history. It has been proposed that Fornax experienced a minor merger event.}
   {Despite recent progress, only the high metallicity end of Fornax field stars ([Fe/H]$>$--1.2 dex) has been sampled in larger number via high resolution spectroscopy. We want to better understand the full chemical evolution of this galaxy by better sampling the whole metallicity range, including more metal poor stars.}
   {We use the VLT-FLAMES multi-fibre spectrograph in high-resolution mode to determine the abundances of several $\alpha$, iron-peak and neutron-capture elements in a sample of 47 individual Red Giant Branch stars in the Fornax dwarf spheroidal galaxy. We combine these abundances with accurate age estimates derived from the age probability distribution from the colour-magnitude diagram of Fornax.}
   {Similar to other dwarf spheroidal galaxies, the old, metal-poor stars of Fornax are typically $\alpha$-rich while the young metal-rich stars are $\alpha$-poor. In the classical scenario of the time delay between SNe II and SNe Ia, we confirm that SNe Ia started to contribute to the chemical enrichment at [Fe/H] between --2.0 and --1.8 dex. We find that the onset of SNe Ia took place between 12--10 Gyrs ago. The high values of [Ba/Fe], [La/Fe] reflect the influence of SNe Ia and AGB stars in the abundance pattern of the younger stellar population of Fornax.}
   {Our findings of low [$\alpha$/Fe] and enhanced [Eu/Mg] are compatible with an initial mass function that lacks the most massive stars and with star formation that kept going on throughout the whole history of Fornax. We find that massive stars kept enriching the interstellar medium in $\alpha$-elements, although they were not the main contributor to the iron enrichment.
}


   \maketitle
%

\section{Introduction}

\par Dwarf spheroidal galaxies (dSphs) in the Local Group are close enough to the Milky Way (MW) to be resolved into stars. The dSphs studied so far have different and complex star formation histories \citep{Tolstoy2009} and thus they are useful for studying the various processes that drive the evolution of galaxies. Moreover, the Milky Way satellites also help to shed light on the formation and evolution of our own Galaxy.\\

\par Discovered by \cite{Sha1938}, Fornax is one of the most luminous and most massive dSphs \citep{Mat1998r}. It hosts several globular clusters \citep{Sha1938,Hodge1961a}, which is unusual for a present day dSph. Stellar (overdense) substructures have also been reported in Fornax \citep{Col2004,Col2005b,Ol2006,deBoer2013}. Comparing the properties of the two innermost shell-like structures to the field stars suggests that they were formed by re-accretion of previously expelled, pre-enriched gas rather than by a recent infall event of pristine gas \citep{deBoer2013}.\\

\par Different populations have been observed in Fornax: an old population ($\geq$ 10 Gyr) has been identified by \cite{BW2002} who found RR Lyrae stars, and by \cite{Savi2000} who reported an extended Horizontal Branch. A large amount of Carbon stars \citep[e.g.][]{AM1980,AM1985,Azz1999} reveals the presence of a prominent intermediate age population. Finally, \cite{Buo1985} discovered young main sequence stars ($\leq$ 1 Gyr) in the very center of Fornax, and \cite{Stet1998} reported a population gradient within Fornax, the youngest populations being more centrally concentrated.
Such a feature appears to be common in the Local Group dSphs \citep{Har2001}. These findings were confirmed by \cite{Batt2006} and \cite{Am2012c} who divided Fornax in three chemo-dynamical components (metal-poor, intermediate and metal-rich) that have different spatial distributions and kinematics.\\

\par From deep colour-magnitude diagrams (CMDs), it is possible to derive the star formation history (SFH) of Fornax \citep{Stet1998,Buo1999,Savi2000,Gall2005,Col2008,deBoer2012b,delPino2013}. Using the {\it Talos} code \citep{deBoer2012a} based on the synthetic CMD method, \cite{deBoer2012b} confirmed that the intermediate age (1--10 Gyr) population is dominant in Fornax and showed in particular that its prominent RGB is mostly related to a $\approx$ 4 Gyr old stellar population at [Fe/H] $\approx$ --1.0 dex. The quantitative, spatially resolved SFH and chemical enrichment history provided by \cite{deBoer2012b} can be summarized as an initial period of star formation ($\approx$ 8--12 Gyr) increasing the metallicity from [Fe/H] $\approx$ --2.5 dex to $\approx$ --1.5 dex, followed by a gradual chemical enrichment up to [Fe/H] $\approx$ --0.8 dex 3 Gyr ago. and by a younger, centrally concentrated episode of star formation.\\ 

\par Low resolution Ca II triplet (CaT) spectroscopy has provided radial velocities and metallicity estimates for an increasing number of stars since the pioneering work of \cite{Tolstoy2001} (33 stars). Later, \cite{Pont2004} added 117 stars, then \cite{Batta2008a} and \cite{Kirby2010} (the latter from medium resolution spectroscopy) studied respectively 870 and 675 stars. With this large dataset, a detailed metallicity distribution function (MDF),  including spatial information can be derived for Fornax: it shows that Fornax is more metal-rich than the other dSphs (with the exception of the Sagittarius (Sgr) dwarf), with a metallicity peak around --0.9 dex. However, like the other classical dSphs, Fornax has an old, metal-poor population and harbors very metal-poor stars. Fornax' higher (bulk) metallicity does not come as a surprise, given its larger mass and the mass-metallicity relation \citep[e.g.,][]{Revaz2009,Kirby2011b}. From CaT spectroscopy, \cite{Batt2006} also showed that the oldest population not only is more spatially extended but also has has hotter kinematics than the intermediate and younger populations.\\

\par Because it allows us to determine the chemical composition ($\alpha$, iron-peak and neutron-capture elements) of individual stars, high resolution spectroscopy puts further constraints on the processes driving the evolution of galaxies. \cite{She2003} and \cite{Tolstoy2003} were the first to analyze high resolution spectra of stars in dSphs and to interpret the results in terms of SFH and chemical evolution, their original sample contains three RGB stars in Fornax. \cite{Let2006} analyzed 9 stars belonging to three of the Fornax globular clusters and later on \cite{Let2010} studied 81 stars in the center of Fornax and reported in particular high values for [Ba/Fe] and [La/Fe] (up to $\sim$+0.7 dex, to be compared for instance to the $\sim$+0.4 dex found by \cite{Lem2012,Venn2012} in Carina or by \cite{Tolstoy2009} in Sculptor). One very metal-poor star in Fornax ([Fe/H] = --3.7 dex) belongs to the sample of \cite{Taf2010}. From medium resolution spectra, \cite{Kirby2010} derived the iron and $\alpha$ abundances of 675 Fornax stars. In a very recent paper, \cite{Hen2014} analyzed three $\alpha$ elements and found a metal-poor position ($\approx$ --1.9 dex) for the Mg ``knee'', where [Mg/Fe] turns down after the onset of SNe Ia.\\

\par In order to reproduce the enhanced [Ba/Fe] values reported by \cite{Let2010}, \cite{Tsuji2011} considered the possibility of a different initial mass function (IMF) in their chemical evolution models. The models show that the IMF lacks stars in its high mass end, above 25 M$_{\odot}$. Using three simple analytic models (''Leaky Box'', ''Pre-Enriched'', and ''Extra Gas''), \cite{Kirby2011} concluded that gas infall significantly affects the star formation (SF) of Fornax over its entire life time. \cite{Lan2003,Lan2008} proposed chemical evolution models for eight Local Group dSphs and Blue Compact dwarf Galaxies (but not Fornax). For all the dSphs, a low SF efficiency together with a high wind efficiency were required to match the models to empirical data. From chemo-dynamical Tree/SPH simulations \cite{Revaz2012} reached the opposite conclusion and ruled-out strong supernova-driven winds. Moreover, \cite{Revaz2009,Revaz2012} nicely reproduced the observational characteristics of Fornax, with SF occurring in a series of short bursts of similar amplitude that mimic a continuous SFH. The same holds for the cosmological simulations of \cite{Stark2013}. \cite{Hen2014} found from chemical evolution models that a bursty star formation (3 bursts at [0--2.6], [2.8--13.2], [13.7--14] Gyr), including a star formation efficiency varying from one burst to another, is required to reproduce a metal-poor knee compatible with the other characteristics of Fornax.\\

\par In this paper we analyze new high resolution spectroscopic measurements for 47 RGB stars $\approx$0.4 degrees offset from the centre of Fornax. Their elliptical radius is comprised between $\approx$ 18$^{\prime}$ and 36$^{\prime}$ (\cite{Irwin1995} report a value of 71$^{\prime}$$\pm$4$^{\prime}$ for the tidal radius of Fornax). Our sample is complementary to the one of \cite{Let2010}, aiming to extend the sample toward lower metallicities and to check the radial behaviour of elemental abundances. In Sect.~\ref{data} we present the data and the data reduction process. The determination of the atmospheric parameters (effective temperature T$_{\rm eff}$, surface gravity $log~g$, microturbulent velocity v$_{t}$) and of the ages of the stars is described in Sect.~\ref{atm_param}. The abundance results for individual elements are discussed in Sect.~\ref{results} and interpreted in terms of SFH and chemical evolution in Sect.~\ref{interp}.\\


\section{Sample selection, observations and data reduction}
\label{data}

\subsection{Sample selection and observations.}

\par
\begin{figure}[htp]
   \centering
   \includegraphics[angle=-90,width=\columnwidth]{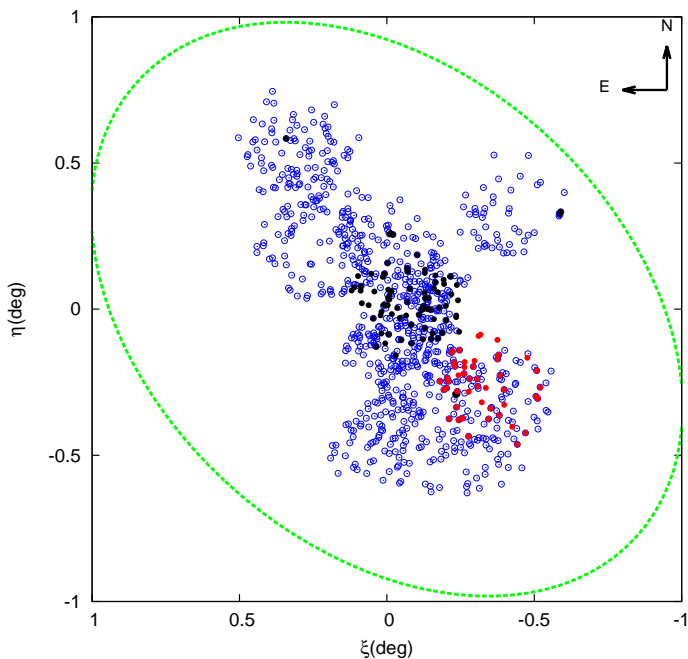}
      \caption{Spatial distribution of the spectroscopic observations across the Fornax dwarf spheroidal galaxy. The blue open circles show the stars observed in the VLT/FLAMES Ca II triplet survey \citep{Batta2008a,Stark2010}. The black dots represent the stars observed using high resolution spectroscopy \citep{Let2006,Let2010}. Our sample is shown as red dots. The green, dashed ellipse is the tidal radius of Fornax, as given by \citep{Batt2006}.}
         \label{Spatial}
   \end{figure}

 We obtained high resolution FLAMES/GIRAFFE spectra \citep{Pas2002} of individual RGB stars in the Fornax dSph Galaxy (program 080.B-0784).
Among the fields surveyed with the CaT (see Fig.~\ref{Spatial}), this specific field was selected in order to maximize the number of metal-poor stars ([Fe/H]$<$--1.5 dex) within a single 25 arcmin diameter FLAMES/GIRAFFE field. At the end we obtain only about a dozen of such stars, illustrating again the difficulty to observe the metal-poor population ($<$--1.5 dex) in Fornax, as \cite{Batt2006} already pointed out. The majority of the targets (46) were selected for membership from low resolution spectra in the CaT region \citep{Batt2006} while the remaining targets (12) were simply selected from their position on the (I, V-I) CMD \citep[][ESO/WFI]{Batt2006}. We cannot exclude some contamination of our sample from AGB stars. Observations were carried out during three consecutive nights between 4 and 6 November 2007. The observing log is listed in Table~\ref{obslog}. 

\par
\begin{figure}[htp]
   \centering
   \includegraphics[angle=90,width=\columnwidth]{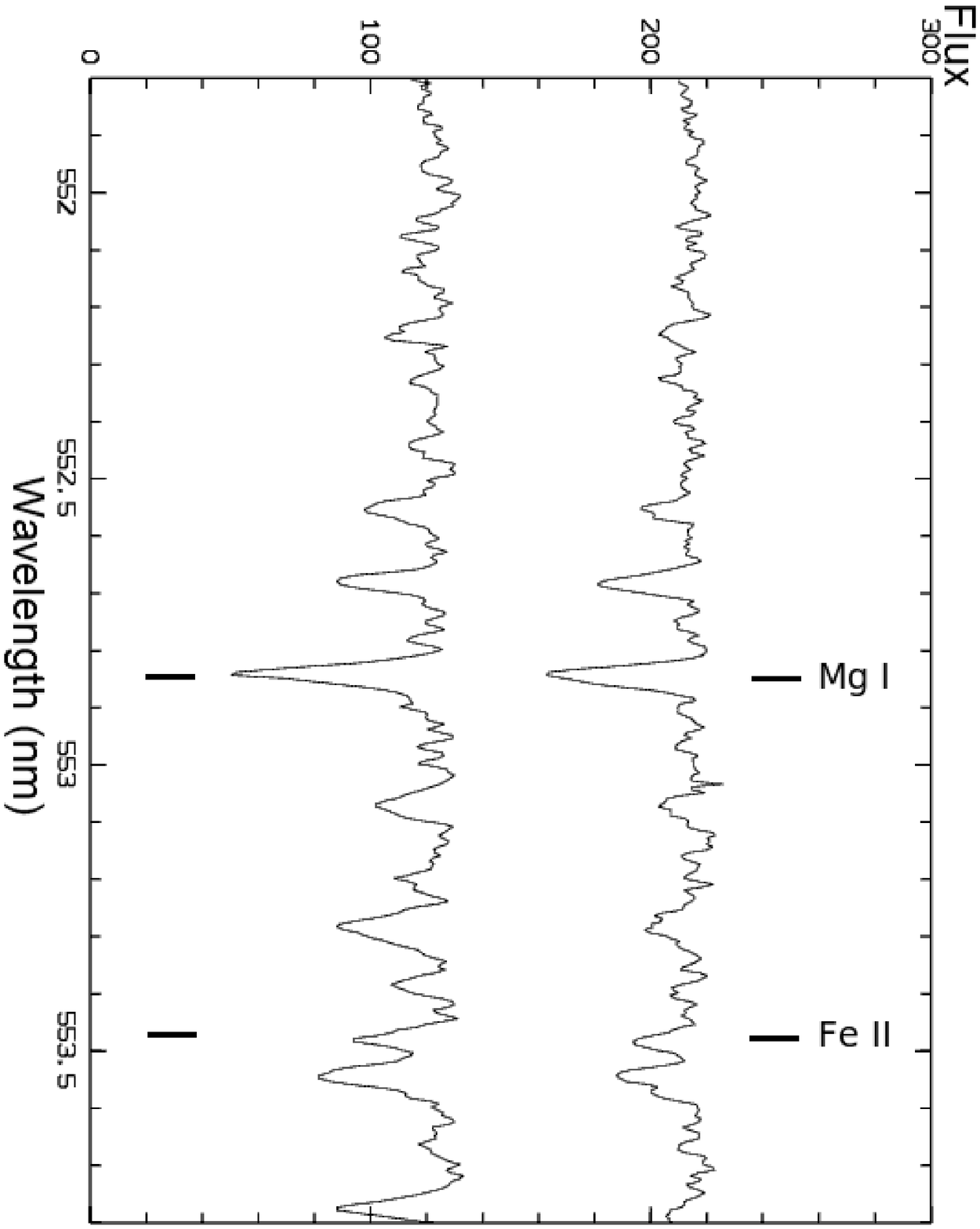}
      \caption{Representative spectra of two stars of our sample, centered on the Mg line at 552.841 nm. (Top) mem0647: S/N = 28, V = 18.604 mag; (bottom) mem0607: S/N = 45, V = 18.576 mag.} 
         \label{Spectra}
   \end{figure}

 We used GIRAFFE in the Medusa mode, having 58 out of the 132 fibres placed on targets while 19 fibres were put on blank sky positions to provide for sky subtraction of the target spectra. The limited number of sufficiently bright RGB stars in the outer regions of Fornax prevented us from increasing the number of targets. In order to perform a canonical (LTE) analysis, the equivalent widths (EW) of a sufficient number of neutral and ionized iron lines needed to be measured; the observations had therefore to provide a large enough wavelength coverage. This is achieved by observing three different wavelength settings with the HR10, HR13 and HR14 gratings whose properties are listed in Table~\ref{grat}. 
This also enabled the analysis of a good number of useful $\alpha$-element and heavy element lines. Several exposures for each HR setting were necessary to obtain a sufficient S/N, which ranges in most cases from $\approx$ 20 to $\approx$ 60 (see Table \ref{atmparam}). To illustrate the quality of our data we show in Fig.~\ref{Spectra} the spectra of two stars (mem0647, S/N=28; mem0607, S/N=45) centered on the Mg line at 552.8 nm (HR10 grating).\\

\begin{table*}[htp]
\caption{Observing log.}
\label{obslog}
\centering
\begin{tabular}{clcccccc}
\hline\hline
ESO archive observation name & Setting & Plate & Exp. time & Airmass & DIMM seeing  & DIMM seeing \\
                             &         &       &    sec    &         & at beginning &   at end    \\
\hline
GIRAF.2007-11-04T00:40:07.005.fits &  H627.3 & Medusa1 & 3600.0 & 1.55 &  0.99 &  1.01 \\ 
GIRAF.2007-11-04T01:49:01.718.fits &  H627.3 & Medusa2 & 3600.0 & 1.23 &  1.00 &  1.01 \\ 
GIRAF.2007-11-04T02:49:58.015.fits &  H627.3 & Medusa2 & 3600.0 & 1.09 &  1.01 &  1.14 \\ 
GIRAF.2007-11-04T03:51:40.705.fits &  H627.3 & Medusa2 & 3600.0 & 1.03 &  1.12 &  1.58 \\ 
GIRAF.2007-11-04T05:02:04.385.fits &  H627.3 & Medusa1 & 3600.0 & 1.02 &  1.50 &  2.12 \\ 
GIRAF.2007-11-04T06:02:57.430.fits &  H627.3 & Medusa1 & 3600.0 & 1.09 &  2.12 &  2.85 \\ 
GIRAF.2007-11-04T07:14:09.603.fits & H651.5A & Medusa2 & 4500.0 & 1.26 &  2.34 &  2.45 \\ 
GIRAF.2007-11-05T00:49:54.336.fits & H651.5A & Medusa1 & 3600.0 & 1.47 &  0.81 &  0.83 \\ 
GIRAF.2007-11-05T01:50:47.003.fits & H651.5A & Medusa1 & 3600.0 & 1.22 &  0.83 &  0.93 \\ 
GIRAF.2007-11-05T03:00:33.829.fits & H651.5A & Medusa2 & 3600.0 & 1.07 &  0.85 &  0.79 \\ 
GIRAF.2007-11-05T04:01:27.485.fits & H651.5A & Medusa2 & 3600.0 & 1.02 &  0.79 &  1.18 \\ 
GIRAF.2007-11-05T05:02:20.450.fits & H651.5A & Medusa2 & 3600.0 & 1.03 &  1.18 &  1.08 \\ 
GIRAF.2007-11-05T06:12:43.418.fits &  H548.8 & Medusa1 & 4500.0 & 1.11 &  1.14 &  1.19 \\ 
GIRAF.2007-11-05T07:28:38.583.fits &  H548.8 & Medusa1 & 5000.0 & 1.33 &  1.19 &  1.44 \\ 
GIRAF.2007-11-06T00:53:24.560.fits &  H548.8 & Medusa1 & 4500.0 & 1.43 &  0.96 &  1.23 \\ 
GIRAF.2007-11-06T02:09:20.617.fits &  H548.8 & Medusa1 & 4500.0 & 1.16 &  1.23 &  1.08 \\ 
GIRAF.2007-11-06T03:33:24.494.fits &  H548.8 & Medusa2 & 4500.0 & 1.03 &  1.40 &  1.57 \\ 
GIRAF.2007-11-06T04:49:19.582.fits &  H548.8 & Medusa2 & 4500.0 & 1.02 &  1.57 &  1.71 \\ 
GIRAF.2007-11-06T06:13:39.929.fits &  H627.3 & Medusa1 & 4500.0 & 1.12 &  1.59 &  1.19 \\ 
GIRAF.2007-11-06T07:29:33.117.fits &  H627.3 & Medusa1 & 4500.0 & 1.35 &  1.19 &  0.98 \\ 
\hline
\end{tabular}
\end{table*}

\begin{table}[htp]
\caption{FLAMES/GIRAFFE gratings used}
\label{grat}
\centering
\begin{tabular}{cccc}
\hline\hline
Grating & HR10 & HR13 & HR14 \\
\hline
$\lambda_{min}$ ($\AA$) &  5339 &  6120   &  6308   \\
$\lambda_{max}$ ($\AA$) &  5619 &  6405   &  6701   \\
Resolution at center    & 19800 & 22500   & 17740   \\
exposure time           & 7h 38min & 8h 30min & 6h 15min \\
\hline
\end{tabular}
\end{table}

\subsection{Data reduction}

We used the ESO pipeline to perform the basic data reduction, extraction and wavelength calibration of the spectra. The sky subtraction was performed via a routine \footnote{M. Irwin, see \cite{Batta2008a} for a complete description} which uses the sky-dedicated fibers to produce an average sky-spectrum. This average sky spectrum is first rescaled to match the actual sky features in each object-dedicated fiber, and then subtracted from each object spectrum. We computed the barycentric correction to radial velocity with the IRAF {\it rvcorrect} and {\it dopcor} tasks and co-added individual spectra with the IRAF {\it scombine} task. We used a flux weighted average, with asymmetric sigma clipping to remove cosmic rays.

\subsection{Membership}

\begin{figure}[htp]
   \centering
   \includegraphics[width=\columnwidth]{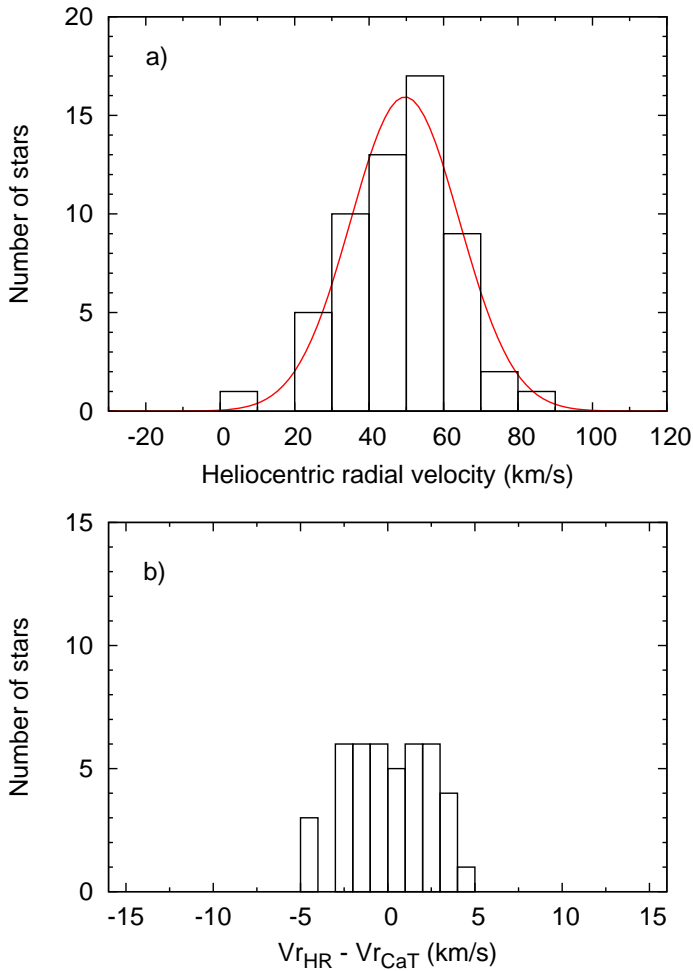}
      \caption{(a) The distribution of heliocentric radial velocities for our whole sample, including a gaussian fit which gives V$_{c}$=49.65 km s$^{-1}$$\pm$1.46 km s$^{-1}$. (b) Comparison between radial velocities derived from our HR data and from CaT \citep{Batta2008a}.}
         \label{radvel}
   \end{figure}

We derived the radial velocities (and the EW of the absorption lines) with DAOSPEC \citep{StetPan2008}: all the lines detected by DAOSPEC are cross-correlated by the software with a list of lines provided by the user. For this sample, the accuracy is in general better than $\pm$1 km s$^{-1}$, with a median error on the individual velocities measurement of 0.64 km s$^{-1}$ (see Table \ref{targets}). The radial velocity distribution of our sample is shown in Fig.~\ref{radvel}a. From a Gaussian fit to the data, we found a systemic radial velocity peak of 49.65 km~s$^{-1}$ for the Fornax dSph, with a dispersion of $\sigma$=14.60 km~s$^{-1}$. Computing the straight average and standard deviation would give V$_{r}$=48.78~km~s$^{-1}$, $\sigma$=14.5~km~s$^{-1}$ 

\par When we compare our radial velocities to those derived from the DART low resolution CaT sample \citep{Batta2008a}, we find that there are no systematic differences (for the stars in common) between the two measurements and that the values agree very well within their error bars (see Fig.~\ref{radvel}b). Under the restrictive assumption that the likely Fornax members fall within 2$\sigma$ from the mean radial velocity (i.e. in the [20.45~km~s$^{-1}$--78.85~km~s$^{-1}$] velocity range), we found only one star inconsistent with membership of the Fornax dSph (rgb0614 at 3.10$\sigma$) and two stars only marginally consistent as they fall just above our threshold value (mem0538 at 2.05$\sigma$ and rgb0509 at 2.35$\sigma$). The former star was excluded from further analysis while we kept the two latter ones.\\ 

\par The value of our radial velocity peak (49.65$\pm$1.46 km s$^{-1}$, $\sigma$=14.6$\pm$1.51 km s$^{-1}$) is a bit lower than previous determinations. For instance, \cite{Batt2006} report a value of 54.1$\pm$0.5 km s$^{-1}$ ($\sigma$=11.4$\pm$0.4 km s$^{-1}$) from CaT measurements over the whole galaxy, \cite{Walk2009} derived a value of 55.2$\pm$0.1 km s$^{-1}$ ($\sigma$=11.7$\pm$0.9 km s$^{-1}$) from magnesium triplet measurements over the whole galaxy and \cite{Let2010} found 55.9 km s$^{-1}$ ($\sigma$=14.2 km s$^{-1}$) from high resolution spectra in the center of Fornax. In the CaT sample of \cite{Batt2006}, we selected according to the membership criterion quoted above the 129 RGB stars spatially overlapping our HR sample. For these stars, we found a mean radial velocity of 50.99$\pm$0.94 km s$^{-1}$ ($\sigma$=12.70$\pm$0.96 km s$^{-1}$), indicating that our slightly lower radial velocities are typical from this region of Fornax and suggesting that the kinematics of Fornax 
is dependent on position.\\ 

\par A fraction (10 stars) of our member candidates have been further discarded as we were unable to fit a proper stellar atmosphere model to these stars. As the S/N of their spectra is sufficient, they could be either AGB stars in Fornax or foreground stars: as we will see in Sect.~\ref{subsec_atmparam}, T$_{\rm eff}$ and $log~g$ are derived from the photometry and computed for stars located at the distance of the Fornax dSph. Our computed atmospheric parameters are therefore not suitable if they lie in the foreground of Fornax. For seven stars, spectra in the CaT region are available and five of them (mem0528, mem0571, mem0610, mem0621, mem0738) are classified as foreground contaminants according to the $\lambda$=8806.8 \AA{} MgI line criterion defined by \cite{Batt2012}. There is no indication that the 2 other stars are CH stars. The stars that have been discarded have nominal metallicity estimates from CaT between --1.4 and --1.7 dex, except one star with [Fe/H]$_{CaT}$=--2.55 dex. Three other stars have no CaT estimates of the metallicity but are clearly metal-rich and therefore likely foreground stars.\\ 

\par
\begin{figure}[htp]
   \centering
   \includegraphics[angle=-90,width=\columnwidth]{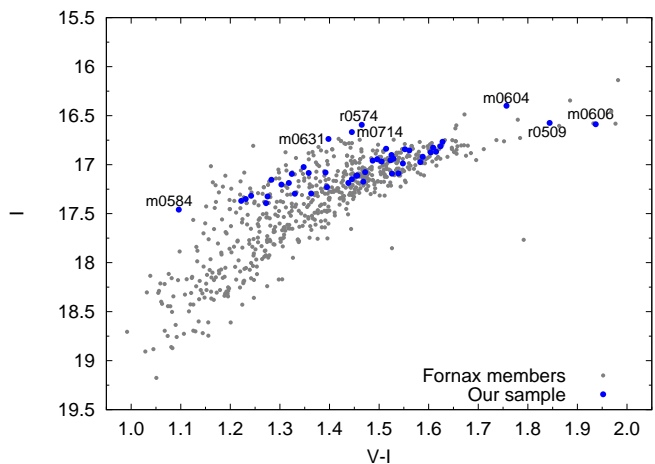}
      \caption{I vs. (V-I) CMD: our FLAMES/GIRAFFE sample is shown in blue dots and other Fornax members (from CaT) are shown in grey dots. Some stars located close to the tip or at the blue end of the RGB are labelled. CTIO VI photometry was provided by \cite{deBoer2012b}. Data in I band are not available beyond an elliptical radius of 0.4 deg and were supplemented by ESO/WFI from \cite{Batta2008a}}. 
         \label{CMD}
   \end{figure}

 Our sample is then reduced to 47 RGB stars which are likely members of the Fornax dSph and for which we could perform a complete abundance analysis. They are overlaid on the V vs. (V-I) CMD in Fig.~\ref{CMD}. For 37 of them, metallicity estimates from CaT are available. Target coordinates, photometry and radial velocities are listed in Table~\ref{targets}. 

\section{Determination of stellar parameters and abundances}
\label{atm_param}

\subsection{Atmospheric parameters}
\label{subsec_atmparam}

\par As we mentioned above, we measured the equivalent widths using DAOSPEC \citep{StetPan2008}, a software optimized for the GIRAFFE HR spectra. At the resolution of our data, the line profile is dominated by the instrumental effects rather than by astrophysical processes, and DAOSPEC fits lines with saturated Gaussians. Lines with EW$>$250~m\AA~have not been considered as they will most likely depart from a Gaussian profile \citep[e.g.,][]{Saka2011,Venn2012}, and a large majority of the measured lines have EW$<$200~m\AA. The EWs determined using DAOSPEC are listed in Table~\ref{EW}.\\

\par Our linelist is similar to the one used in the previous DART \citep[\textbf{D}warf \textbf{A}bundances and \textbf{R}adial velocity \textbf{T}eam,][]{Tolstoy2006} papers based on GIRAFFE medium resolution spectroscopy \citep{Let2006,Let2010,Lem2012}. It is based on the line list of \cite{She2003}, supplemented with lines taken in \cite{Pomp2008}.\\

\par We used OSMARCS atmosphere models in spherical symmetry \citep{Gus2008}: the grid of atmosphere models covers the following range of parameters: T$_{\rm eff}$ = [3600~\textendash~5500] K, $log~g$ = [0.0~\textendash~3.5] dex, [Fe/H] = [--3.0~\textendash~+1.0] dex with [$\alpha$/Fe] increasing when [Fe/H] decreases (different levels of $\alpha$-enhancements ranging from 0.0 to +0.4 dex are considered for different metallicity bins).

Abundances are computed with Calrai, a LTE spectrum synthesis code originally developed by \cite{Spite1967} and regularly updated since then. The abundances are first derived for each individual line. The uncertainties on the DAOSPEC EW measurements are propagated into uncertainties on individual abundances. In most cases we could measure several lines for a given element; the uncertainties on the individual abundances (per line) are further propagated into error estimates on the abundances for the given element by weighting each line by 1/($\sigma^{2}$).\\

\par Stellar temperatures T$_{\rm eff}$ and surface gravities $log~g$ are determined from photometric data in the BV(I)JHK bands. We have BV photometry from CTIO/MOSAIC \citep{deBoer2012a} for our whole sample while the CTIO I band is available only for a small subset of our sample (the I-band coverage is complete only out to an elliptical radius of 0.4 deg, while the B,V photometry extends further out till an elliptical radius of 0.8 deg). We also have ESO/VISTA JHK magnitudes for our whole sample. The photometry is listed in Table~\ref{targets}.\\

\par We adopted as photometric temperature the average T$_{\rm eff}$ derived from the five different colors (B-V), (V-I), (V-J), (V-H) and (V-K), following the calibration for giants from \cite{Rami2005}. The temperatures from different colors are in very good agreement (see Table~\ref{atmparam_all}) and indeed they fall in most cases within 100~K. This error is larger than that propagated error on Teff based on the color errors which are of the order of 0.01 mag and translate into errors of $\approx$ 10~K on the individual temperatures derived from each color. We adopted the \cite{Schle1998} reddening laws with an extinction of 0.03 mag \citep[also from][]{Schle1998}. T$_{\rm eff}$ were first evaluated with the metallicities derived from the CaT when available (assuming [Fe/H] = --1.5 dex when it was not the case). In an iterative process, they were later updated with the final [Fe/H] value of our targets.\\

\par RGB stars in our sample have T$_{\rm eff}$ ranging from 3640 to 4370~K, with a large fraction of them having T$_{\rm eff}$~$<$~4000~K. Four stars have T$_{\rm eff}$~$<$~3800~K. Molecular features (mostly CN at these wavelengths and for this range of stellar parameters) could have an impact on our results. Atomic lines with possible blends with molecular lines have been removed from the linelist, using synthetic spectra \citep{Let2010}. 
As molecular features are stronger when T$_{\rm eff}$ decreases, they could still affect the results for the coolest stars in our sample. Therefore, we checked that for a given [Fe/H] (namely, --1.0$<$[Fe/H]$<$--0.5 dex), the abundances ratios derived are largely independent from T$_{\rm eff}$, at the possible exception of [Na/H] (see Fig.~\ref{abvsTeff}). Please note that even on this restricted [Fe/H] range, the chemical evolution of Fornax is still embedded in the plots. We also checked that our method enables us to recover the chemical composition of Arcturus, a Milky Way RGB star in the same metallicity range: a comparison with \cite{vdS2013} and references therein indicates that our results (based on the EW method) are identical when they also used the EW method and in good agreement when they used the spectral synthesis method (See Table~\ref{Arcturus}), despite the different tools ({\it turbospectrum} vs {\it fitline}). the different linelist and different wavelength coverage (they used the HR11 FLAMES/GIRAFFE setting while we used HR10). Finally, we analyzed two stars with T$_{\rm eff}$~$<$~4000~K in the GAIA benchmark sample ($\alpha$~Cet and $\gamma$~Sge). From NARVAL spectra downgraded to a resolution of 20~000 \citep{Blanco2014} and considering the spectral ranges covered by the FLAMES/GIRAFFE HR10/HR13/HR14 gratings, we were able to recover atmospheric parameters and metallicities very similar to those reported by \cite{Jofre2013}. Results are given in Table \ref{GAIAbenchmark}.\\

\par Distance modulus is an important parameter to be considered when deriving the bolometric magnitude and, in turn, the surface gravity from photometry. The most recent determination of the Fornax distance modulus \citep{Pietr2009} was derived from the J \& K near-infrared (NIR) magnitudes of the tip of the red giant branch (TRGB), in good agreement with the NIR values derived by \cite{Gull2007} from TRGB (20.75$\pm$0.19 mag) or red clump (RC) stars (20.74$\pm$0.11 mag). From K-band magnitudes of RC stars, \cite{Pietr2003} previously reported a distance modulus of 20.858$\pm$0.013 mag. These results are also in good agreement with the luminosity of TRGB and horizontal branch (HB) stars in the optical bands. \cite{Savi2000} report a distance modulus of 20.70$\pm$0.12 mag and \cite{Rizzi2007} a value of 20.71$\pm$0.07 mag from the TRGB method while \cite{Bersier2000} found a slightly smaller value of 20.65$\pm$0.11 mag. For HB stars, the results of \cite{Buo1999}, \cite{Savi2000}, \cite{Rizzi2007} are also in good agreement at respectively 20.76$\pm$0.10 mag, 20.76$\pm$0.04 mag, 20.72$\pm$0.06 mag. The same outcome applies to the distances derived from RC stars as \cite{Rizzi2007} found a distance modulus of 20.73$\pm$0.09 mag. Finally, \cite{Greco2005} also derived a similar value (20.72$\pm$0.10 mag) from field RR-Lyrae stars. It is worth mentioning that the quoted studies made use of different E(B-V) values ranging from 0.03 to 0.05 mag.\\

\par Using our temperature estimates, a distance modulus of $\mu_{v}$=20.84$\pm$0.04 mag \citep{Pietr2009} and the bolometric correction from \cite{Alonso1999}, we computed the surface gravity according to the standard formula:
\begin{equation*}
 log~g_{\star} = log~g_{\odot} + log~\dfrac{\mathcal{M}_{\star}}{\mathcal{M}_{\odot}} + 4 \times log~\dfrac{T_{\rm eff\star}}{T_{\rm eff\odot}} + 0.4 \times (M_{Bol\star} - M_{Bol\odot})
\end{equation*}

\par Following the BaSTI isochrones prescriptions \citep{Piet2004}, we assumed a stellar mass of 0.9 M$_{\sun}$ for the old, metal-poor stars ([Fe/H]$<$--1.8 dex), and 1.3 M$_{\sun}$ for the youngest, metal-rich population ([Fe/H]$>$--0.7 dex). The stars with --1.8 dex$<$~[Fe/H]~$<$--0.7 dex were assigned a stellar mass of 1.1 M$_{\sun}$. Switching from a 0.9 M$_{\sun}$ to an 1.3 M$_{\sun}$ for the stellar mass would increase $log~g$ by 0.15--0.20 dex with no impact on T$_{\rm eff}$. We mentioned that some stars in our sample might be AGB stars. Given the mass of a white dwarf, the lowest possible mass for an AGB star would be 0.6 M$_{\sun}$ (but the bulk of the population would probably have masses similar to the ones used in this study). Assigning a mass of 0.6 M$_{\sun}$ to all the stars in our sample (which currently have been assigned masses of 0.9, 1.1 or 1.3 M$_{\sun}$, depending on their metallicity) would decrease $log~g$ by 0.26$\pm$0.1 dex and have a limited influence ($\lesssim$0.1 dex) on 
neutral species but a noticeable influence on the abundances of ionized elements ($\approx$0.15 dex).\\ 

\begin{table*}[ht!]
\centering
\caption{The atmospheric parameters T$_{\rm eff}$, $log~g$, V$_{t}$ and [Fe/H] for stars in our sample. In this table we also list the age of the star and the S/N for each setting.}
\begin{tabular}{ccccccccc}
\hline\hline
 Target & T$_{\rm eff}$ & {\it log g} & V$_{t}$ & [Fe/H] & Age & S/N (H10) & S/N (H13) & S/N (H14) \\ 
        &       K       &     dex     &  km/s   &   dex  & Gyr &           &           &           \\ 
 \hline
Fnx-mem0514 & 3830 & 0.40 & 1.70 & -0.70 &  5.30$\pm$1.44 & 28 & 38 & 22 \\ 
Fnx-mem0522 & 3825 & 0.45 & 1.70 & -0.90 &  6.17$\pm$2.33 & 24 & 24 & 20 \\ 
Fnx-mem0532 & 4000 & 0.75 & 1.50 & -0.50 &  2.96$\pm$0.89 & 29 & 23 & 27 \\ 
Fnx-mem0539 & 3975 & 0.60 & 1.90 & -0.70 &  5.50$\pm$0.93 & 29 & 43 & 16 \\ 
Fnx-mem0543 & 3900 & 0.50 & 2.00 & -1.00 &  8.45$\pm$2.96 & 23 & 38 & 47 \\ 
Fnx-mem0556 & 3970 & 0.65 & 1.60 & -0.65 &  4.36$\pm$0.86 & 28 & 35 & 34 \\ 
Fnx-mem0572 & 3960 & 0.65 & 1.80 & -0.90 &  6.62$\pm$2.25 & 28 & 34 & 25 \\ 
Fnx-mem0574 & 3830 & 0.45 & 1.90 & -0.80 &  6.28$\pm$2.52 & 36 & 45 & 37 \\ 
Fnx-mem0584 & 4360 & 0.85 & 1.60 & -2.20 & 12.51$\pm$1.08 & 25 & 34 & 32 \\ 
Fnx-mem0595 & 3890 & 0.50 & 1.90 & -0.90 &  5.75$\pm$1.78 & 27 & 38 & 27 \\ 
Fnx-mem0598 & 3755 & 0.35 & 2.20 & -0.85 &  $>$12.25      & 42 & 48 & 70 \\ 
Fnx-mem0604 & 3730 & 0.20 & 1.90 & -0.75 &  $>$5.75       & 39 & 52 & 37 \\ 
Fnx-mem0606 & 3640 & 0.25 & 1.90 & -0.65 &  $>$5.17       & 48 & 69 & 64 \\ 
Fnx-mem0607 & 3915 & 0.55 & 1.90 & -0.95 &  6.58$\pm$2.40 & 45 & 50 & 43 \\ 
Fnx-mem0612 & 4170 & 0.80 & 1.40 & -1.70 &  9.33$\pm$2.19 & 18 & 30 & 27 \\ 
Fnx-mem0613 & 3855 & 0.50 & 1.50 & -0.45 &  1.71$\pm$0.67 & 35 & 34 & 27 \\ 
Fnx-mem0626 & 3980 & 0.65 & 1.90 & -1.00 &  6.73$\pm$1.78 & 29 & 44 & 48 \\ 
Fnx-mem0629 & 3950 & 0.60 & 2.00 & -1.00 &  7.47$\pm$1.77 & 25 & 23 & 26 \\ 
Fnx-mem0631 & 4035 & 0.55 & 1.90 & -0.85 &  2.01$\pm$0.37 & 47 & 40 & 48 \\ 
Fnx-mem0633 & 3835 & 0.45 & 1.90 & -0.95 &  8.41$\pm$3.16 & 40 & 43 & 67 \\ 
Fnx-mem0634 & 3980 & 0.75 & 1.30 & -0.30 &  1.58$\pm$0.58 & 27 & 55 & 21 \\ 
Fnx-mem0638 & 3835 & 0.50 & 1.70 & -0.70 &  5.46$\pm$1.47 & 40 & 36 & 69 \\ 
Fnx-mem0647 & 4070 & 0.70 & 1.80 & -1.45 &  9.13$\pm$1.73 & 28 & 50 & 39 \\ 
Fnx-mem0654 & 4260 & 0.80 & 1.80 & -1.90 & 12.99$\pm$0.47 & 37 & 30 & 59 \\ 
Fnx-mem0664 & 4330 & 0.85 & 1.70 & -2.30 &  $>$13.92      & 28 & 37 & 53 \\ 
Fnx-mem0675 & 4150 & 0.85 & 1.70 & -0.80 &  9.22$\pm$3.45 & 28 & 47 & 42 \\ 
Fnx-mem0678 & 3870 & 0.60 & 1.80 & -0.70 &  6.75$\pm$2.00 & 25 & 34 & 35 \\ 
Fnx-mem0682 & 3845 & 0.50 & 1.90 & -0.80 &  7.57$\pm$2.62 & 44 & 51 & 39 \\ 
Fnx-mem0704 & 4330 & 0.85 & 1.90 & -2.55 &  $>$13.75      & 54 & 48 & 42 \\ 
Fnx-mem0712 & 4230 & 0.75 & 1.80 & -2.10 & 13.15$\pm$0.30 & 33 & 23 & 20 \\ 
Fnx-mem0714 & 4090 & 0.50 & 1.70 & -1.80 & 11.32$\pm$1.99 & 31 & 53 & 22 \\ 
Fnx-mem0715 & 3945 & 0.65 & 2.00 & -1.00 &  7.87$\pm$2.74 & 34 & 59 & 35 \\ 
Fnx-mem0717 & 3875 & 0.55 & 1.40 & -0.40 &  $>$3.08       & 24 & 37 & 19 \\ 
Fnx-mem0732 & 4080 & 0.60 & 2.10 & -2.40 &  $>$13.75      & 53 & 48 & 57 \\ 
Fnx-mem0747 & 4080 & 0.75 & 1.80 & -1.40 &  8.28$\pm$1.44 & 38 & 31 & 28 \\ 
Fnx-mem0754 & 3850 & 0.45 & 1.50 & -0.75 &  5.66$\pm$1.23 & 26 & 42 & 39 \\ 
Fnx-mem0779 & 4235 & 0.70 & 1.30 & -2.10 & 12.88$\pm$0.51 & 28 & 44 & 35 \\ 
Fnx-rgb0507 & 4065 & 0.65 & 1.60 & -0.75 &  3.54$\pm$0.90 & 24 & 21 & 21 \\ 
Fnx-rgb0509 & 3740 & 0.30 & 2.20 & -1.15 &  8.21$\pm$2.38 & 18 & 41 & 21 \\ 
Fnx-rgb0522 & 4220 & 0.70 & 1.60 & -1.85 & 11.14$\pm$1.82 & 40 & 51 & 24 \\ 
Fnx-rgb0539 & 3885 & 0.55 & 1.60 & -0.60 &  3.99$\pm$0.74 & 19 & 24 & 23 \\ 
Fnx-rgb0553 & 3855 & 0.60 & 1.60 & -0.50 &  4.63$\pm$1.22 & 16 & 23 & 08 \\ 
Fnx-rgb0556 & 3915 & 0.65 & 1.70 & -0.55 &  4.39$\pm$0.83 & 39 & 35 & 47 \\ 
Fnx-rgb0561 & 3940 & 0.55 & 1.80 & -0.95 &  5.73$\pm$1.61 & 40 & 51 & 44 \\ 
Fnx-rgb0574 & 3950 & 0.40 & 2.00 & -0.95 &  3.37$\pm$0.81 & 62 & 62 & 54 \\ 
Fnx-rgb0590 & 4215 & 0.90 & 1.80 & -1.50 &  9.65$\pm$2.38 & 19 & 52 & 47 \\ 
Fnx-rgb0596 & 4367 & 0.87 & 1.40 & -2.68 &  $>$13.67      & 37 & 40 & 40 \\ 
\hline
\end{tabular}
\label{atmparam}
\end{table*}

\par
\begin{figure}[htp]
   \centering
   \includegraphics[angle=-90,width=\columnwidth]{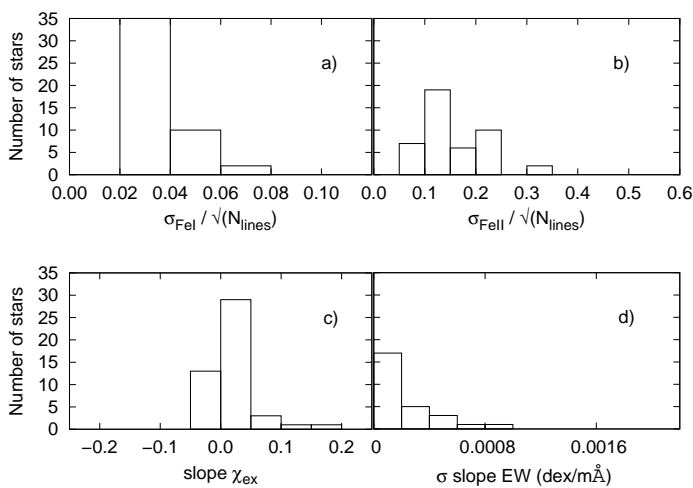}
      \caption{The distribution of errors on the stellar atmospheric diagnostics for the stars in our sample: (a) The error on the mean [FeI/H] and (b) [FeII/H] which are measured from the dispersion around the mean divided by the square root of the number of lines measured. (c) The slope of [FeI/H] versus the excitation potential $\chi_{ex}$. (d) The propagated error on the measurement of the slope of [FeI/H] versus the line strength.
              }
         \label{disp}
   \end{figure}

 An alternative approach to determine surface gravity $\log~g$ is by imposing the ionization balance between FeI and FeII. Our FeII linelist contains 7 lines, 5 of which are in common with \cite{Let2010}. Four of the lines in common with \cite{Let2010} are listed among the most reliable FeII lines for high resolution analysis of FGK giants by \cite{Jofre2013} (see their Table 5). Two other lines are not analyzed in the quoted study. The remaining line (at 6247.56 \AA) is recommended only for K dwarfs by \cite{Jofre2013}, however including it or not has a negligible influence on our final [FeII/H] values. We also followed the prescriptions of that paper regarding the more metal-poor stars, which in practice consisted in discarding the 5414.08 \AA{} line for the 2 metal-poor stars of our sample in which it could be measured.\\
We could then measure only a maximum of 7 FeII lines in our spectra, but this number dropped down to 3--4 lines in most cases. As a result the average errors on [FeII/H] reach $\pm$0.25 dex (see Fig.~\ref{disp}b), to be compared to the errors on [FeI/H] (Fig.~\ref{disp}a) that do not exceed $\pm$0.08 dex. This uncertainty on [FeII/H] translates into an uncertainty on $log~g$ up to 0.3 dex and the ionization balance cannot be accurately determined. The situation is even worse in the case of Ti, for which we could measure even fewer lines, 1--9 TiI lines and 1--3 TiII lines. Instead of relying on a rather uncertain ionization balance to derive individual spectroscopic gravities, we decided to stick to a homogeneous photometric $log~g$ scale. Surface gravity has a minor effect on abundances derived from neutral ions as a (large) variation of 0.5 in $log~g$ translates in a variation $\lesssim$0.1 dex in the abundances of neutral species (see Table~\ref{atm_errors}). On the other hand the impact on the ionized elements is noticeable as it reaches $\approx$ 0.20 dex. We also checked the photometric T$_{\rm eff}$ values by ensuring that [FeI/H] does not depend on the excitation potential $\chi_{ex}$ as can be seen in Fig.~\ref{disp}c showing that the slopes of [FeI/H] versus $\chi_{ex}$ are minimal.\\

\par The microturbulent velocity v$_{t}$ is determined via an iterative process until the requirement that [FeI/H] does not depend on EW is met. In practice, one minimizes the slope between [FeI/H] and EW. Varying v$_{t}$ around its correct value modifies the value of slope and it is found empirically that v$_{t}$ varies linearly and symmetrically as a function of the slope. This enables us to estimate the uncertainty on the slope between [FeI/H] and EW. With an error on the slope $\leq\pm$0.0008 dex/m\AA~(see Fig.~\ref{disp}d), the uncertainty on microturbulent velocity does not exceed 0.2 km/s. Following \cite{Mag1984}, we have used the theoretical EWs (computed from the atomic parameters of the line and the atmospheric parameters of the star) rather than the measured values in order to avoid systematic biases on v$_{t}$ caused by random errors on the EW measurements. We tested the dependence on the cut-off value of the equivalent width by only considering lines with EW$<$180 m\AA. This has no impact on T$_{\rm eff}$ as it is derived from photometry and a very limited influence ($\lesssim$0.1 km/s) on v$_{t}$ and hence on the chemical composition. The stellar parameters for the stars in our sample can be found in Table~\ref{atmparam}.\\

\subsection{Error budget}
\label{secErrBudget}

\par DAOSPEC returns an error estimate on the EW measurements ($\delta$$_{dao}$) that was propagated throughout the abundance determination. However the uncertainties on the atomic parameters of the lines should be taken into account in addition to the uncertainties on the EW measurements. Therefore we also used the error on the mean value of the abundance, defined for a given element X as $\frac{\sigma(X)}{\sqrt{N_X}}$ where N$_{X}$ is the number of lines measured for this element. For those of the elements for which abundances are determined from a limited number of lines, the abundance dispersion may be underestimated. To avoid the problem of small number statistics, we did not allow that any abundance can be measured more accurately than [FeI/H] by setting the [FeI/H] dispersion as a lower limit, resulting in a lower limit for the error estimate of $\frac{\sigma(FeI)}{\sqrt{N_{X}}}$. Finally, we adopted as the error on [X/H] the maximum of these three values ($\delta$$_{dao}$, $\frac{\sigma(X)}{\sqrt{N_X}}$, $\frac{\sigma(FeI)}{\sqrt{N_{X}}}$). It includes all the errors due to measurements. We note that the error on the mean value ($\frac{\sigma(X)}{\sqrt{N_X}}$,  $\frac{\sigma(FeI)}{\sqrt{N_{X}}}$) is always larger than $\delta$$_{dao}$ as it already includes a contribution from the measurement errors. The errors on the abundance ratios [X/Fe] were subsequently computed as the quadratic sum of the errors on [X/H] and [Fe/H]. Averaged over the whole sample, they lead to representative error bars shown on the figures.\\

\par To assess the impact of uncertainties on the atmospheric parameters on the final abundance results, we selected two stars, mem-0607 and mem-0647. A good number of elements could be measured in these stars and their metallicities are representative of the bulk of our sample. For these two stars, we computed the variations of the abundances within conservative error bars of respectively $\Delta$T$_{\rm eff}$=$\pm$150 K, $\Delta$$log g$=$\pm$0.5 dex, $\Delta$v$_{t}$=$\pm$0.3 km s$^{-1}$. Their sum in quadrature would give an overestimated value for the total error, as this method by construction ignores covariances between the different atmospheric parameters \citep[e.g.,][]{McWil1995,Johnson2002}. Therefore we followed the prescriptions from \cite{Cayrel2004}, also adopted by e.g., \cite{vdS2013}: for our two representative stars, we changed T$_{\rm eff}$ by its estimated error ($\pm$ 150 K), derived the other atmospheric parameters ($log~g$, v$_{t}$ and [Fe/H]) corresponding to these new temperatures and compared them to the nominal values. We adopted as the error on [X/Fe] due to the sensitivity on the abundance parameters:\\

$ \begin{array}{rcr}
  \sigma_{param} & = & MAX \bigl[|[X/Fe]_{T+150K}-[X/Fe]_{nominal}|, \\
                 &   & |[X/Fe]_{T-150K}-[X/Fe]_{nominal}|\bigl]\\
  \end{array}$\\

\noindent and as the total error:\\

$\sigma_{total}$ = $\sqrt{(\sigma_{meas}^{2})+(\sigma_{param}^{2})}$\\

The uncertainties on the individual abundances and abundance ratios due to the determination of atmospheric parameters are listed respectively in Table~\ref{atm_errors} and table \ref{atm_errors_XFe}. 

\begin{table*}[ht!]
\centering
\caption{Errors on abundances due to sensitivity on the stellar parameters, computed for mem0607 and mem0647.}
\begin{tabular}{r|rrrrrr|rrrrrr}
\hline\hline
   & \multicolumn{6}{c|}{mem0607} & \multicolumn{6}{c}{mem0647} \\
\hline
 & \thead{$\Delta$T$_{\rm eff}$=} & \thead{$\Delta$T$_{\rm eff}$=} & \thead{$\Delta$$log g$=} & \thead{$\Delta$$log g$=} & \thead{$\Delta$v$_{t}$=} & \multicolumn{1}{c|}{$\Delta$v$_{t}$=} & \thead{$\Delta$T$_{\rm eff}$=} & \thead{$\Delta$T$_{\rm eff}$=} & \thead{$\Delta$$log g$=} & \thead{$\Delta$$log g$=} & \thead{$\Delta$v$_{t}$=} & \thead{$\Delta$v$_{t}$=} \\
 \multicolumn{1}{c|}{Element}       &  \thead{--150} & \thead{+150} & \thead{--0.5} & \thead{+0.5} & \thead{--0.3} & \multicolumn{1}{c|}{+0.3} & \thead{--150} & \thead{+150} & \thead{--0.5} & \thead{+0.5} & \thead{--0.3} & \thead{+0.3} \\
        &      \thead{K}        &     \thead{K}         &   \thead{dex}  &   \thead{dex}   & \thead{km s$^{-1}$} & \multicolumn{1}{c|}{km s$^{-1}$} &     \thead{K}        &     \thead{K}         &   \thead{dex}  &   \thead{dex}   & \thead{km s$^{-1}$} & \thead{km s$^{-1}$} \\
\hline
$\Delta$[FeI/H]  & -0.07 &  0.02 & -0.17 & 0.12 & 0.13 & -0.10 &  0.11 &  0.23 &  0.06 &  0.09 &  0.29 & -0.18 \\{}
$\Delta$[FeII/H] &  0.26 & -0.22 & -0.26 & 0.29 & 0.08 & -0.06 &  0.24 & -0.26 & -0.24 &  0.27 &  0.08 & -0.07 \\{}
$\Delta$[NaI/H]  & -0.14 &  0.15 &  0.00 & 0.01 & 0.01 & -0.01 &       &       &       &       &       &       \\{}
$\Delta$[MgI/H]  &  0.01 &  0.01 & -0.05 & 0.06 & 0.01 & -0.02 & -0.07 &  0.08 &  0.04 & -0.06 &  0.16 & -0.16 \\{}
$\Delta$[SiI/H]  &  0.15 & -0.12 & -0.10 & 0.13 & 0.04 & -0.02 &       &       &       &       &       &       \\{}
$\Delta$[CaI/H]  & -0.10 &  0.18 &  0.04 & 0.16 & 0.17 & -0.06 & -0.20 &  0.19 & -0.03 &  0.01 &  0.11 & -0.07 \\{}
$\Delta$[ScII/H] & -0.01 &  0.01 & -0.20 & 0.21 & 0.03 & -0.02 &  0.00 & -0.04 & -0.20 &  0.21 &  0.03 & -0.03 \\{}
$\Delta$[TiI/H]  & -0.26 &  0.27 & -0.09 & 0.06 & 0.09 & -0.07 & -0.32 &  0.27 & -0.10 &  0.04 &  0.03 & -0.05 \\{}
$\Delta$[TiII/H] &  0.03 & -0.02 & -0.20 & 0.22 & 0.06 & -0.06 &  0.02 & -0.08 & -0.21 &  0.20 &  0.04 & -0.06 \\{}
$\Delta$[CrI/H]  & -0.25 &  0.27 & -0.13 & 0.11 & 0.22 & -0.16 & -0.29 &  0.27 & -0.04 &  0.00 &  0.27 & -0.31 \\{}
$\Delta$[NiI/H]  & -0.01 &  0.03 & -0.14 & 0.18 & 0.10 & -0.08 & -0.05 &  0.08 & -0.13 &  0.15 &  0.04 & -0.08 \\{}
$\Delta$[ YII/H] &       &       &       &      &      &       &       &       &       &       &       &       \\{}
$\Delta$[BaII/H] &       &       &       &      &      &       & -0.08 &  0.02 & -0.20 &  0.18 &  0.31 & -0.31 \\{}
$\Delta$[NdII/H] & -0.05 &  0.05 & -0.18 & 0.20 & 0.09 & -0.05 &       &       &       &       &       &       \\{}
$\Delta$[LaII/H] & -0.11 &  0.10 & -0.22 & 0.21 & 0.12 & -0.08 &       &       &       &       &       &       \\{}
$\Delta$[EuII/H] & -0.04 &  0.02 & -0.21 & 0.21 & 0.10 & -0.07 & -0.02 & -0.03 & -0.20 &  0.21 &  0.04 & -0.02 \\
\hline
\end{tabular}
\label{atm_errors}
\end{table*}

\begin{table*}[ht!]
\centering
\caption{
Typical measurement errors ($\sigma$$_{meas}$) on [X/Fe] computed for the metal-rich ([Fe/H]$>$--1.4 dex, col. 2) and metal-poor ([Fe/H]$<$--1.4 dex, col. 6) subsamples. The final error on [X/Fe] (col. 5, col. 9) is derived in combining the measurement errors with the errors due to the sensitivity on the stellar parameters ($\sigma$$_{meas}$, cols. 3-4, 7-8). $\sigma$$_{param}$ is given as the maximum of ([X/Fe]$_{\sigma(param,~T+150K)}$--[X/Fe]$_{nominal}$) and ([X/Fe]$_{\sigma(param,~T-150K)}$--[X/Fe]$_{nominal}$).
}
\begin{tabular}{r|rccr|rccr}
\hline\hline
                & \thead{$\sigma_{meas}$} & $\sigma_{param~(T+150K)}$ & $\sigma_{param~ (T-150K)}$ & $\sigma_{total}$ & $\sigma_{meas}$ & $\sigma_{param~ (T+150K)}$ & $\sigma_{param~ (T-150K)}$ & $\sigma_{total}$ \\
Elemental ratio &       (MR)      &     mem0607               &     mem0607               &                  &       (MP)      &     mem0647                                  &     mem0647               &                 \\
\hline
[FeI/H]   & 0.04 &  0.03 & -0.08 & 0.09 & 0.05 &  0.26 & -0.04 & 0.26 \\{}
[FeII/H]  & 0.12 & -0.17 &  0.14 & 0.21 & 0.10 & -0.09 &  0.18 & 0.21 \\{}
[NaI/Fe]  & 0.17 &  0.11 & -0.06 & 0.20 & 0.09 &       &       &      \\{}
[MgI/Fe]  & 0.19 &  0.00 &  0.06 & 0.20 & 0.19 & -0.19 &  0.01 & 0.27 \\{}
[SiI/Fe]  & 0.19 & -0.11 &  0.17 & 0.25 & 0.12 &       &       &      \\{}
[CaI/Fe]  & 0.09 &  0.20 &  0.01 & 0.22 & 0.11 & -0.10 & -0.12 & 0.16 \\{}
[ScII/Fe] & 0.20 &  0.05 & -0.04 & 0.21 & 0.17 & -0.17 & -0.03 & 0.24 \\{}
[TiI/Fe]  & 0.09 &  0.23 & -0.19 & 0.25 & 0.19 &  0.00 & -0.28 & 0.34 \\{}
[TiII/Fe] & 0.19 & -0.02 &  0.01 & 0.19 & 0.20 & -0.21 &  0.01 & 0.29 \\{}
[CrI/Fe]  & 0.19 &  0.21 & -0.17 & 0.28 & 0.17 &  0.01 & -0.22 & 0.28 \\{}
[NiI/Fe]  & 0.09 &  0.01 &  0.01 & 0.09 & 0.13 & -0.13 & -0.03 & 0.18 \\{}
[ YII/Fe] & 0.17 &       &       &      &      &       &       &      \\{}
[BaII/Fe] & 0.17 &       &       &      & 0.18 & -0.17 &  0.03 & 0.25 \\{}
[NdII/Fe] & 0.18 &  0.07 & -0.05 & 0.19 & 0.17 &       &       &      \\{}
[LaII/Fe] & 0.18 &  0.12 & -0.12 & 0.22 & 0.11 &       &       &      \\{}
[EuII/Fe] & 0.19 &  0.04 & -0.05 & 0.20 & 0.16 & -0.16 & -0.04 & 0.23  \\
\hline
\end{tabular}
\label{atm_errors_XFe}
\end{table*}

\par
\begin{figure}[htp]
   \centering
   \includegraphics[angle=-90,width=\columnwidth]{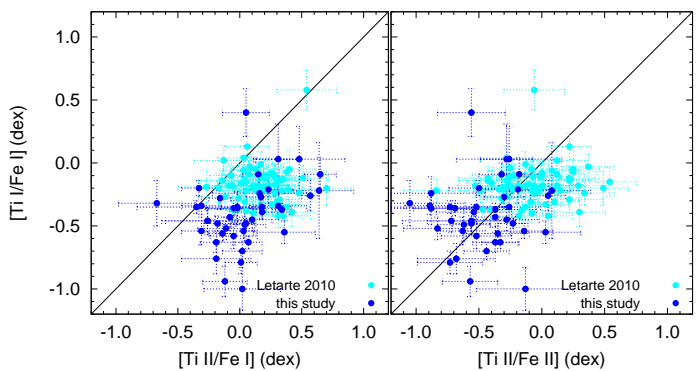}
      \caption{The two ionization states for Titanium, Ti I and Ti II, are compared for both our HR sample (blue) and the HR sample of \cite{Let2010} (cyan). The left panel shows a clear offset between [TiI/FeI] and [TiII/FeI] while on the right panel, [TiI/FeI] and [TiII/FeII] fall on a more common scale, showing that ionization balance is probably not very well achieved in these stars.
}
         \label{NLTE}
   \end{figure}

 In our results, FeI is systematically lower than FeII and similarly TiI is systematically lower than TiII. To summarize this effect, we show in Fig.~\ref{NLTE} that [TiI/Fe] differs from [TiII/Fe] by $\approx$--0.4 dex, i. e. the same order of magnitude already mentioned by \cite{Let2010} and \cite{Lem2012} for RGB stars observed with the same FLAMES setup. This discrepancy could be due to NLTE effects, with a departure from ionization equilibrium affecting predominantly FeI and TiI. However, any bias in the T$_{\rm eff}$ scale would shift the ionization equilibrium in a similar way and affect above all FeI and TiI lines that have a lower excitation potential on average. However, as we already mentioned, our T$_{\rm eff}$ are derived from five different colors and values are all in good agreement (see Table~\ref{atmparam_all}) so this hypothesis does not appear to be very likely. \cite{Berg2011} examined the ionization balance of Ti in late type stars, unfortunately our RGB stars are out of the range of parameters covered by this study. However \cite{Berg2011} report that NLTE effects on TiI are significant for any set of atmospheric parameters and that they are dominated by overionization effects. Consequently \cite{Berg2011} strongly recommend to disregard the use of TiI for abundance studies but mention that [TiII/FeII] seems to give relatively safe results.

\subsection{Ages}

\par Determining individual ages for RGB stars in our sample first requires the determination of the SFH of Fornax. This is achieved using the {\it Talos} code \citep{deBoer2012a} that compares observed CMDs with grids of synthetic CMDs. In order to determine an accurate SFH, {\it Talos} uses not only all the photometric filters available in our dataset but also the MDF of the RGB derived from CaT measurements. From the SFH, a global synthetic CMD of Fornax is built.
\par To determine the age of an individual RGB star, all the stars with the same magnitude (in all the photometric bands) and the same metallicity within the observational uncertainties are extracted from the synthetic CMD. The mean age and the standard deviation of this subsample are then adopted respectively as the age of the individual RGB star and its associated uncertainty. The synthetic CMD of Fornax is largely oversampled to ensure that a sufficient number of synthetic stars (typically $>$100) are available to compute a reliable error bar.\\

\par The ages from \cite{deBoer2012b} were computed with Dartmouth isochrones \citep{Dotter2008}. In this paper, the ages have been derived from a somewhat different SFH using an improved age resolution and BaSTI isochrones \citep{Piet2004}. The BaSTI isochrones have AGB sequences included and cover a sufficient range in magnitude so that we could also determine the ages for the brightest stars of our sample. This new computation of the SFH has a very mild influence on the results concerning the young and intermediate-age populations. The main difference is that the star formation rate is a bit increased at the oldest epochs, leading to a more regular slope in the AMR, as seen in Sculptor \citep{deBoer2012a}. Stars with ages younger than the main AMR locus typically show very blue colours in the CMD and are most likely AGB stars. A more complete description of the differences arising in the SFH from the two sets of isochrones will be given in a forthcoming paper (de Boer et al, in prep.).\\

\par We note that ages derived in this way are valid only for the adopted distance and extinction values. To assess the impact of the uncertainty on the distance of Fornax, we determined the ages of stars in the annulus 3 \citep[as defined by][]{deBoer2012b} for a shorter distance modulus: 20.74 mag instead of 20.84 mag. Shifting the CMD by 0.1 mag systematically increases the ages by $\sim$0.8 Gyr and their uncertainties by $\sim$0.4 Gyr on average and affects mainly the young and intermediate age stars. Ages are determined using the SFH results at different distances from the centre, thereby taking into account the effect of the population gradient in Fornax. However, if the SFH changes on spatial scales smaller than the annuli used in the SFH study, the ages might be biased. There is currently no obvious sign of stellar population difference within the region studied in this work. Finally, the SFH used in this work was determined with a single [$\alpha$/Fe] value corresponding to the sample of \cite{Let2010}. Within our new sample of Fornax stars, the [$\alpha$/Fe] ratio has changed notably only for metallicities between roughly --2.5$<$[Fe/H]$<$--1.5 dex. However, the effect of the change in $\alpha$-elements abundance would be small compared to the effect of the uncertainty on metallicity only, and most pronounced on the RGB. In particular, changing the $\alpha$-elements abundance would have a negligible influence on the shape and position of the Main Sequence Turn-Off, which strongly dominates the age fitting. Therefore, no significant changes are expected for the age derivation due to [$\alpha$/Fe] differences.\\

\par Two stars in the HR sample (mem0598 (this study) at [Fe/H]=--0.85, [Mg/Fe]=--0.13 and ET228 \citep{Let2010} at [Fe/H]=--0.87, [Mg/Fe]=--0.19) and one star in the MR sample \citep{Kirby2010}, at [Fe/H]=--1.18, [Mg/Fe]=--0.07 dex) are attributed very old ages ($\geq$12 Gyr) while their [Fe/H] and [Mg/Fe] are those of younger stars. A look at the CMD indicates that they are redder than the main RGB (considering only stars in the same annulus), yet show metallicities more metal-rich than the red edge of the RGB. Therefore, they are very probably foreground stars (We checked that mem0598 does not seem to be a CH star). On the other hand, mem0598 (the only star for which we have spectra in the CaT region) is not considered as dwarf foreground contaminant according to the $\lambda$=8806.8 \AA MgI line criterion \citep{Batt2012}.\\


\section{Results}
\label{results}

\subsection{Iron and iron-peak elements}

\subsubsection{The metallicity distribution function of Fornax}

\par
\begin{figure}[htp]
   \centering
   \includegraphics[width=\columnwidth]{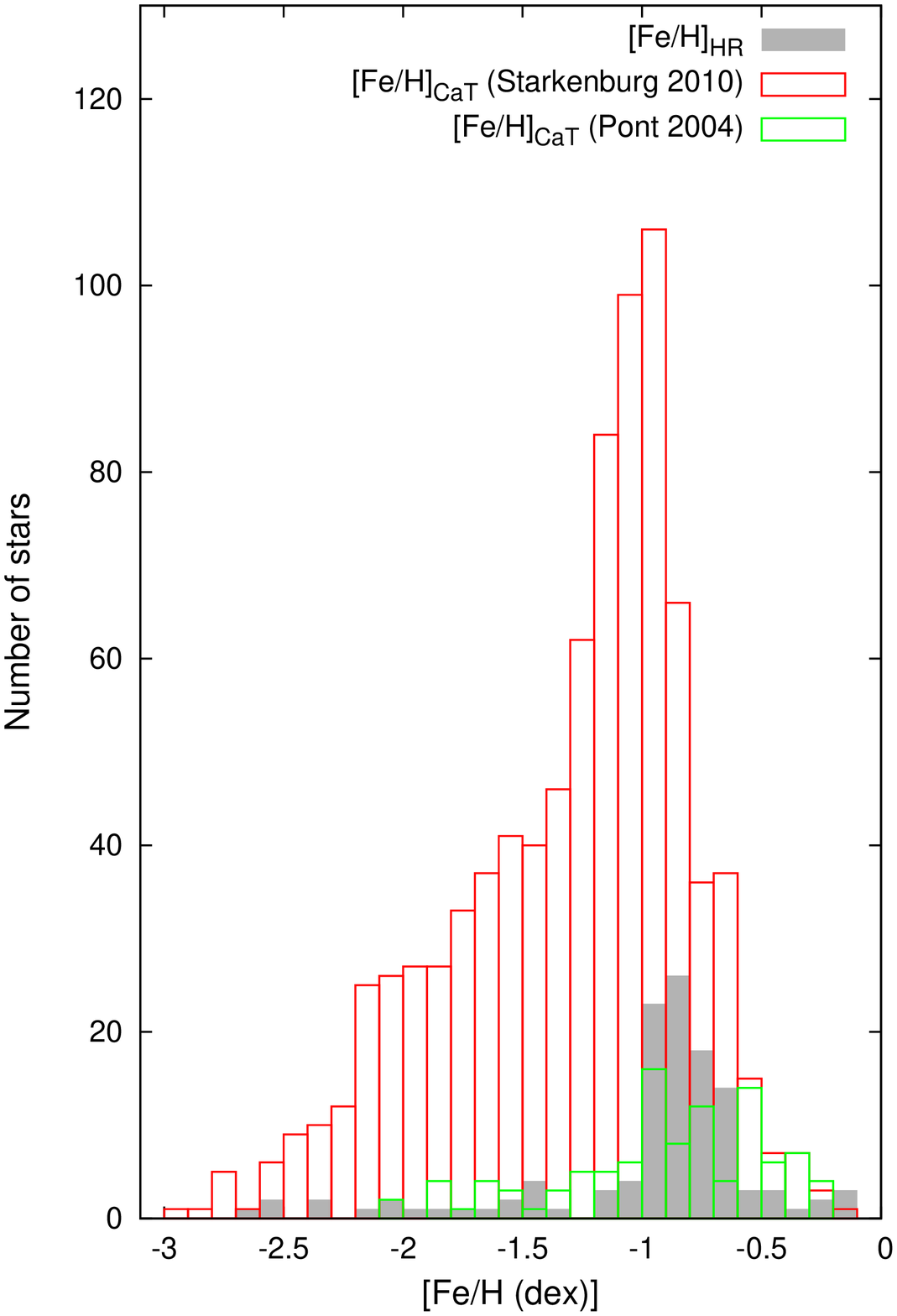}
      \caption{The [Fe/H] distribution of the Fornax high-resolution sample (\cite{Let2006,Let2010,Taf2010}, this paper) in grey. The complete set of CaT measurements \citep{Pont2004,Batta2008a} is shown respectively in red and green for comparison.
}
         \label{MDF}
   \end{figure}

 The iron abundances are determined from between 15 and 56 Fe~I lines (see Table~\ref{EW}) which are in good agreement as the dispersion around the mean values fall between 0.05 and 0.10 dex for almost all of the stars. Our sample spans almost the whole metallicity range of Fornax from [Fe/H]$\approx$--0.3 dex to [Fe/H]$\approx$--2.7 dex (see Fig.~\ref{MDF}). The metallicity distribution function (MDF) of RGB stars in Fornax derived from CaT measurements is also shown in Fig.~\ref{MDF} for comparison. The low-resolution spectroscopy in the CaT region (from \cite{Batta2008a} have been recomputed with the new calibration provided by \cite{Stark2010} while the values of the \cite{Pont2004} sample are taken directly from their paper\footnote{The reader should keep in mind that all these studies used different calibrations of the CaT.}.
\par As we already mentioned, our field was selected to contain the maximum number of metal-poor stars ([Fe/H]$<$--1.5 dex) within one FLAMES/GIRAFFE field. The fraction of metal-poor stars is higher in the outskirts of Fornax but their density is low in these regions. It turns out that the density of metal-poor stars is highest in fields still relatively close to the center. In these fields, the mean metallicity remains quite high, most of the stars in our sample fall therefore in the peak of the MDF. As a result, our sample appears slightly biased as it misses (in relative numbers) RGB stars between --2.0 and --1.0 dex. The CaT sample might also incorporate some foreground stars that match the velocity range for Fornax membership, but their number should be quite low: combining the Mg I line at $\lambda$=8806.8 \AA{} and the CaT, \cite{Batt2012} can remove the Milky Way dwarf stars contaminating the samples of RGB kinematic members in the dSphs. They estimate the level of contamination to be 5\%. In a similar approach using the surface gravities measured in the CaT domain, \cite{Kordo2013} found 86 foreground stars over 1060 targets ($\approx$8\%) in the Fornax line of sight.

\subsubsection{Comparison with [Fe/H] estimates from the CaII triplet}

\par
\begin{figure}[htp]
   \centering
   \includegraphics[angle=-90,width=\columnwidth]{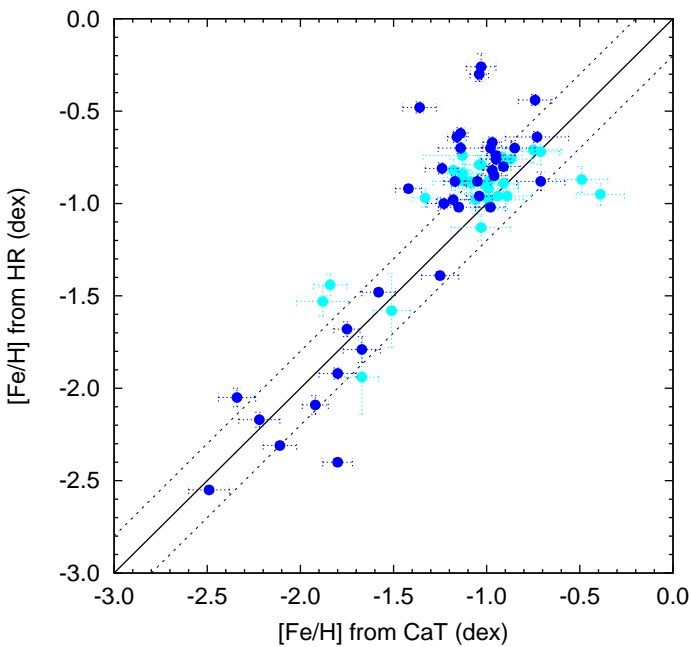}
      \caption{For our sample of RGB stars in the Fornax dSph, we show the comparison between [Fe/H] measured from high resolution spectroscopy and from the CaT (\cite{Batta2008a}, recomputed with the new calibration of \cite{Stark2010}). The solid line indicates where [Fe/H]$_{CaT}$=[Fe/H]$_{HR}$ while the dashed lines indicate a difference of $\pm$0.2 dex. Blue points represent our data while the cyan points are the HR data from \cite{Let2010}.}
         \label{CaT-HR}
   \end{figure}

 For 37 out of the 47 stars in our sample, both the CaT metallicity and our high resolution determination of [Fe/H] are available.
Results agree well in general, they fall within $\pm$0.2 dex in most cases (see Fig.~\ref{CaT-HR}.) and differences never exceed 0.4 dex. This value is similar  to the one obtained by \cite{Stark2010}, see for instance their Fig.~10.
Differences between [Fe/H] derived from HR spectroscopy or from the CaT calibration are noticeable above [Fe/H]= --1.0 dex, they become substantial in the metal-rich regime ([Fe/H] $\geq$ --0.5 dex) and indeed \cite{Stark2010} do not advocate the use of the calibration above [Fe/H]= --0.5 dex. To understand the origin of the differences, it is worth mentioning first that the atmospheric parameters of the stars in our HR sample are derived from the new CTIO BVI photometry provided by \cite{deBoer2012b} supplemented by VISTA NIR photometry, while the CaT estimates are based on different data, namely VI photometry from ESO/WFI \citep{Batta2008a}. However, differences in the photometric datasets probably do not account for much of the differences, as the HR sample of \cite{Let2010} is derived from the same VI photometry as the CaT estimates (together with NIR photometry for 60\% of their sample) but still shows a similar pattern to ours. We also note that we computed the atmospheric parameters for our young, metal-rich stars assuming a mass of 1.3 M$_{\odot}$ while the CaT calibration uses a constant mass of 1.0 M$_{\odot}$ over the whole sample. However, we already mentioned above that differences on the adopted mass have a mild impact on the surface gravities (and, in turn, on the metallicities) derived for the stars. Finally, a low [Ca/Fe] is a common feature to the outliers in our sample; indeed all of them have [Ca/Fe] $\lesssim$ 0.0 dex, i.e, at the bottom edge or outside the range of stellar models used to calibrate the CaT vs [Fe/H] relation (+0.0 $<$ [$\alpha$/Fe] $<$ +0.4 dex). However this does not seem to be a sufficient condition as some other stars in the same [Ca/Fe] range are a good match between HR and CaT. To conclude, although the mismatch in individual cases is difficult to pin-point down to one particular uncertainty in the analysis, a combination of factors can be at play. Moreover, only three stars show a significant offset, among which two belong to the very metal-rich end of the MDF and therefore do not affect the rest of the analysis.

\subsubsection{Nickel and chromium}

\par
\begin{figure}[htp]
   \centering    
   \includegraphics[angle=0,width=\columnwidth]{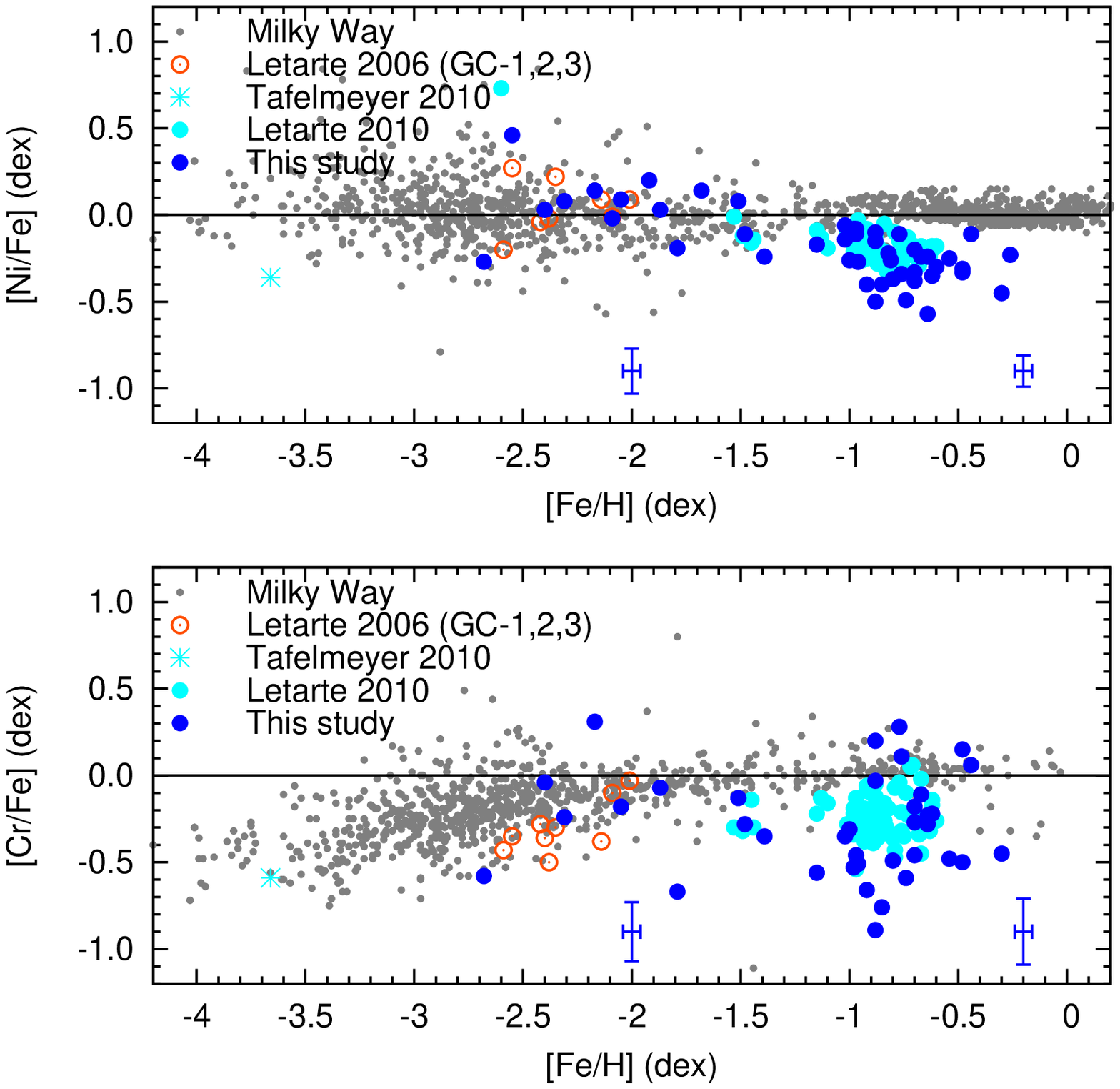}
      \caption{The distribution of [Ni/Fe] (top) and [Cr/Fe] (bottom) for our sample of RGB stars in the Fornax dSph as blue filled circles. The cyan filled circles are the data of \cite{Let2010} and the cyan triangle the metal-poor star of \cite{Taf2010}. We also show in orange filled circles the stars in Fornax globular clusters from \cite{Let2006}. Milky Way halo stars from \cite{Venn2004} and \cite{Frebel2010} and references therein are in small grey dots. Representative error bars are given for the metal-poor ([Fe/H]$<$--1.4 dex) and metal-rich ([Fe/H]$>$--1.4 dex) regimes.}
         \label{ironpeak}
   \end{figure}

 Iron peak elements (besides iron, we show here nickel and chromium) are mainly formed through explosive nucleosynthesis in SNe Ia. We present in Fig.~\ref{ironpeak} the evolution of the [Ni/Fe] and [Cr/Fe] ratios against [Fe/H]. Nickel abundances are derived from 6--12 lines, except for the most metal-poor stars in our sample, where they rely only on 1--4 lines. If Ni follows the MW trend in the metal-poor regime, it becomes underabundant for [Fe/H] $\geq$ --1.5 dex, a feature already observed for RGB stars in the Large Magellanic Cloud (LMC) disk stars \citep{Pomp2008,vdS2013}, in the Sagittarius and Sculptor dSphs or in the center of Fornax \citep{Sbo2007,Tolstoy2009,Let2010}. In this metallicity range, Sagittarius, Sculptor and Fornax are also deficient in $\alpha$-elements, and indeed nickel underabundances have also been reported by \cite{Niss1997,Niss2010,Schu2012} in a sample of $\alpha$-poor halo stars. 

\par In contrast with nickel, most of our chromium abundances are derived from only one line at 6330.09 \AA{} (for a handful of intermediate metallicity stars, the line at 5409.80 \AA{} could also be measured and both lines give very consistent results, but the latter becomes too strong in the metal-rich regime). As a result, Cr shows a lot more scatter than Ni. \cite{Berg2010} have shown that the decrease of [Cr/Fe] at low metallicities observed for both MW and Fornax stars (and other systems as well) is an artifact due to strong non local thermodynamic equilibrium (NLTE) effects taking place in the formation of CrI lines. NLTE corrections are of the order of +0.1 dex at solar metallicities and up to +0.5 dex for metal-poor stars. Unfortunately, the study of \cite{Berg2010} deals with a large number of Cr lines but not with the 6330.09 \AA{} one.\\

\par The SNe Ia yields for iron-peak elements are still weakly constrained and strongly depend on the adopted model for the SNe Ia. It is therefore difficult to interpret the Ni and Cr underabundances with respect to iron above [Fe/H] $\approx$ --1.5 dex. As both Ni and Fe are produced via the same production mechanism in SNe Ia \citep{Tra2005}, one would expect that the [Ni/Fe] ratio remains constant around the solar ratio. One possible explanation could be that the SNe Ia yields are in fact slightly different for each iron-peak element and depend on the metallicity of the SNe Ia progenitor \citep{Tim2003,Tra2005}. \cite{Li2013} explain that the difference is related to a different weight of the Ni sources in Fornax and in the MW with respect to the solar ratios. They found that in the MW, the main Ni sources are the primary-like and secondary-like processes\footnote{The nucleosynthesis processes that do not require pre-existing seed nuclei are called primary processes while the secondary processes require pre-existing seed nuclei to produce new elements. These definitions have been initially established to study the neutron-capture processes but have been extended by e.g., \cite{Li2013b} to light and iron-peak elements; in such cases one refers to primary-like and secondary-like processes} in massive stars. They fall above (primary-like processes) or at (secondary-like processes) the solar ratio, but are compensated by a SNe Ia contribution smaller than in the Sun, leading to [Ni/Fe]$\approx$0.0 dex. In contrast, \cite{Li2013} report that SNe Ia dominate as a Ni source in Fornax. Moreover, both the SNe Ia and the primary-like process contributions reach the solar ratio, while the contribution of the secondary-like process falls well below the solar ratio (by about 0.5 dex), leading to a subsolar value for [Ni/Fe]. The results of \cite{Li2013} are then compatible with a Fornax IMF depleted in the most massive stars, at least for the metallicity range considered in their study.\\

\subsubsection{The Ni--Na relationship}

\par
\begin{figure}[htp]
   \centering
   \includegraphics[angle=0,width=\columnwidth]{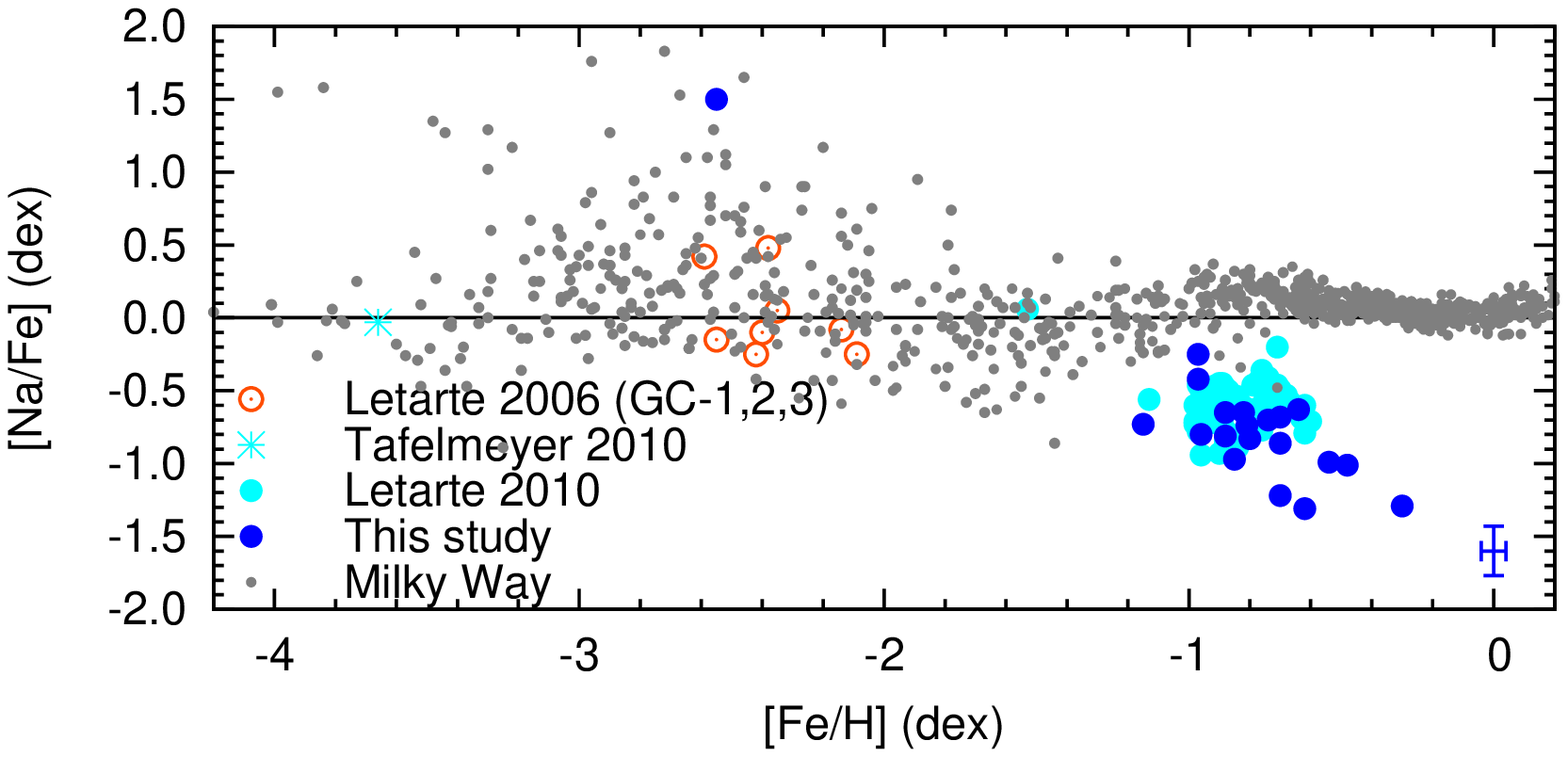}
      \caption{Same as Fig.~\ref{ironpeak} for [Na/Fe].}
         \label{NaFe}
   \end{figure}

\begin{figure}[htp]
   \centering
   \includegraphics[angle=0,width=\columnwidth]{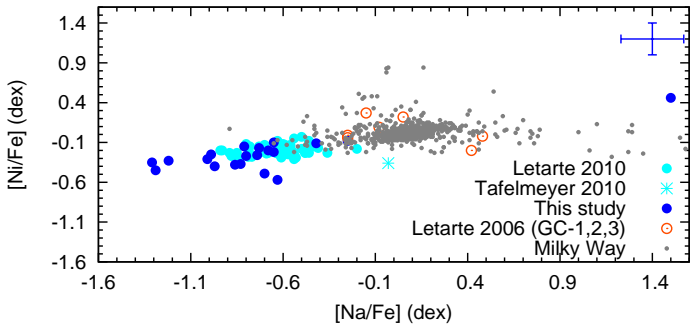}
      \caption{Same as Fig.~\ref{ironpeak} for [Ni/Fe] vs [Na/Fe]. \cite{Venn2004} indicate that the Ni-Na correlation observed in the Milky Way is primarily due to thin and thick disk stars, while it is not observed for halo stars that therefore lie off the relation.}
         \label{NiNa}
   \end{figure}

 The halo stars of \cite{Niss1997,Niss2010,Schu2012} not only show a nickel underabundance, but they are also deficient in sodium, leading to a [Na/Fe]--[Ni/Fe] correlation. The same outcome applies to Fornax stars (see Figs.~\ref{NaFe},\ref{NiNa}) falling in the same metallicity range (--1.6$<$[Fe/H]$<$+0.4 dex). Sodium is mainly produced through carbon burning in massive stars that will ultimately end as SNe II, but a fraction of Na is also produced via hot-bottom proton burning by the Ne-Na cycle \citep{WooWea1995}. It is therefore expected that [Na/Fe] decreases with [Fe/H] as SNe Ia take over. On the other hand, \cite{SM2002} and \cite{MS2005b} mentioned that the low [Na/Fe] ratios are compatible with a paucity of material ejected from SNe II and/or with a low SNe II/SNe Ia ratio. In contrast to sodium, nickel is also produced in large quantities in SNe Ia \citep{Tsuji1995} where the production of Ni does not depend on Na \citep[e.g.][]{Iwa1999}.\\ 

\par The current explanation for the [Na/Fe]--[Ni/Fe] correlation is that the yields of Ni and Na depend on the neutron excess in SNe II, affected itself by the [$\alpha$/Fe] ratio \citep{Venn2004}. \cite{Let2010} showed that the Fornax RGB stars tightly stick to the MW Ni--Na correlation. In this study we are able to extend this tight correlation further down to lower [Na/Fe] and [Ni/Fe] values as is shown in Fig.~\ref{NiNa}.
\par As the origin of the Ni--Na correlation is related to SNe II, it should be observed in the metal-poor regime, where only SNe II have contributed to the chemical enrichment. However \cite{Venn2004} found no correlation in the metal-poor regime for Milky Way stars and there are (to our knowledge) no studies reporting the Ni--Na correlation at low metallicities. In Fornax, the metal-poor star from \cite{Taf2010} clearly stands out of the relation, even after correction for the NLTE effects. The same applies to most of the globular cluster stars of \cite{Let2006} (--2.5 $<$ [Fe/H] $<$ --2.0 dex) but they probably have a different chemical evolution history. 
We measured sodium in only one field star in this metallicity range and it has a very high Na abundance. Correcting for the NLTE effects \citep{Lind2011}, has a minor effect of $\Delta_{NLTE}$=--0.09\footnote{Data obtained from the INSPECT database, version 1.0 (www.inspect-stars.net)} dex. Because only one weak Na line could be measured in this star, this result should be treated with caution until other stars with a similarly high Na abundance are found in Fornax; it is however interesting to note that it matches very well the Ni-Na relationship.\\  

\par The metal-rich populations of dSphs are in most cases also $\alpha$-poor and this low [$\alpha$/Fe] ratio is usually associated with a large contribution of the SNe Ia. Therefore one would expect that the correlation breaks in the metal-rich regime, which is not observed \citep[see also][]{Venn2004}. Given the lines available in the selected FLAMES settings, both \cite{Let2010} and this study were able to measure Na in stars with metallicities from $\approx$ --1.5 dex up to $\approx$ --0.5 dex and even in one star at --0.3 dex. Two stars in our sample ([Fe/H]$\approx$--0.7 dex) clearly stand out of the relationship. This could be due either to larger error bars (Na, especially, is derived from only one line in these stars) or to a peculiar abundance pattern for these stars. We thus have to explain why the SNe II signature of the Ni--Na correlation remains visible at higher metallicities. It could be due to the fall-back of ''fresh'' gas contaminated only by SNe II ejectas \citep[e.g.][]{Revaz2012}, mixing with gas enriched by SNe Ia with low yields for Ni, due to lower metallicity progenitors in Fornax than in the MW. Because the correlation has only been observed in the SNe Ia--dominated regime, it could also very well be that the current explanation based on SNe II is simply not correct.\\

\subsection{$\alpha$-elements}
\label{sect_alp}

\par The $\alpha$-elements can be divided in two groups: those that are produced through hydrostatic He burning in massive stars and released when the star explodes as a SNe II (O, Mg) and those that are mainly produced during the SNe II explosion itself (Si, Ca, Ti). SNe Ia produce only negligible quantities of O and Mg but they can also produce large quantities of Si, Ca and Ti \citep{Tsuji1995}. Therefore [$\alpha$/Fe] (and especially [O/Fe] and [Mg/Fe]) depend on the ratio of SNe II to SNe Ia that have enriched the interstellar medium (ISM). Stars formed when SNe II only enriched the ISM have a high [$\alpha$/Fe]. SNe Ia have a longer timescale than SNe II, and their dominant iron production will cause a decrease of [$\alpha$/Fe] when they start to contribute to the ISM enrichment \citep{Tin1979,Matt1990,Gil1991,Matt2003}. A complementary explanation would be a bottom-heavy initial mass function where the low [$\alpha$/Fe] result from a lack of the most massive stars.\\

\par
\begin{figure*}[htp!]
   \centering
   \includegraphics[angle=0]{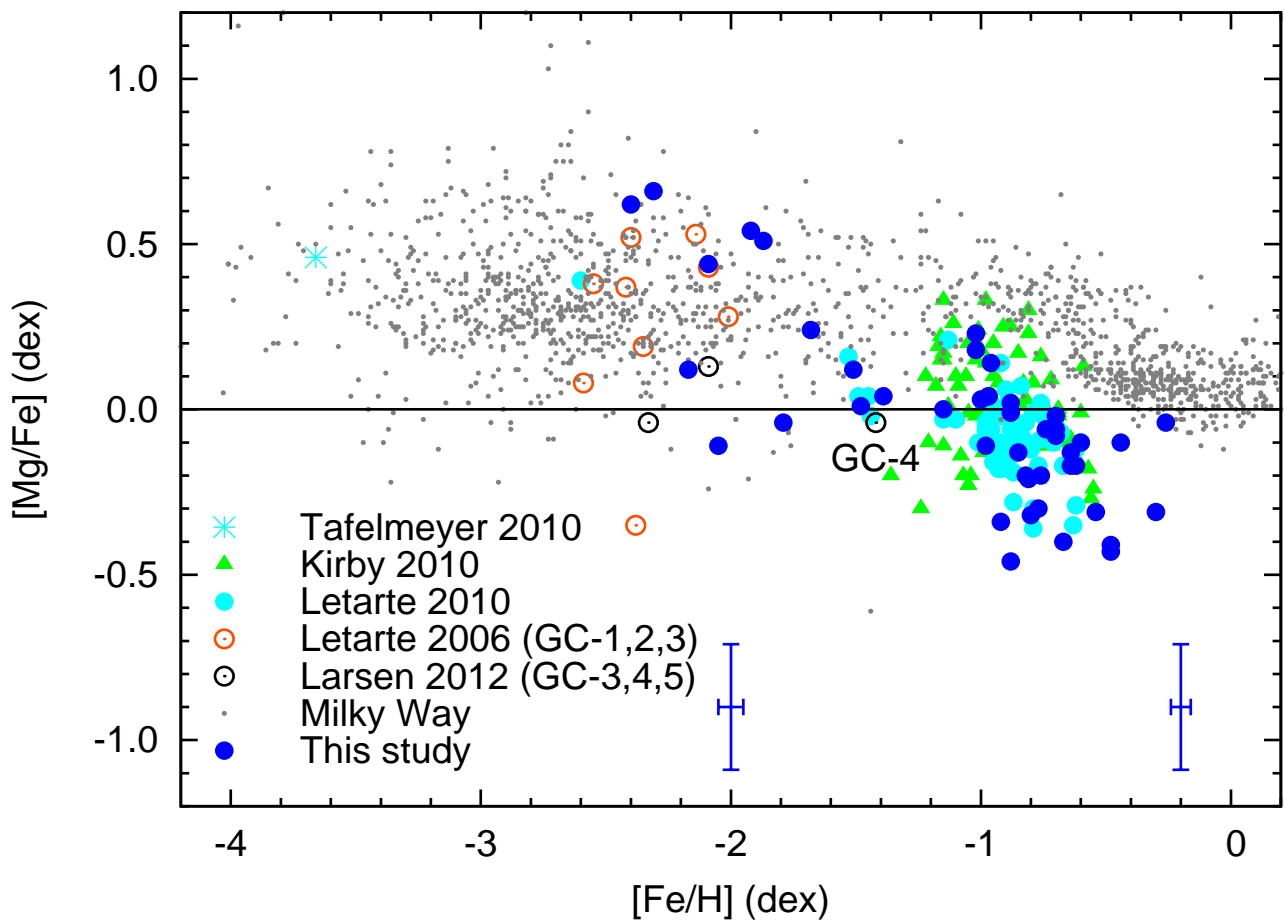}
      \caption{Same as Fig.~\ref{ironpeak} for [Mg/Fe].}
         \label{MgFe}
   \end{figure*}

\begin{figure}[htp]
   \centering
   \includegraphics[angle=0,width=\columnwidth]{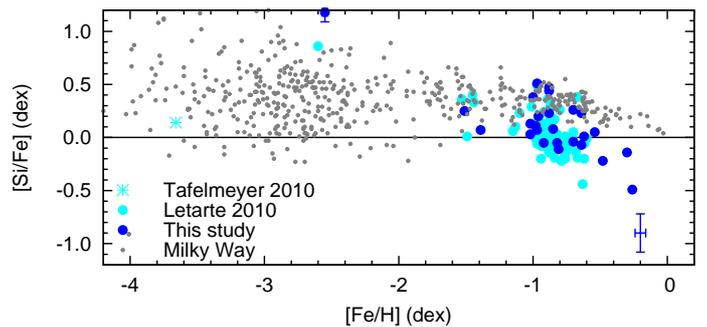}
      \caption{Same as Fig.~\ref{ironpeak} for [Si/Fe].}
         \label{SiFe}
   \end{figure}

\begin{figure}[htp]
   \centering
   \includegraphics[angle=0,width=\columnwidth]{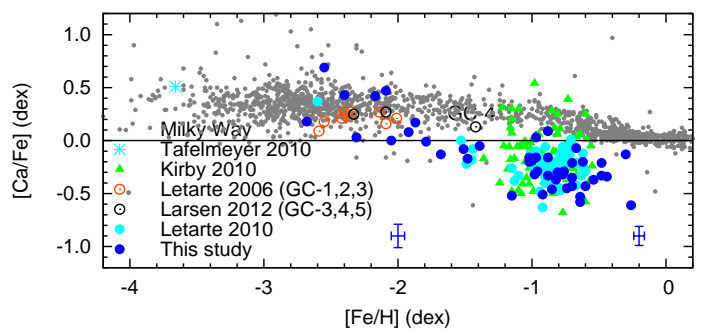}
      \caption{Same as Fig.~\ref{ironpeak} for [Ca/Fe].}
         \label{CaFe}
   \end{figure}

 Figs.~\ref{MgFe},~\ref{SiFe},~\ref{CaFe} display the individual $\alpha$-element ratios [Mg/Fe], [Si/Fe] and [Ca/Fe] for our sample of FLAMES/GIRAFFE data\footnote{Two O lines (6300~\AA, 6363~\AA) are available in our settings. However, the 6300~\AA{} line matches a telluric absorption line at the typical radial velocity of Fornax and the 6363~\AA{} line is weak and close to the detection limit with DAOSPEC at the FLAMES/GIRAFFE resolution. Moreover the continuum is not well defined in this region because of CN lines and a Ca~I autoionisation feature.
Therefore we did not derive the O abundances.}. [Mg/H] is determined from a single, well defined line at $\lambda$ = 5528.41 \AA. 4 to 12 Ca lines could be measured, and [Si/H] is determined for most of our sample from the single $\lambda$ = 6155.14 \AA{} line. For $\lesssim$ 10 stars, we could also measure the $\lambda$ = 6237.33 \AA{} line and the abundances derived are always in excellent agreement. We also show the similar FLAMES/GIRAFFE sample of \cite{Let2010} and the Fornax metal-poor star in \cite{Taf2010}. For the three Fornax stars in \cite{She2003} that are are in common with \cite{Let2010}, we chose the values derived by the latter study as they were obtained with the exact same instrument settings. We note that the median errors on [Mg/Fe] are slightly larger in our sample (0.19 dex) than in the sample of \cite{Let2010} who reported a value of 0.16 dex. \cite{Let2006} observed stars in three Fornax globular clusters and \cite{Lars2012} obtained the mean chemical composition of three Fornax globular clusters from integrated light spectroscopy. Finally, \cite{Kirby2010} provided iron and $\alpha$-element abundances for a large number of stars from medium resolution (R$\approx$7000 at 8500 \AA) spectroscopy. From this dataset we retained only the stars with an uncertainty in [Fe/H] and [Mg/H] lower than 0.2 dex and that were not included in higher resolution samples. \cite{Hen2014} analyzed 431 FLAMES/GIRAFFE HR spectra in the CaT region (HR 21 grating, R$\approx$ 16000). Using the {\it SPACE} code based on a library of Generalized Curve Of Growths, they could derive Mg abundances (together with Si and Ti) for 58 stars. They did not published their data but their results are very similar to ours.\\

\par In the metal-poor regime, Fornax is similar to the other MW dSphs and to the MW itself, it shows high (supersolar) values for [$\alpha$/Fe]. Like in the other dSphs, Mg reaches higher values than Ca. The scatter is also quite large, following here again a pattern similar to the MW. This scatter is certainly partially related to uncertainties in the determination of the abundances, as the number of lines available decreases towards the metal-poor end. 
We note that [Ca/Fe] lies at the bottom edge or below the MW distribution in the metal-poor region where SNe II are supposed to dominate the chemical enrichment in $\alpha$-elements, both for the globular clusters stars and for most of the field stars.\\
\indent In contrast at higher metallicities, [$\alpha$/Fe] are in most cases underabundant with respect to the MW. It should be noted that if [Mg/Fe] is almost systematically lower than in the MW, a noticeable fraction of the sample has [Ca/Fe] and especially [Si/Fe] overlapping the MW values. As in the other dSphs, [Si/Fe] keeps higher values than [Mg/Fe] and [Ca/Fe]. However, unlike in Sculptor \citep{Tolstoy2009} and Carina \citep{Lem2012,Venn2012}, [Ca/Fe] reaches even lower values (--0.7 dex) than [Mg/Fe] (--0.5 dex). Given the timescale of star formation in Fornax \citep{deBoer2012b}, SNe Ia (that produce Ca but almost no Mg) had largely enough time to contribute to the chemical enrichment of the galaxy. Therefore a low [Ca/Fe] (with respect to [Mg/Fe]) could be the sign of either a low number of SNe Ia (in comparison to SNe II) or a sensitivity of SNe Ia yields to the metallicity of their progenitors. However this result should be treated with caution because NLTE effects described in Sect.~\ref{secErrBudget} in the case of TiI/TiII probably also affect CaI and MgI, possibly by a different amount.

\subsection{Neutron-capture elements} 

\par Neutron-capture elements are divided in s-process (for slow-process) and r-process (for rapid-process) elements, depending on the timescale of the neutron capture events with respect to the timescale for $\beta^{--}$ decay. The s-process takes place in relatively low-mass stars (2--4~M$_{\odot}$) while the r-process is believed to take place in SNe II explosions \citep{Sne2008}, although neutron-star neutron-star mergers are an equally attractive site \citep[e.g.,][]{Frei1999}. Moreover, the yields of s-process elements in AGB stars are metallicity-dependent: neutrons are captured by only a few iron-peak nuclei at low metallicity and therefore heavy s-process elements (Ba, La, Pb) are preferentially produced while the same amount of neutrons is shared within many more iron-peak seeds at high metallicity, where light s-process (Y, Zr) are mainly produced.\\

\par Our neutron-capture measurements are based one 1--2 lines in the case of Ba, La, Nd. Only a single line was available for Y and Eu, and it could be measured only in 5 stars in the case of Y. The Ba lines are usually very strong even at relatively low metallicities, therefore the Ba abundances could not be determined for some of the metal-rich stars. The hyperfine structure (HFS) of a line tends to desaturate it, leading to a possible overestimate of the abundance of the corresponding element. Following \cite{Let2010}, we applied a line-by-line HFS correction to the La $\lambda$=6320.43 \AA{} and Eu $\lambda$=6645.13 \AA{} lines. This correction only depends on the EW of the lines. The HFS corrections for Ba were found to be negligible and therefore were not applied.

\begin{figure}[htp!]
   \centering
   \includegraphics[angle=0,width=\columnwidth]{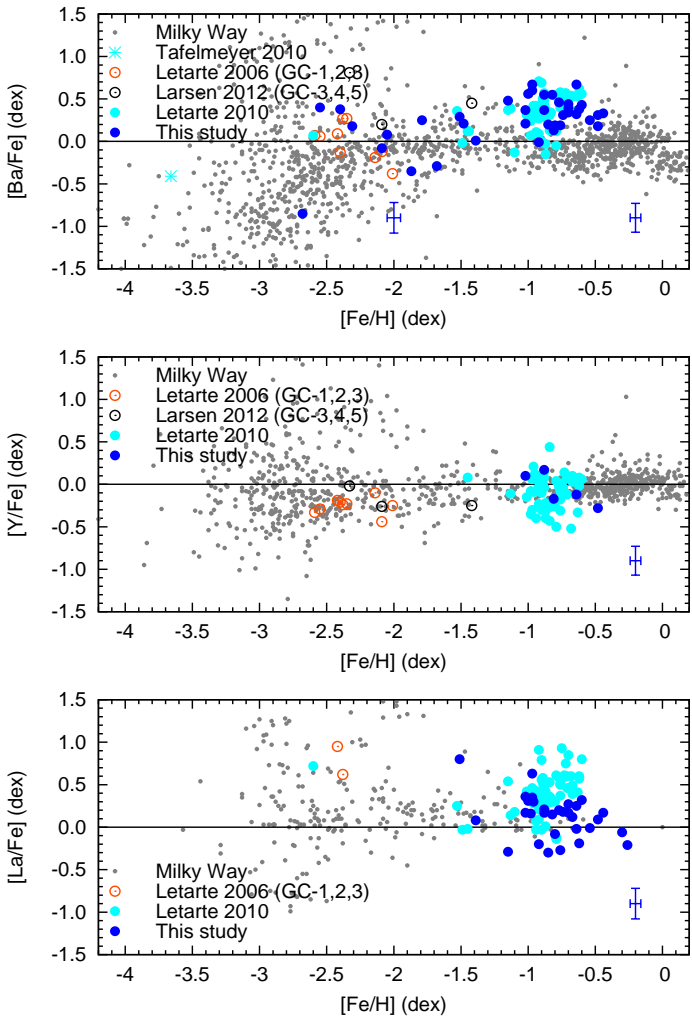}
      \caption{Same as Fig.~\ref{ironpeak} for [Ba/Fe](top),  [Y/Fe](middle) and [La/Fe](bottom), with additional Milky Way data from \cite{Simm2004} for [La/Fe].}
         \label{sprocess}
   \end{figure}

\par
\begin{figure}[hbp]
   \centering
   \includegraphics[angle=0,width=\columnwidth]{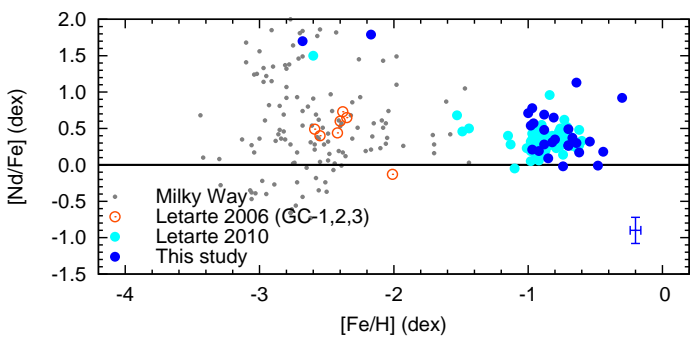}
      \caption{Same as Fig.~\ref{ironpeak} for [Nd/Fe].}
         \label{NdFe}
   \end{figure}

\begin{figure}[htp]
   \centering
   \includegraphics[angle=0,width=\columnwidth]{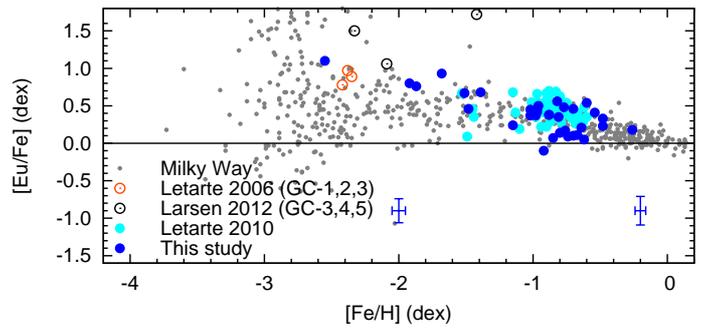}
      \caption{Same as Fig.~\ref{ironpeak} for [Eu/Fe], with additional Milky Way data from \cite{Simm2004}.}
         \label{EuFe}
   \end{figure}

 Fig.~\ref{sprocess} shows the evolution of the s-process (Ba, Y, La) elements with [Fe/H], in decreasing order of s-process contribution in the Sun. Nd is believed to be produced at 42\% via the r-process in the Sun \citep{Sne2008}. The evolution of [Nd/Fe] is shown in Fig.~\ref{NdFe}. Eu in the Sun is almost entirely produced via the r-process (\cite{Sne2008} quote the value of 97\%), its evolution is shown in Fig.~\ref{EuFe}. At low metallicities, the neutron-capture elements follow well the MW halo trend. Below [Fe/H]=--1.5 dex, the scatter in [Ba/Fe] increases and, more importantly, [Ba/Fe] turns down. This is a well-known feature in the MW halo \citep[e.g.][]{Fran2007} although it occurs at much lower metallicities ([Fe/H]$<$--3.0 dex) in the halo \citep[but see][]{Simm2004}. As the early evolution of neutron-capture elements is driven by the r-process, it could mean that the sources of the r-process (which are supposed to be SNe II) are either less common or less efficient in Fornax than in the MW halo.

\par
\begin{figure}[htp]
   \centering
   \includegraphics[angle=0,width=\columnwidth]{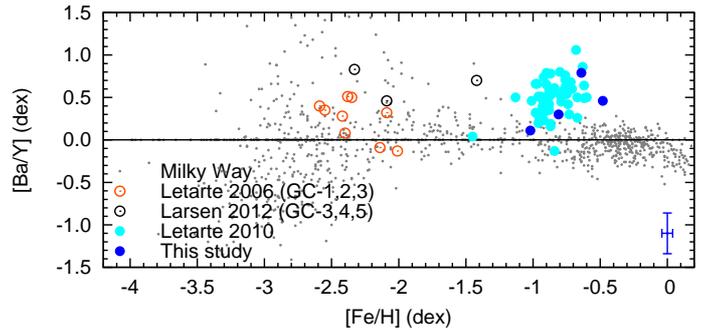}
      \caption{Same as Fig.~\ref{ironpeak} for [Ba/Y].}
         \label{BaY}
   \end{figure}

 When low and intermediate mass AGB stars begin to contribute to the chemical enrichment of Fornax via the s-process, both [Ba/Fe] and [La/Fe] rise quickly, reach the MW halo values and even exceed it, reaching [Ba/Fe] $\approx$+0.5 dex and [La/Fe] almost up to +1.0 dex at [Fe/H]$>$--1.0 dex. The production of Ba and La has then been more efficient in Fornax than in the Milky Way, and the same outcome applies to the LMC \citep{vdS2013}. In contrast, [Y/Fe] remains around or even below the MW trend. Fig.~\ref{BaY} confirms that Ba is a lot more abundant than Y, at levels that largely exceed the MW values. The metallicity dependence of the AGB yields enables to explain this different behaviour: at solar metallicities, AGB stars produce mostly light s-process elements like Y, while the heavier elements Ba, La are preferably produced by metal-poor AGB stars. Comparing the trends of the different s-process elements and examining the values reached by the [La/Y] ratios \citep{Bus1999,Bus2001} then indicates that the AGB stars in Fornax are predominantly metal-poor.

\par
\begin{figure}[htp]
   \centering
   \includegraphics[angle=0,width=\columnwidth]{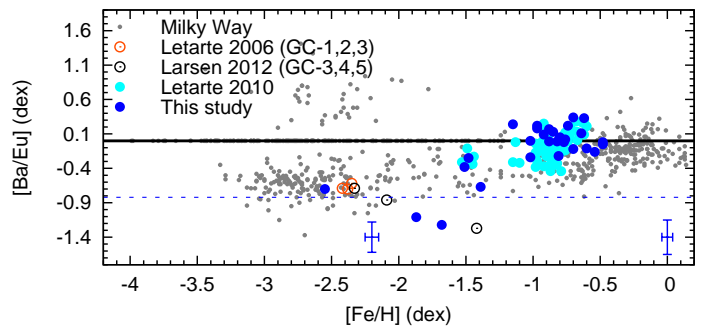}
      \caption{Same as Fig.~\ref{ironpeak} for [Ba/Eu]. The horizontal blue line indicates the solar r-process ratio [Ba$_{r}$/Eu$_{r}$]=--0.82 dex from Table 1 in \cite{Sne2008}.}
        \label{BaEu}
   \end{figure}

 Fig.~\ref{BaEu} illustrates the relative contributions of Ba produced via the s-process to the Ba produced via the r-process. Again, the low values at early times indicate that both Ba and Eu were produced via the r -process \citep{Sne2008}. The metallicity at which [Ba/Eu] turns up, i. e. where the s-process starts to contribute to the production of Ba, is poorly constrained but seems to be around --1.5 dex. This is a more metal-rich value than in Sculptor \citep[][(--1.8 dex)]{Tolstoy2009} but more metal-poor than in the LMC \citep[][(--0.8 dex)]{Pomp2008,vdS2013}, indicating that the chemical enrichment of Fornax was slower than in the LMC but quicker than in Sculptor.

\par The evolution of [Eu/Fe] displayed in Fig~\ref{EuFe} indicates a decrease at higher metallicities, a bit similar to the decrease of the $\alpha$-elements but [Eu/Fe] remains supersolar. This trend is not surprising, as Eu is an almost purely r-process element and, as such, presumably produced in SNe II which also produces large quantities of $\alpha$-elements via hydrostatic or explosive nucleosynthesis. It is interesting to note that [Eu/Fe] in Fornax and in the Milky Way are essentially indistinguishable in the metal-rich regime while below [Fe/H] = --1.0 dex, most of the Fornax HR sample falls on the upper side of the Milky Way [Eu/Fe] distribution.

\par
\begin{figure}[htp]
   \centering
   \includegraphics[angle=-0,width=\columnwidth]{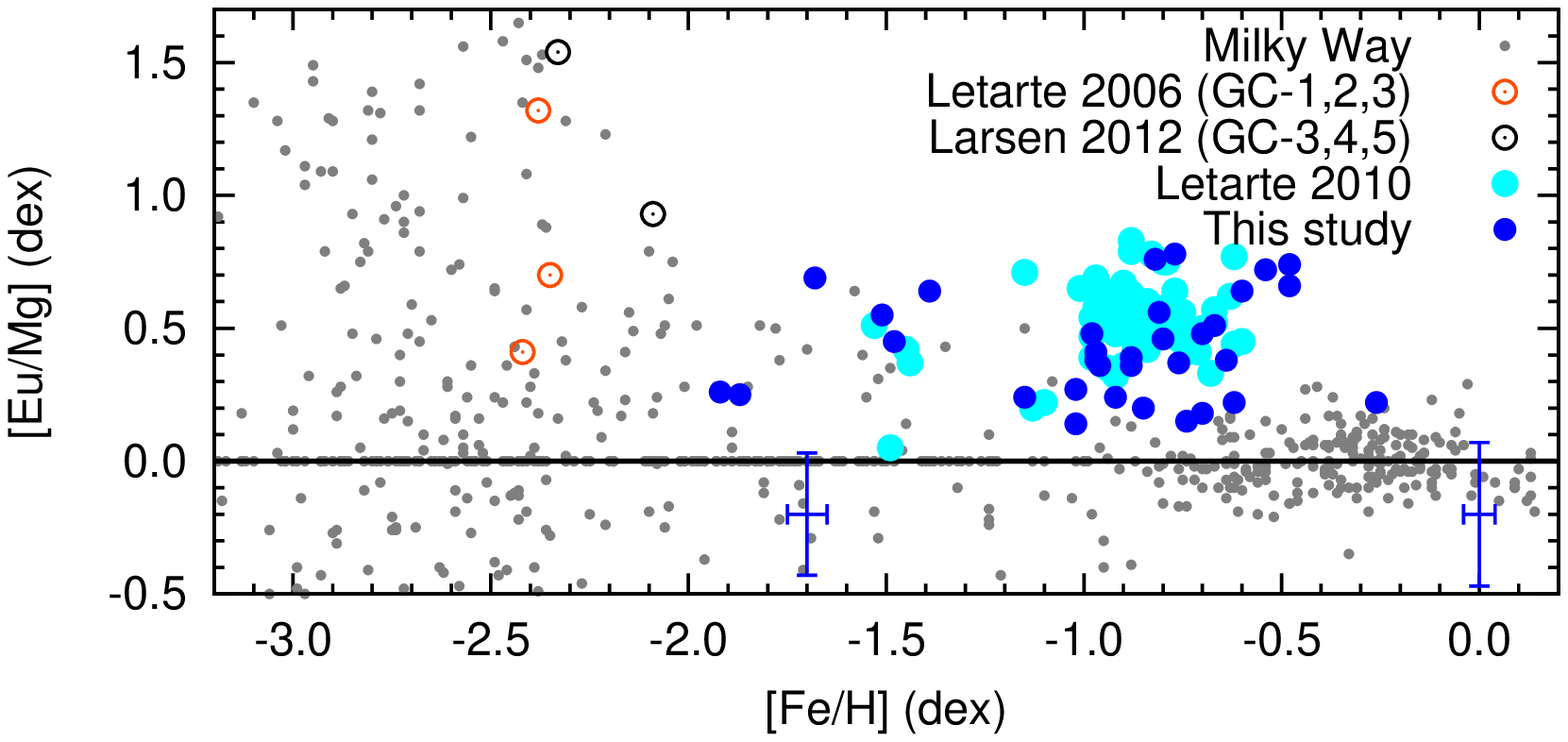}
      \caption{Same as Fig.~\ref{ironpeak} for [Eu/Mg].}
        \label{EuMg}
   \end{figure}

 Plotting the evolution of [Eu/Mg] as a function of [Fe/H] indicates that [Eu/Mg] in Fornax falls largely above its MW value and shows a large dispersion (See Fig.~\ref{EuMg}). Such high values for [Eu/Mg] (and similarly for [Eu/O]) have already been reported by \cite{McWil2013} for the Sgr dwarf spheroidal. Based on the study of extremely metal-poor stars in the MW halo, it has been proposed that a large fraction of the r-process elements are produced only by a rare sub-class of SNe II that would not count for more than a few percents of all SNe II. The details of this rare r-process are not known. However, the differences in [Eu/Mg] and [Eu/O] between the MW on one side and Sagittarius and Fornax on the other side indicate that at least one of the parameters controlling this process is different between these galaxies. It seems at first sight unlikely that paramaters such as the angular momentum or the binarity of the SNe II progenitors differ notably between the MW and Fornax/Sagittarius. For instance, \cite{Minor2013} show that Fornax is consistent with having binary population similar to that of the Milky Way field binaries to within 68\%. Moreover, the metallicity range for which these high values of [Eu/O] and [Eu/Mg] have been observed in the dwarf spheroidals is similar to the one of the MW thick disk, thus ruling out the metallicity as the origin of the differences. \cite{McWil2013} propose that the yields of the r-process are sensitive to the mass of the SNe II progenitors, in the sense that the r-process elements are synthesized preferentially in low mass SNe II. If this explanation is correct, the high [Eu/O] and [Eu/Mg] ratios naturally arise from a SNe II mass distribution deprived from its most massive elements.

\section{Discussion}
\label{interp}

\subsection{The classical scenario}

\par In the traditional interpretation \citep[e.g.][]{Tin1979,Matt1990,Gil1991,Matt2003}, the low [$\alpha$/Fe] in the dSphs are believed to be the direct consequence of the time delay between SNe II and SNe Ia: given the short timescale of SNe II, the enrichment of the ISM by the massive stars starts almost immediately after the beginning of star formation, while the SNe Ia contribute only later to chemical enrichment, leading to low [$\alpha$/Fe] values. Because of this time-delay, this scenario is sometimes referred to as the "[$\alpha$/Fe] clock".\\ 
\indent In the Sculptor dSph, star formation decreased after an early burst and stopped around 7 Gyr ago. Star formation possibly stopped because of strong galactic winds removing a large fraction of the gas \citep[e.g.][]{Lan2003}. From a chemical evolution model in full cosmological context, \cite{Romano2013} indicate that gas removal via internal (SNe feedback) and/or external causes (tidal interactions, ram pressure stripping) is mandatory to reproduce the observed properties of Sculptor. \cite{Revaz2009,Revaz2012} found that SNe feedback and galactic winds are inefficient to quench star formation, external causes are then needed to strip the gas. Sculptor displays a well-defined "knee" (the inflexion point where [$\alpha$/Fe], especially [Mg/Fe], begins to turn down) at 10.9$\pm$1.0 Gyr and at a metallicity of $\approx$--1.8 dex \citep{Tolstoy2009,deBoer2012a}. \\ 

\par
\begin{figure}[htp]
   \centering
   \includegraphics[width=\columnwidth]{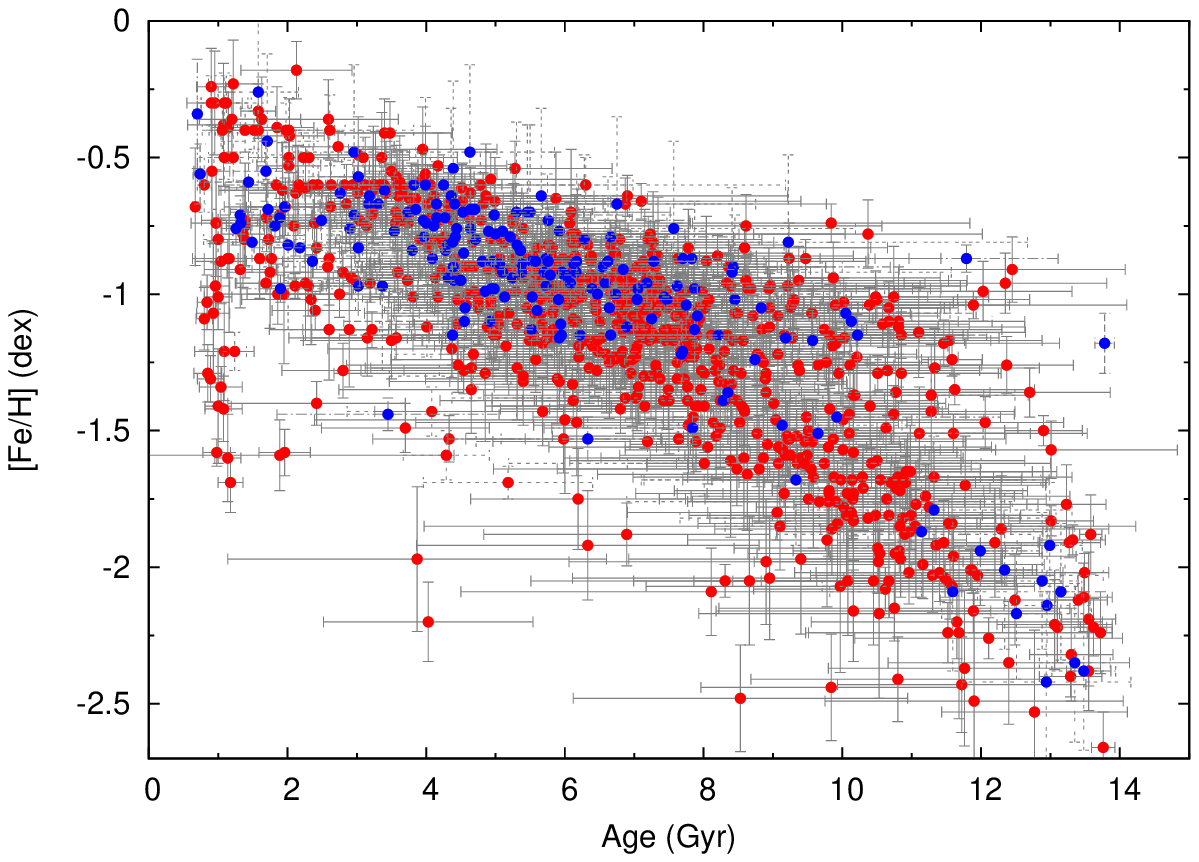}
      \caption{The Age-Metallicity Relation of stars on the upper RGB in the Fornax dSph. Medium \citep{Kirby2010} and high resolution spectroscopy \citep[this study, ][]{Let2006,Let2010} is shown as blue points, while the metallicity estimates from CaT spectroscopy \citep{Pont2004,Stark2010} are shown in red.}
         \label{AMR}
   \end{figure}

\begin{figure*}[htp!]
   \centering
   \includegraphics[angle=-90,width=0.8\textwidth]{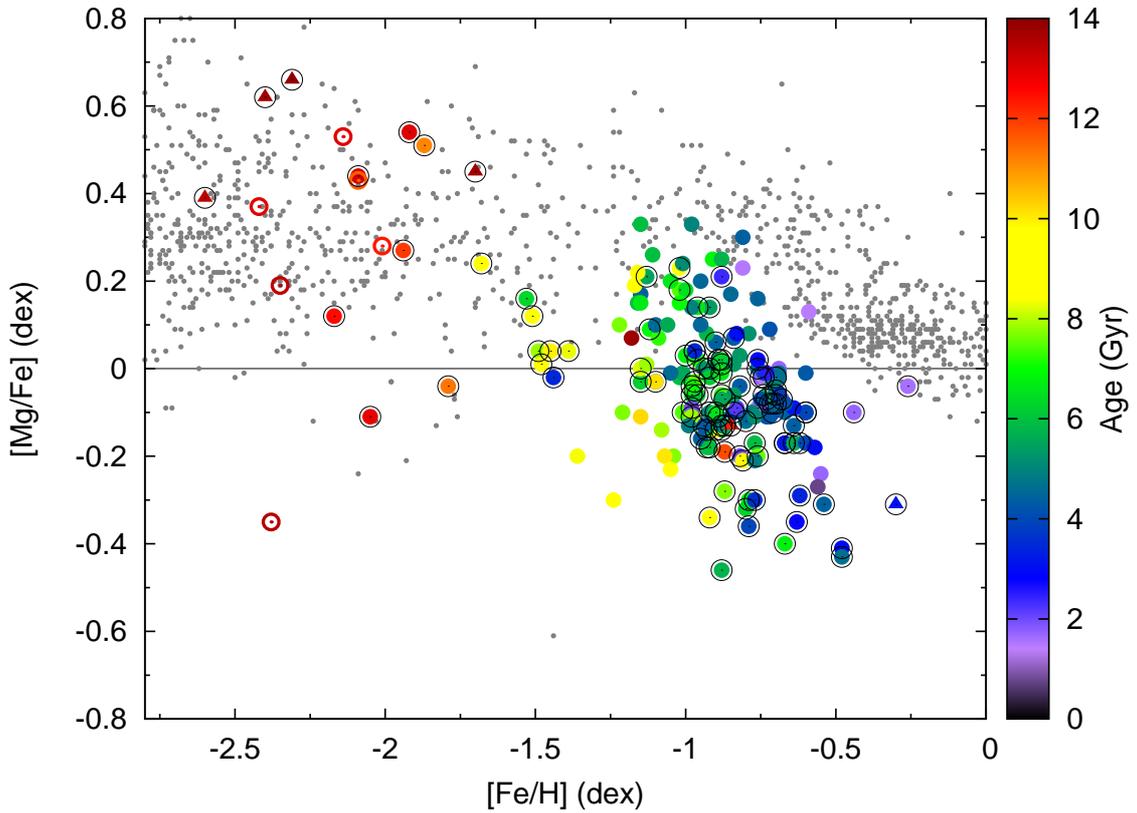}
      \caption{[Mg/Fe] vs [Fe/H] from MR/HR spectroscopic measurements of Fornax RGB stars (coloured filled circles/triangles). The colours represent the age in Gyr, derived from the SFH. Stars in the Milky Way are shown for comparison (small grey points). Field stars for which the probability distribution for age could be determined are shown as filled circles, while filled triangles show stars for which no statistical age estimate could be derived. For those stars, only an age estimate is given, based on the closest distance from synthetic CMD satisfying the magnitude and metallicity constraints. Globular cluster stars are shown as open circles. Stars belonging to the high resolution samples \citep[this study,][]{Let2010} are surrounded by black open circles.}
         \label{MgFe_age}
   \end{figure*}

\begin{figure*}[hbp!]
   \centering
   \includegraphics[angle=-90,width=0.8\textwidth]{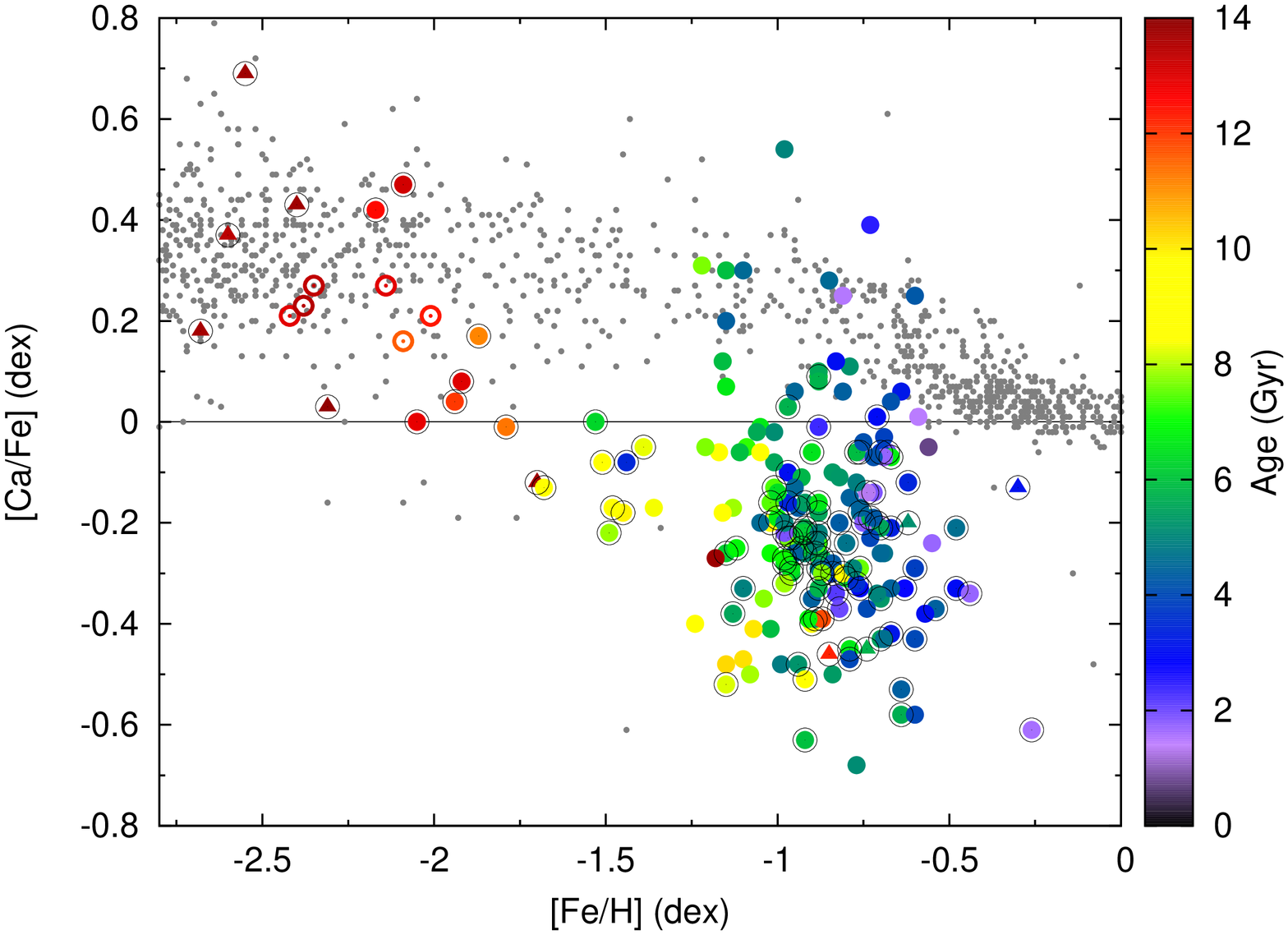}
      \caption{Same as Fig.~\ref{MgFe_age} for [Ca/Fe].}
         \label{CaFe_age}
   \end{figure*}

\begin{figure}[hbp!]
   \centering
   \includegraphics[width=\columnwidth]{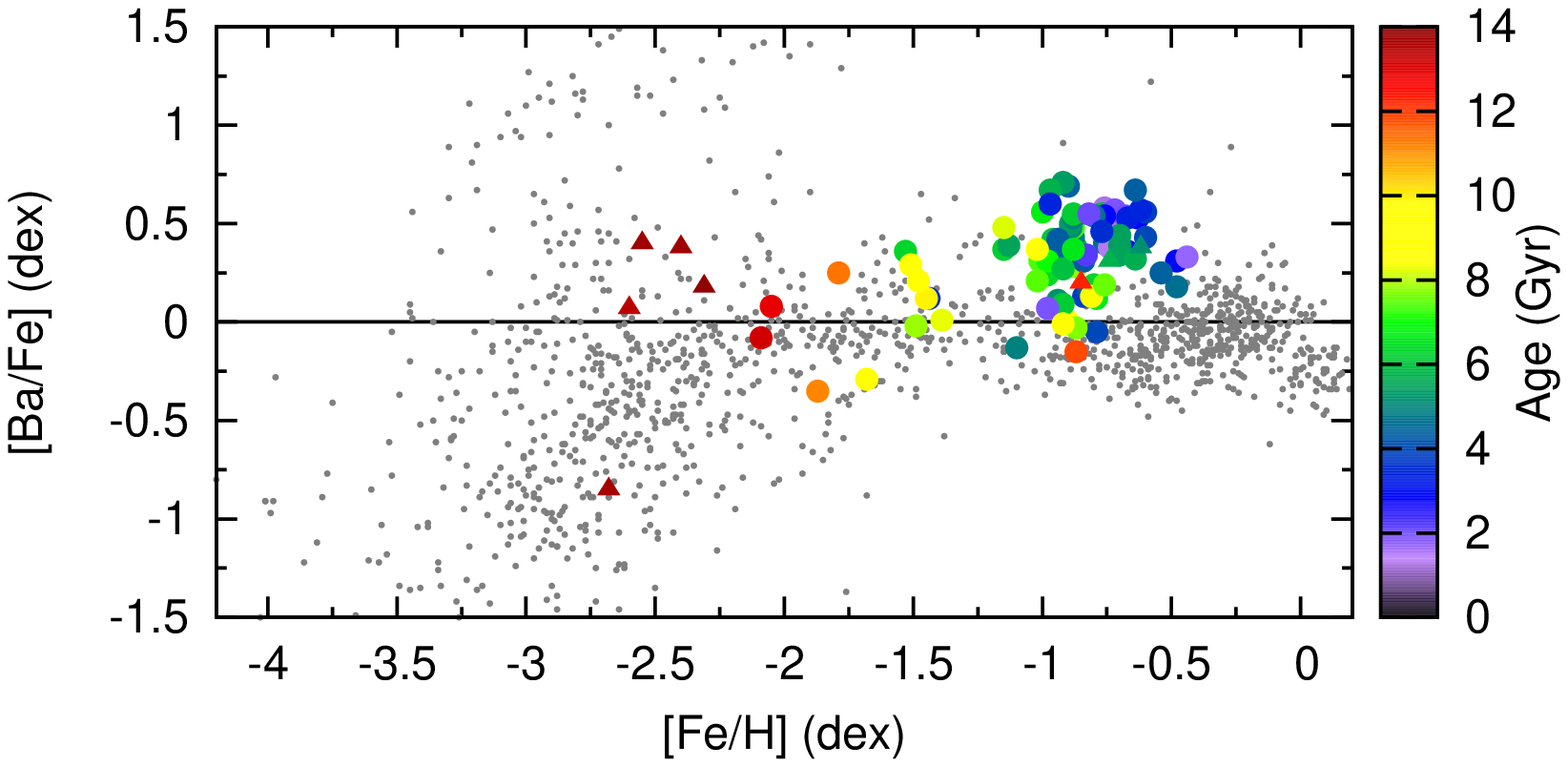}
      \caption{Same as Fig.~\ref{MgFe_age} for [Ba/Fe].}
         \label{BaFe_age}
   \end{figure}

 In this scenario, it appears from the age-metallicity relation (Fig.~\ref{AMR}) and from the evolution of [Mg/Fe] and [Ca/Fe] with [Fe/H] (Figs~\ref{MgFe_age},\ref{CaFe_age}) that the knee in Fornax occurred approximately between 12--10 Gyrs ago at [Fe/H] between --2.0 and --1.8 dex. This range of values perfectly matches the value of --1.9 dex for the Mg knee recently found by \cite{Hen2014}. They also report that the knees for Si and Ti are less clearly defined but fall anyway below --1.8 dex. The rise of the s-process seems to follow the [$\alpha$/Fe] knee (see Fig.~\ref{BaFe_age}), at least within our age resolution. These values are similar to those found for the Sculptor dSph. Sculptor is less massive and more metal-poor than Fornax and thus one would expect that Fornax would produce and retain metals more efficiently to reach a higher metallicity before SNe Ia begin to contribute to the chemical enrichment. It is possible that Sculptor was able to produce metals but could not retain them because of its relatively low mass. In contrast, Fornax, {\raise.17ex\hbox{$\scriptstyle\mathtt{\sim}$}}10 times more massive than Sculptor, should produce more metals (SF efficiency scales with galaxy mass) and keep a larger fraction of them. A merger scenario (see Sect.~\ref{merger}) could explain why Fornax early enrichment scenario is similar to Sculptor despite the fact that it is {\raise.17ex\hbox{$\scriptstyle\mathtt{\sim}$}}10 times more massive.\\

\par The qualitative arguments discussed just above, as well as most of the chemical evolution models, assume a relatively uniform SF efficiency over time. \cite{Hen2014} indicate however that, in the absence of a merger, a continuous SFH (or even a bursty SFH with constant SF efficiency for all the bursts) does not allow to produce a metal-poor knee together with the other characteristics of Fornax. They found in contrast that a bursty SFH (3 bursts at [0--2.6], [2.8--13.2], [13.7--14] Gyr) including (short) periods with no star formation at all and a SF efficiency varying from one burst to another is mandatory to recover a knee at low metallicity, together with a metal-rich MDF and a $\alpha$-poor plateau at high metallicities. Such breaks in star formation are not seen even in the most recent photometric studies \citep[e.g.][]{deBoer2012b,delPino2013}. It is however very likely that the very short breaks (0.5 Gyr) proposed by \cite{Hen2014} cannot be recovered by photometric SFH studies, even if both studies quoted above use deep CMDs going down to the MSTO and a CaT calibration from \cite{Stark2010} improved towards low metallicities.\\

\par
\begin{figure}[htp]
   \centering
   \includegraphics[angle=0,width=\columnwidth]{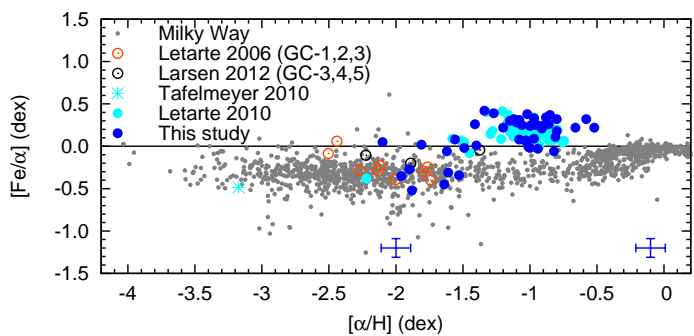}
      \caption{The distribution of [Fe/$\alpha$] as a function of [$\alpha$/H] for our sample of RGB stars in the Fornax dSph as filled circles (cyan). We computed [$\alpha$/H] as $\dfrac{[Mg/H] + [Ca/H]}{2}$.}
        \label{FeAlp}
   \end{figure}

 As we already mentioned, Fornax is more massive than Sculptor and most of the other dSphs, its potential is therefore deeper and it is more difficult for galactic winds to remove the gas, as was pointed out by \cite{Li2013}. (We note in passing that the wind efficiency of the \cite{Hen2014} model falls at the lower end of the values previously used to model dSph galaxies). The steady decline of [Mg/Fe], the fact that the low-mass AGBs yields have not been lost also argue against strong galactic winds. Indeed \cite{deBoer2012b} have shown that Fornax experienced a continuous SFH from 14 to 0.25 Gyr, dominated by an intermediate-age population (1--10 Gyr). \cite{Li2013} also showed that the contribution of massive stars to [Mg/H] increases monotonically with [Fe/H], supporting the idea that the galactic winds in Fornax were not strong enough to stop the star formation. This can also be seen in Fig.~\ref{FeAlp} where [Fe/$\alpha$] plotted against [$\alpha$/H] shows a flat, extended plateau at the high [$\alpha$/H] end of the distribution, proving that massive stars keep enriching the ISM in $\alpha$-elements, even if they are not anymore the main source for the Fe-enrichment.\\

The underlying cause for a possible change in the SF efficiency is still unknown. Many studies have investigated how star formation is affected by tidal interactions and/or ram-pressure stripping \citep[e.g.][]{Mayer2006,Pena2008,Nichols2014}, pericentric passage \citep[e.g.,][for Carina]{Pas2011}, or dark matter profile \citep[e.g.][]{Sawa2011}. One can also invoke the fall back of gas (see here below) or possible merger events (see Sect.~\ref{merger}). \cite{Nichols2014} have shown that tidal interactions during perigalacticon passages are sufficient to quench star formation, even in gas-rich galaxies. Therefore tidal forces keeping gas at low densities (with the help of supernovae) seem to be a more promising explanation than galactic winds.

\par Figs~\ref{MgFe_age}, \ref{CaFe_age} indicate that, for a given age bin, both [Mg/Fe] and [Ca/Fe] show a large scatter: extremely $\alpha$-poor stars can have comparable [Fe/H], with roughly the same age, as supersolar [$\alpha$/Fe] stars. However, from a comparison with Figs \ref{MgFe}, \ref{CaFe} it appears that this scatter is most likely coming from the medium resolution sample. This scatter is not seen in the high resolution samples \citep[this study,][]{Let2010}. Despite the general trend of [$\alpha$/Fe] decreasing with increasing [Fe/H], we note (even when considering only the HR sample) that some of the youngest stars ($<$~4~Gyr) still have --0.2$<$[$\alpha$/Fe]$<$0.0 dex, similar or higher than that the low [$\alpha$/Fe] values reached by some older stars ($>$~6~Gyr). This may suggest inhomogeneous mixing of the ISM in Fornax and possibly the fall-back of previously expelled gas. 

\par \cite{deBoer2013} studied the inner overdensity of Fornax and concluded that it was likely formed from Fornax gas because it matches the age-metallicity relation of the Fornax field stars, thus compatible with a scenario in which it would be formed from previously expelled gas. They also discovered a new stellar overdensity of very young, metal-rich stars and therefore probably formed from the gas left over after the previous star formation episode.\\

\subsection{A top-light IMF ?}

\par There is mounting evidence of steep (i.e. top-light) IMF for the dwarf galaxies \citep[e.g.][]{Tolstoy2003,Krou2013}. As far as Fornax is concerned, \cite{Tsuji2011} (from a detailed model of the galaxy) and \cite{Li2013} (comparing the relative contributions of different processes to the nucleosynthesis in the Milky Way and in Fornax) reached the same conclusion. From the analysis of RGB stars in the Sagittarius dSph, \cite{McWil2013} conclude that the low [$\alpha$/Fe] observed in this galaxy cannot be explained by the classical SNe Ia time-delay scenario but result rather from a top-light IMF, which deprived the galaxy of the most massive stars ($>$30M$_{\odot}$). Indeed, the production of oxygen and, to a lesser extent, of magnesium is dominated by the most massive SNe II \citep{WooWea1995}. Therefore the low values of [Mg/Fe] reported by \cite{McWil2013} in Sagittarius and that are also found at high metallicities in Fornax are compatible with an IMF that lacks the most massive SNe II. \cite{McWil2013} also proposed that the r-process elements are synthesized preferentially in low mass SNe II. A direct consequence of this hypothesis is that the supersolar [Eu/Mg] we find in Fornax can be easily explained by a top-light IMF. Several other results in this paper (e.g. the low [Ni/Fe], the high values of [Ba/Fe], [La/Fe]) are also compatible with a top-light IMF as they reflect the strong influence of SNe Ia and AGB stars in driving the abundance pattern of Fornax.\\ 

\par On the other hand, it should be noted that the studies quoted above could only use the Fornax data available at that time, i.e. globular clusters stars in the metal-poor regime \citep{Let2006} and field stars in the metal-rich, $\alpha$-poor regime \citep{She2003,Let2010}. Only a handful of field stars had [Fe/H]$<$--1.2 dex. It would be interesting to see in particular how the fitting of components of \cite{Li2013} would be modified by taking into account our new sample that includes almost 15 stars with [Fe/H]$<$--1.4 dex. From the results provided in this paper, we expect that the contribution of the primary-like yields would increase. If we revisit the plots of \cite{Tsuji2011}, their model Fnx-1 (with a mass-cut at 25 M$_{\odot}$) could probably be ruled-out as it does not fit the data for [Mg/Fe] in the metal-poor regime. We note however that [Eu/Fe] and [Ba/Fe] (mostly formed via the r-process in the metal-poor regime) are quite well reproduced by this model.
\par The same holds for the studies of \cite{Tolstoy2003} and \cite{McWil2013}, that cover a rather narrow range of metallicities. For instance, the stars in \cite{McWil2013} have [Fe/H] close to the mean value of Sagittarius, i.e. at a stage where the galaxy is already well evolved. Therefore, it might very well be that the top-light IMF takes place only at some point in the evolution of the galaxy.\\

\par Moreover, a steep IMF is the natural outcome of systems where gas is not present in sufficient quantities to form the largest molecular clouds that in turn will give birth to the most massive stars \citep{Oey2011}. The presence of a system of globular clusters both in Fornax and in Sagittarius indicates that gas should have been present in sufficient quantities at least in the very early times of these galaxies.
The low values of [$\alpha$/Fe] reported here (and by \cite{Let2010}, \cite{Kirby2010}, and \cite{Hen2014} in recent papers) could be explained by both the top-light IMF and the time-delay scenarios. However, the heavy element ratios, especially [Eu/Mg] at low [Mg/Fe] favour another mechanism, such as different SNe II yields in Fornax. As these mechanisms are not mutually exclusive, better constraints (in particular on the SNe yields) are mandatory to determine their relative weights in the chemical evolution of Fornax.\\

\subsection{Fornax as the result of a minor merger ?}
\label{merger}

\par Several papers have recently revisited the hypothesis that Fornax has undergone a relatively recent minor merger. Recent studies of the Star Formation History in Fornax \citep[e.g.][]{Col2008,deBoer2012b,delPino2013} all report a significant peak in star formation approximately 4 Gyr ago. The mechanism that triggered this peak is not known. It has been proposed that Fornax experienced a minor merger with another sub-halo. This scenario gives an explanation for the shell-like structures \citep{Col2004,Col2005b,deBoer2013} in the inner regions of Fornax. However, it has been realized that i) these features are $\approx$ 1.5~Gyr old and therefore do not match the timescale of a merger event 4 Gyr ago; ii) given their chemical composition, these structures are most likely formed from Fornax gas rather than from gas with an external origin \citep{Ol2006,Col2008,deBoer2012b,deBoer2013}. Moreover, the probability of a relatively recent collision between two sub-halos orbiting the Milky Way is much smaller than it was at the early times of the Local Group \citep[e.g.][]{deRijcke2004} \\

\par \cite{Am2012c} divided Fornax in three chemo-dynamical components (metal-poor, intermediate and metal-rich) where the more metal-rich components are also more spatially concentrated and have colder kinematics. They also report a misalignment between the angular momentum of the metal-poor and intermediate metallicity populations, interpreted as counter-rotation. Putting all the pieces together, \cite{Am2012c} proposed that Fornax is the result of the late merger of a bound pair. In a numerical simulation where the merging companion weights 5\% of the Fornax mass, \cite{Yo2012} could recover qualitatively the shell-like substructure, with a merging event occurring between 3.5 and 2.1 Gyr ago.\\

\par It is tempting to associate the three subpopulations of \cite{Am2012c} with the three age groups that stand out in Figs.~\ref{MgFe_age}-\ref{CaFe_age}. In this picture, their metal-poor population (red+yellow) would be older than 8 Gyr, while their intermediate population (green) would have ages ranging from 5 to 7 Gyr and the metal-rich population (blue) ages would range between 2 and 4 Gyr. In particular, the peaks of the MDFs of our age groups correspond nicely to the peaks of the MDFs of the metallicity groups of \cite{Am2012c}. However the distribution of the stars within these groups is not a perfect match, as the classification of \cite{Am2012c} is based on kinematics and [Fe/H], while we show here that stars at a given [Fe/H] span quite a large range in age and [$\alpha$/Fe]. Adding these criteria (in particular the age) would probably help in better disentangling these groups.\\

The merging of a bound pair or the accretion of a gas-rich system could also be a way to explain a higher SF efficiency in Fornax in the last few Gyrs. It is however not sure if this enhanced SFR observed in massive, interacting galaxies \citep[e.g.][]{Ell2010,Scu2012} also holds for dwarf galaxies. Such a late addition of mass could simultaneously explain the similarities between Fornax and Sculptor in the metal-poor regime if the former was originally less massive.\\

 
\section{Summary and conclusions}

\par We determined from FLAMES/GIRAFFE high resolution spectroscopy the abundances of $\alpha$, iron-peak and neutron-capture elements in a sample of 47 individual Red Giant Branch stars in the Fornax dwarf spheroidal galaxy. 
We also determined accurate ages for the whole sample by computing the age probability distribution of stars located at the same position as our targets in a synthetic colour-magnitude diagram of Fornax.\\
\par As it has already been observed in several other dwarf spheroidal galaxies, the old, metal-poor population of Fornax is $\alpha$-rich and the younger, more metal-rich population is $\alpha$-poor. In the classical interpretation of the time delay between SNe II and SNe Ia, our results confirm that SNe Ia started to contribute to the chemical enrichment at [Fe/H] between --2.0 and --1.8 dex. We find that the onset of SNe Ia took place around 12--10 Gyrs ago. The value of [Fe/H] at which the turnover in [$\alpha$/Fe] occurs in Fornax is surprisingly similar to the one derived for the Sculptor dwarf spheroidal galaxy, despite the fact that Fornax is much more massive than Sculptor.\\
\par Our results (e.g., low [Mg/Fe], supersolar [Eu/Mg]) suggest a top-light initial mass function in Fornax, as it has been recently reported for Sagittarius, a dwarf spheroidal galaxy with a mass similar to Fornax. The high values we found for [Ba/Fe] and [La/Fe] in the metal-rich regime reflect the strong influence of SNe Ia and AGB stars in the abundance pattern of the younger and more metal-rich stars of Fornax. 
Despite this possible top-light IMF, our results indicate that massive stars kept enriching the interstellar medium in $\alpha$-elements, although they were no longer the main contributor to the iron enrichment, confirming that star formation kept going on over almost the whole history of Fornax, and the gas was continually enriched. The time-delay and the top-light IMF mechanisms are not mutually exclusive, they are probably going on simultaneously in Fornax. A better knowledge of the SNe yields is however mandatory to determine their relative contributions in the chemical evolution of Fornax.\\
\par Our results are compatible with the three subpopulations described by \cite{Am2012c}, although age and [$\alpha$/Fe] should also be taken into account when matching a star to any of these groups. Our results do not allow us to confirm or rule out the minor merger or the merging of a bound pair scenarios that have been proposed in the case of Fornax. They are however compatible with the fall-back of previously expelled gas.

\begin{acknowledgements}
We thank the anonymous referee for his/her detailed comments that helped to improve the quality of the paper. PJ acknowledges the support of the Swiss National Science Foundation. GB acknowledges financial support by the Spanish Ministry of Economy and Competitiveness (MINECO) under the Ramon y Cajal Program (RYC-2012-11537). ES gratefully acknowledges the Canadian Institute for Advanced Research (CIFAR) Global Scholar Academy for support. The authors thank ISSI (Bern) for support of the team “Defining the Life-Cycle of Dwarf Galaxy Evolution: the Local Universe as a Template”.
\end{acknowledgements}


\begin{appendix}
\section{Tables}

\begin{table}[ht!]
\caption{Abundance ratios for Arcturus (median S/N) determined by \cite{vdS2013} compared to our results for Arcturus obtained in the same way as for our Fornax sample. We also indicate the number of lines and the method that have been used (EW: equivalent width or SS: spectral synthesis).}
\begin{tabular}{r|cc}
\hline\hline
Elements      &  Abundance ratios by  &  Abundance ratios:  \\ 
              &    \cite{vdS2013}     &    this study       \\
              &       (dex)           &       (dex)         \\ 
\hline
 {[}Ba/Fe]    &       -0.20$\pm$0.07 (2)  - SS &        0.05$\pm$0.12 (1)  - EW \\ 
 {[}Ca/Fe]    &        0.04$\pm$0.02 (10) - EW &        0.05$\pm$0.10 (5)  - EW \\ 
 {[}Cr/Fe]    &       -0.08$\pm$0.05 (3)  - SS &       -0.26$\pm$0.09 (1)  - EW \\ 
 {[}Eu/Fe]    &        0.40$\pm$0.07 (2)  - SS &        0.25$\pm$0.10 (1)  - EW \\ 
 {[}Fe I/H]   &       -0.65                    &       -0.64$\pm$0.07 (58) - EW \\ 
 {[}Fe II/H]  &        -                       &       -0.55$\pm$0.10 (7)  - EW \\ 
 {[}Na/Fe]    &        0.08$\pm$0.04 (3)  - SS &        0.01$\pm$0.12 (1)  - EW \\ 
 {[}Ni/Fe]    &        0.07$\pm$0.03 (6)  - EW &        0.07$\pm$0.11 (12) - EW \\ 
 {[}Sc/Fe]    &        0.23$\pm$0.04 (4)  - SS &        0.02$\pm$0.11 (1)  - EW \\ 
 {[}Si/Fe]    &        0.32$\pm$0.06 (2)  - EW &        0.23$\pm$0.10 (2) - EW \\ 
 {[}Ti I/Fe]  &        0.35$\pm$0.03 (8)  - EW &        0.28$\pm$0.11 (11) - EW \\
 {[}Ti II/Fe] &        0.30$\pm$0.04 (3)  - EW &        0.29$\pm$0.10 (4)  - EW \\
\hline
\end{tabular}
\label{Arcturus}
\end{table}

\begin{table}[ht!]
\caption{Atmospheric parameters for $\alpha$~Cet and $\gamma$~Sge in the GAIA benchmark sample (GBS) determined by \cite{Jofre2013} compared to our results obtained in the same way as for our Fornax sample.}
\begin{tabular}{c|cc|cc}
\hline\hline
Star  & \multicolumn{2}{c|}{$\alpha$~Cet} & \multicolumn{2}{c}{$\gamma$~Sge} \\
      & this study & GBS & this study & GBS \\ 
\hline
T$_{\rm eff}$ & 3800   &   3796 & 3800   & 3807    \\
{\it log g}   & 1.3    &   1.36 & 1.4    & 1.43    \\
V$_{t}$       & 0.6    &   0.68 & 1.0    & 1.05    \\
{[}Fe/H]      & --0.44 & --0.45 & --0.08 & --0.16  \\
\hline
\end{tabular}
\label{GAIAbenchmark}
\end{table}

\begin{table*}[ht!]
\centering
\caption{Positional, photometric and kinematic data for our targets. The BVI photometry comes from CTIO/MOSAIC, supplemented by ESO/WFI when the I band was not available. JHK photometry comes from ESO/VISTA.}
\begin{tabular}{cccccccccc}
\hline\hline
Target       &     RA     &       dec    &   B    &   V    &   I    &   J    &   H    &   K    & V$_{r}$     \\ 
             &    hms     &       dms    &  mag   &  mag   &  mag   &  mag   &  mag   &  mag   & km/s        \\ 
\hline
 Fnx-mem0514 & 2:38:40.14 & -34:39:11.80 & 20.056 & 18.450 & 16.897 & 15.520 & 14.753 & 14.547 & 39.3$\pm$0.36 \\
 Fnx-mem0522 & 2:38:47.86 & -34:39:39.40 & 20.149 & 18.535 & 16.872 & 15.645 & 14.896 & 14.685 & 40.1$\pm$0.94 \\ 
 Fnx-mem0528 & 2:37:32.45 & -34:39:53.30 & 19.743 & 18.536 & 17.290 & 16.036 & 15.429 & 15.290 & 55.9$\pm$0.42 \\ 
 Fnx-mem0532 & 2:38:01.37 & -34:40:11.30 & 20.218 & 18.722 & 17.288 & 16.135 & 15.438 & 15.272 & 48.5$\pm$0.94 \\ 
 Fnx-mem0538 & 2:38:03.84 & -34:40:38.70 & 19.893 & 18.561 &   -    &   -    &   -    &   -    & 79.5$\pm$0.00 \\ 
 Fnx-mem0539 & 2:37:32.86 & -34:40:41.80 & 20.096 & 18.570 & 17.139 & 15.951 & 15.239 & 15.080 & 25.0$\pm$0.24 \\ 
 Fnx-mem0543 & 2:38:02.43 & -34:41:00.40 & 20.140 & 18.544 & 17.018 & 15.814 & 15.092 & 14.909 & 51.1$\pm$0.30 \\ 
 Fnx-mem0546 & 2:38:46.31 & -34:41:14.50 & 20.132 & 18.693 & 17.286 & 16.268 & 15.636 & 15.486 & 65.2$\pm$0.57 \\ 
 Fnx-mem0556 & 2:38:45.53 & -34:41:44.90 & 20.159 & 18.666 & 17.171 & 16.063 & 15.344 & 15.151 & 50.4$\pm$0.72 \\ 
 Fnx-mem0571 & 2:38:49.44 & -34:42:29.20 & 19.802 & 18.530 & 16.991 & 16.006 & 15.409 & 15.214 & 38.9$\pm$1.06 \\ 
 Fnx-mem0572 & 2:38:44.58 & -34:42:33.10 & 20.185 & 18.665 & 17.172 & 16.042 & 15.335 & 15.138 & 47.4$\pm$0.38 \\ 
 Fnx-mem0574 & 2:38:26.46 & -34:42:36.60 & 20.291 & 18.606 & 16.979 & 15.794 & 15.028 & 14.820 & 38.5$\pm$0.24 \\ 
 Fnx-mem0584 & 2:37:23.65 & -34:43:20.90 & 19.832 & 18.653 & 17.518 & 16.386 & 15.824 & 15.700 & 65.0$\pm$1.16 \\ 
 Fnx-mem0595 & 2:38:35.50 & -34:44:02.40 & 20.190 & 18.567 & 16.999 & 15.839 & 15.122 & 14.937 & 48.7$\pm$0.19 \\ 
 Fnx-mem0598 & 2:38:00.01 & -34:44:21.80 & 20.340 & 18.582 & 16.928 & 15.558 & 14.758 & 14.556 & 37.9$\pm$0.38 \\ 
 Fnx-mem0604 & 2:38:52.25 & -34:44:55.00 & 20.047 & 18.254 & 16.458 & 15.229 & 14.433 & 14.216 & 44.7$\pm$0.29 \\ 
 Fnx-mem0606 & 2:38:30.20 & -34:45:02.70 & 20.464 & 18.623 & 16.647 & 15.316 & 14.483 & 14.265 & 51.3$\pm$1.11 \\ 
 Fnx-mem0607 & 2:38:22.41 & -34:45:06.00 & 20.123 & 18.576 & 17.014 & 15.871 & 15.170 & 14.968 & 50.3$\pm$0.38 \\ 
 Fnx-mem0610 & 2:38:36.71 & -34:45:20.70 & 20.006 & 18.537 & 16.526 & 15.358 & 14.754 & 14.523 & 57.6$\pm$0.29 \\ 
 Fnx-mem0612 & 2:38:50.89 & -34:45:35.00 & 19.887 & 18.541 & 17.144 & 16.134 & 15.500 & 15.338 & 68.4$\pm$0.77 \\ 
 Fnx-mem0613 & 2:38:59.50 & -34:45:36.00 & 20.029 & 18.493 & 16.826 & 15.642 & 14.894 & 14.716 & 53.5$\pm$0.81 \\ 
 Fnx-mem0621 & 2:37:44.47 & -34:45:58.60 & 19.741 & 18.607 & 17.293 & 16.238 & 15.668 & 15.516 & 53.6$\pm$0.50 \\ 
 Fnx-mem0626 & 2:38:21.97 & -34:46:36.60 & 20.119 & 18.646 & 17.135 & 16.041 & 15.367 & 15.182 & 38.0$\pm$0.43 \\ 
 Fnx-mem0629 & 2:37:20.19 & -34:46:45.70 & 20.200 & 18.691 & 17.207 & 16.030 & 15.306 & 15.149 & 53.2$\pm$0.79 \\ 
 Fnx-mem0631 & 2:38:52.93 & -34:47:01.00 & 19.677 & 18.234 & 16.797 & 15.726 & 15.048 & 14.878 & 43.5$\pm$0.44 \\ 
 Fnx-mem0633 & 2:37:55.85 & -34:47:15.70 & 20.157 & 18.514 & 16.914 & 15.637 & 14.884 & 14.725 & 34.1$\pm$0.19 \\ 
 Fnx-mem0634 & 2:38:55.48 & -34:47:20.70 & 20.238 & 18.739 & 17.232 & 16.127 & 15.415 & 15.282 & 50.3$\pm$0.88 \\ 
 Fnx-mem0638 & 2:38:42.14 & -34:47:40.00 & 20.125 & 18.533 & 16.885 & 15.680 & 14.907 & 14.696 & 35.2$\pm$0.57 \\ 
 Fnx-mem0647 & 2:37:24.24 & -34:48:37.20 & 20.007 & 18.604 & 17.262 & 16.061 & 15.424 & 15.285 & 63.7$\pm$1.05 \\ 
 Fnx-mem0654 & 2:37:22.52 & -34:49:04.40 & 19.949 & 18.690 & 17.429 & 16.329 & 15.753 & 15.636 & 44.3$\pm$1.27 \\ 
 Fnx-mem0664 & 2:38:17.87 & -34:49:53.10 & 19.859 & 18.679 & 17.409 & 16.412 & 15.857 & 15.736 & 23.9$\pm$0.34 \\ 
 Fnx-mem0675 & 2:37:55.73 & -34:50:25.00 &   -    & 18.756 & 17.354 & 16.237 & 15.538 & 15.396 & 66.2$\pm$0.20 \\ 
 Fnx-mem0678 & 2:38:42.95 & -34:50:55.90 & 20.316 & 18.726 & 17.148 & 15.935 & 15.175 & 14.984 & 73.7$\pm$0.91 \\ 
 Fnx-mem0682 & 2:38:09.02 & -34:51:06.70 & 20.268 & 18.634 & 17.047 & 15.770 & 15.054 & 14.851 & 54.9$\pm$0.77 \\ 
 Fnx-mem0704 & 2:38:00.50 & -34:52:33.70 & 19.869 & 18.658 & 17.377 & 16.413 & 15.858 & 15.745 & 59.0$\pm$2.71 \\ 
 Fnx-mem0712 & 2:38:37.11 & -34:53:10.20 & 19.855 & 18.603 & 17.246 & 16.247 & 15.637 & 15.465 & 47.6$\pm$1.25 \\ 
 Fnx-mem0714 & 2:38:50.23 & -34:53:19.20 & 19.641 & 18.211 & 16.727 & 15.607 & 15.040 & 14.875 & 38.8$\pm$1.07 \\ 
 Fnx-mem0715 & 2:38:11.07 & -34:53:20.40 & 20.257 & 18.722 & 17.245 & 16.031 & 15.353 & 15.168 & 44.6$\pm$0.33 \\ 
 Fnx-mem0717 & 2:38:40.56 & -34:53:30.50 & 20.175 & 18.571 & 17.027 & 15.788 & 15.041 & 14.854 & 36.1$\pm$0.33 \\ 
 Fnx-mem0725 & 2:38:15.52 & -34:54:22.40 & 20.062 & 18.861 & 17.573 & 16.588 & 16.010 & 15.851 & 50.1$\pm$0.09 \\ 
 Fnx-mem0732 & 2:37:47.45 & -34:54:52.60 & 19.943 & 18.493 & 16.902 & 15.868 & 15.215 & 15.103 & 25.9$\pm$0.25 \\ 
 Fnx-mem0738 & 2:37:22.05 & -34:55:20.70 & 19.834 & 18.490 & 16.901 & 15.743 & 15.138 & 14.993 & 27.5$\pm$1.04 \\ 
 Fnx-mem0747 & 2:37:34.47 & -34:56:08.70 & 20.120 & 18.725 & 17.356 & 16.218 & 15.593 & 15.440 & 57.5$\pm$0.91 \\ 
 Fnx-mem0754 & 2:38:30.93 & -34:56:53.90 & 20.178 & 18.576 & 16.933 & 15.735 & 15.030 & 14.840 & 32.8$\pm$0.29 \\ 
 Fnx-mem0779 & 2:37:42.52 & -34:58:30.20 & 19.777 & 18.470 & 17.083 &   -    &   -    &   -    & 69.7$\pm$0.59 \\ 
 Fnx-rgb0507 & 2:38:19.28 & -34:36:00.60 & 19.971 & 18.535 & 17.213 & 16.005 & 15.373 & 15.186 & 50.2$\pm$0.33 \\ 
 Fnx-rgb0509 & 2:38:21.30 & -34:36:18.10 & 20.142 & 18.516 & 16.633 & 15.348 & 14.608 & 14.389 & 83.8$\pm$0.46 \\ 
 Fnx-rgb0510 & 2:37:48.40 & -34:36:21.90 & 20.287 & 18.504 & 16.941 & 15.652 & 14.949 & 14.733 & 51.6$\pm$0.71 \\ 
 Fnx-rgb0522 & 2:38:02.66 & -34:37:03.30 & 19.807 & 18.517 & 17.154 & 16.162 & 15.574 & 15.398 & 62.6$\pm$1.45 \\ 
 Fnx-rgb0536 & 2:38:19.73 & -34:38:53.80 & 20.508 & 18.604 & 17.104 & 15.603 & 14.879 & 14.606 & 46.3$\pm$0.85 \\ 
 Fnx-rgb0539 & 2:38:46.77 & -34:39:24.70 & 20.082 & 18.524 & 16.960 & 15.754 & 15.003 & 14.796 & 51.0$\pm$0.38 \\ 
 Fnx-rgb0553 & 2:38:25.17 & -34:41:23.90 & 20.284 & 18.655 & 17.033 & 15.877 & 15.127 & 14.917 & 66.3$\pm$0.81 \\ 
 Fnx-rgb0556 & 2:38:34.99 & -34:41:37.80 & 20.254 & 18.716 & 17.151 & 16.015 & 15.286 & 15.107 & 43.5$\pm$0.80 \\ 
 Fnx-rgb0561 & 2:38:34.98 & -34:42:40.50 & 20.039 & 18.539 & 17.004 & 15.866 & 15.170 & 14.962 & 47.2$\pm$0.39 \\ 
 Fnx-rgb0574 & 2:38:41.08 & -34:44:20.10 & 19.672 & 18.157 & 16.653 & 15.509 & 14.812 & 14.600 & 49.2$\pm$0.20 \\ 
 Fnx-rgb0590 & 2:38:13.90 & -34:46:55.20 & 20.057 & 18.763 & 17.452 & 16.392 & 15.801 & 15.646 & 63.1$\pm$1.01 \\ 
 Fnx-rgb0596 & 2:38:30.85 & -34:47:44.10 & 19.847 & 18.697 & 17.383 & 16.498 & 15.965 & 15.830 & 28.6$\pm$1.15 \\ 
 Fnx-rgb0614 & 2:37:47.30 & -34:54:30.50 & 19.841 & 18.448 & 16.592 & 15.407 & 14.793 & 14.622 &  4.6$\pm$0.00 \\ 
\hline
\end{tabular}
\label{targets}
\end{table*}

\begin{table*}[!ht]
\centering
\caption{The atmospheric parameters (derived from photometry) for stars in our sample. Column 2 indicate the mass adopted for each star. Columns 3 to 7 gather the temperatures calculated from different colours, following the calibrations for giants from \cite{Rami2005}. Column 8 lists the bolometric correction, and column 9 the bolometric magnitude. Column 10 lists the mean temperature and column 11 the corresponding surface gravity $log~g$. The complete list of atmospheric parameters (T$_{\rm eff}$, {\it log g}, V$_{t}$ and [Fe/H]) is given in Table \ref{atmparam}.}
\begin{tabular}{ccccccccccc}
\hline\hline
 Target & mass & T$_{(B-V)}$ & T$_{(V-I)}$ & T$_{(V-J)}$ & T$_{(V-H)}$ & T$_{(V-K)}$ & $BC_{V}$ & M$_{bol}$ &$<$ T$_{\rm eff}$ $>$& {\it log g}\\ 
   &M$_{\odot}$&       K     &       K     &       K     &       K     &       K     &    mag   &    mag    &         K          &     dex     \\ 
 \hline
Fnx-mem0514 & 1.1 & 3806 & 4000 & 3778 & 3778 & 3783 & -1.19 & -3.67 & 3829 & 0.39 \\ 
Fnx-mem0522 & 1.1 & 3790 & 3936 & 3808 & 3799 & 3798 & -1.18 & -3.58 & 3826 & 0.43 \\ 
Fnx-mem0532 & 1.3 & 3941 & 4111 & 3971 & 3973 & 3977 & -0.92 & -3.14 & 3995 & 0.75 \\ 
Fnx-mem0539 & 1.1 & 3899 & 4118 & 3952 & 3947 & 3956 & -0.95 & -3.31 & 3974 & 0.60 \\ 
Fnx-mem0543 & 1.1 & 3811 & 4037 & 3891 & 3884 & 3887 & -1.04 & -3.44 & 3902 & 0.52 \\ 
Fnx-mem0556 & 1.1 & 3943 & 4051 & 3962 & 3953 & 3945 & -0.95 & -3.23 & 3971 & 0.64 \\ 
Fnx-mem0572 & 1.1 & 3902 & 4060 & 3951 & 3946 & 3937 & -0.96 & -3.24 & 3959 & 0.63 \\ 
Fnx-mem0574 & 1.1 & 3715 & 3947 & 3841 & 3828 & 3826 & -1.18 & -3.51 & 3831 & 0.46 \\ 
Fnx-mem0584 & 0.9 & 4297 & 4573 & 4319 & 4292 & 4319 & -0.58 & -2.86 & 4360 & 0.86 \\ 
Fnx-mem0595 & 1.1 & 3782 & 3997 & 3888 & 3888 & 3890 & -1.07 & -3.44 & 3889 & 0.51 \\ 
Fnx-mem0598 & 1.1 & 3638 & 3931 & 3742 & 3726 & 3737 & -1.35 & -3.70 & 3755 & 0.35 \\ 
Fnx-mem0604 & 1.1 & 3602 & 3849 & 3741 & 3727 & 3733 & -1.41 & -4.09 & 3731 & 0.18 \\ 
Fnx-mem0606 & 1.3 & 3554 & 3775 & 3634 & 3620 & 3639 & -1.72 & -4.03 & 3644 & 0.24 \\ 
Fnx-mem0607 & 1.1 & 3868 & 4004 & 3902 & 3907 & 3899 & -1.02 & -3.38 & 3916 & 0.55 \\ 
Fnx-mem0612 & 0.9 & 4151 & 4209 & 4177 & 4150 & 4156 & -0.72 & -3.11 & 4169 & 0.68 \\ 
Fnx-mem0613 & 1.3 & 3900 & 3904 & 3814 & 3823 & 3835 & -1.15 & -3.59 & 3855 & 0.51 \\ 
Fnx-mem0626 & 1.1 & 3956 & 4047 & 3962 & 3973 & 3966 & -0.93 & -3.22 & 3981 & 0.64 \\ 
Fnx-mem0629 & 1.1 & 3910 & 4075 & 3932 & 3916 & 3928 & -0.97 & -3.21 & 3952 & 0.63 \\ 
Fnx-mem0631 & 1.1 & 3996 & 4115 & 4028 & 4024 & 4021 & -0.86 & -3.57 & 4037 & 0.53 \\ 
Fnx-mem0633 & 1.1 & 3760 & 3973 & 3810 & 3804 & 3823 & -1.17 & -3.59 & 3834 & 0.43 \\ 
Fnx-mem0634 & 1.1 & 3922 & 4055 & 3962 & 3947 & 3969 & -0.94 & -3.14 & 3971 & 0.67 \\ 
Fnx-mem0638 & 1.3 & 3820 & 3930 & 3819 & 3807 & 3806 & -1.17 & -3.57 & 3837 & 0.51 \\ 
Fnx-mem0647 & 1.1 & 4112 & 4255 & 4071 & 4069 & 4091 & -0.76 & -3.10 & 4120 & 0.75 \\ 
Fnx-mem0654 & 0.9 & 4222 & 4377 & 4221 & 4217 & 4247 & -0.65 & -2.89 & 4257 & 0.80 \\ 
Fnx-mem0664 & 0.9 & 4296 & 4392 & 4324 & 4298 & 4327 & -0.60 & -2.86 & 4327 & 0.84 \\ 
Fnx-mem0675 & 1.1 & 9999 & 4150 & 4020 & 4007 & 4020 & -0.75 & -2.93 & 4150 & 0.83 \\ 
Fnx-mem0678 & 1.3 & 3825 & 3979 & 3850 & 3841 & 3845 & -1.11 & -3.32 & 3868 & 0.62 \\ 
Fnx-mem0682 & 1.1 & 3772 & 3975 & 3811 & 3828 & 3828 & -1.16 & -3.46 & 3843 & 0.49 \\ 
Fnx-mem0704 & 0.9 & 4255 & 4367 & 4326 & 4297 & 4396 & -0.60 & -2.88 & 4328 & 0.84 \\ 
Fnx-mem0712 & 0.9 & 4228 & 4277 & 4240 & 4199 & 4199 & -0.67 & -3.00 & 4229 & 0.75 \\ 
Fnx-mem0714 & 0.9 & 4095 & 4140 & 4054 & 4076 & 4086 & -0.79 & -3.51 & 4090 & 0.49 \\ 
Fnx-mem0715 & 1.1 & 3881 & 4080 & 3910 & 3926 & 3923 & -0.98 & -3.20 & 3944 & 0.64 \\ 
Fnx-mem0717 & 1.3 & 3817 & 4000 & 3851 & 3854 & 3859 & -1.11 & -3.48 & 3876 & 0.57 \\ 
Fnx-mem0732 & 0.9 & 4105 & 4116 & 4096 & 4021 & 4068 & -0.79 & -3.24 & 4081 & 0.59 \\ 
Fnx-mem0747 & 1.1 & 4036 & 4209 & 4041 & 4052 & 4057 & -0.81 & -3.02 & 4079 & 0.77 \\ 
Fnx-mem0754 & 1.1 & 3811 & 3930 & 3820 & 3844 & 3848 & -1.14 & -3.51 & 3851 & 0.47 \\ 
Fnx-mem0779 & 0.9 & 4186 & 4266 & 9999 & 9999 & 9999 & -0.64 & -3.11 & 4266 & 0.72 \\ 
Fnx-rgb0507 & 1.3 & 4007 & 4246 & 4009 & 4039 & 4025 & -0.83 & -3.24 & 4065 & 0.75 \\ 
Fnx-rgb0509 & 1.1 & 3777 & 3837 & 3697 & 3693 & 3701 & -1.37 & -3.79 & 3741 & 0.31 \\ 
Fnx-rgb0522 & 1.1 & 4196 & 4254 & 4222 & 4212 & 4206 & -0.68 & -3.10 & 4218 & 0.79 \\ 
Fnx-rgb0539 & 1.3 & 3868 & 3986 & 3859 & 3857 & 3853 & -1.09 & -3.50 & 3884 & 0.56 \\ 
Fnx-rgb0553 & 1.3 & 3786 & 3936 & 3854 & 3855 & 3850 & -1.14 & -3.43 & 3856 & 0.58 \\ 
Fnx-rgb0556 & 1.3 & 3891 & 3986 & 3899 & 3899 & 3903 & -1.04 & -3.26 & 3916 & 0.67 \\ 
Fnx-rgb0561 & 1.1 & 3921 & 4030 & 3923 & 3925 & 3912 & -0.98 & -3.38 & 3942 & 0.56 \\ 
Fnx-rgb0574 & 1.1 & 3903 & 4056 & 3938 & 3937 & 3921 & -0.97 & -3.75 & 3951 & 0.42 \\ 
Fnx-rgb0590 & 1.1 & 4200 & 4292 & 4190 & 4199 & 4204 & -0.68 & -2.85 & 4217 & 0.89 \\ 
Fnx-rgb0596 & 0.9 & 4356 & 4288 & 4329 & 4357 & 4364 & -0.59 & -2.83 & 4339 & 0.86 \\ 
\hline
\end{tabular}
\label{atmparam_all}
\end{table*}

\onecolumn

\begin{longtab}
\LTcapwidth=\textwidth
\begin{longtable}{ccccccc}
\caption{Individual abundances for the RGB stars in our sample, listed together with their associated errors and the number of lines used in the analysis for a given element.}\label{abund}\\
\hline\hline
\endfirsthead
\caption{continued.}\\
\hline\hline
\endhead
\hline
\endfoot
    Star    &    [FeI/H]          &     [FeII/H]       &     [Na/H]         &     [Mg/H]         &     [Si/H]         &     [Ca/H]          \\   
            &      dex            &       dex          &       dex          &       dex          &       dex          &       dex           \\  
\hline
Fnx-mem0514 & -0.70$\pm$0.04 (42) &                    & -1.38$\pm$0.16 (2) & -0.76$\pm$0.23 (1) & -0.44$\pm$0.23 (1) & -1.05$\pm$0.08 ~(8) \\ 
Fnx-mem0522 & -0.88$\pm$0.05 (39) & -0.02$\pm$0.18 (3) & -1.53$\pm$0.22 (2) & -0.86$\pm$0.31 (1) & -0.41$\pm$0.31 (1) & -0.79$\pm$0.15 ~(4) \\ 
Fnx-mem0532 & -0.48$\pm$0.03 (29) & -0.02$\pm$0.09 (3) & -1.49$\pm$0.16 (1) & -0.89$\pm$0.16 (1) & -0.70$\pm$0.16 (1) & -0.81$\pm$0.08 ~(4) \\ 
Fnx-mem0539 & -0.70$\pm$0.03 (30) & -0.41$\pm$0.11 (4) & -1.56$\pm$0.14 (1) & -0.72$\pm$0.14 (1) &                    & -0.91$\pm$0.07 ~(7) \\ 
Fnx-mem0543 & -1.02$\pm$0.02 (39) & -0.30$\pm$0.20 (4) &                    & -0.79$\pm$0.10 (1) & -0.99$\pm$0.10 (1) & -1.22$\pm$0.08 ~(7) \\ 
Fnx-mem0556 & -0.64$\pm$0.03 (42) & -0.26$\pm$0.20 (3) &                    & -0.77$\pm$0.22 (1) & -0.71$\pm$0.16 (2) & -1.17$\pm$0.09 ~(6) \\ 
Fnx-mem0572 & -0.88$\pm$0.02 (44) & -0.24$\pm$0.25 (3) & -1.69$\pm$0.16 (1) & -0.89$\pm$0.16 (1) & -0.44$\pm$0.16 (1) & -1.04$\pm$0.09 ~(9) \\ 
Fnx-mem0574 & -0.80$\pm$0.02 (40) & -0.42$\pm$0.07 (5) & -1.63$\pm$0.15 (1) & -1.12$\pm$0.15 (1) &                    & -1.11$\pm$0.06 ~(7) \\ 
Fnx-mem0584 & -2.17$\pm$0.04 (15) &                    &                    & -2.05$\pm$0.16 (1) &                    & -1.75$\pm$0.12 ~(4) \\ 
Fnx-mem0595 & -0.88$\pm$0.03 (30) & -0.14$\pm$0.22 (2) &                    & -1.34$\pm$0.19 (1) & -0.65$\pm$0.19 (1) & -1.21$\pm$0.14 ~(5) \\ 
Fnx-mem0598 & -0.85$\pm$0.02 (45) & -0.68$\pm$0.20 (2) & -1.82$\pm$0.15 (1) & -0.98$\pm$0.15 (1) & -0.77$\pm$0.11 (2) & -1.31$\pm$0.06 ~(7) \\ 
Fnx-mem0604 & -0.74$\pm$0.02 (32) & -0.53$\pm$0.23 (3) & -1.44$\pm$0.10 (1) & -0.80$\pm$0.10 (1) &                    & -1.19$\pm$0.08 ~(8) \\ 
Fnx-mem0606 & -0.62$\pm$0.03 (51) & -0.17$\pm$0.12 (4) & -1.93$\pm$0.23 (1) & -0.79$\pm$0.23 (1) & -0.61$\pm$0.16 (2) & -0.82$\pm$0.12 ~(4) \\ 
Fnx-mem0607 & -0.96$\pm$0.02 (41) & -0.68$\pm$0.09 (6) & -1.76$\pm$0.10 (1) & -0.82$\pm$0.10 (1) & -0.76$\pm$0.19 (2) & -1.25$\pm$0.05 ~(8) \\ 
Fnx-mem0612 & -1.68$\pm$0.04 (25) & -0.64$\pm$0.21 (1) &                    & -1.44$\pm$0.21 (1) &                    & -1.81$\pm$0.07 ~(8) \\ 
Fnx-mem0613 & -0.44$\pm$0.03 (44) &  0.06$\pm$0.19 (1) &                    & -0.54$\pm$0.19 (1) &                    & -0.78$\pm$0.08 ~(5) \\ 
Fnx-mem0626 & -1.00$\pm$0.03 (47) & -0.53$\pm$0.10 (4) &                    & -0.97$\pm$0.21 (1) & -0.62$\pm$0.21 (1) & -1.19$\pm$0.06 (11) \\ 
Fnx-mem0629 & -1.02$\pm$0.02 (27) & -0.31$\pm$0.32 (3) &                    & -0.84$\pm$0.08 (2) & -0.89$\pm$0.11 (1) & -1.18$\pm$0.08 (12) \\ 
Fnx-mem0631 & -0.82$\pm$0.02 (40) & -0.44$\pm$0.10 (3) & -1.47$\pm$0.15 (1) & -1.02$\pm$0.15 (1) & -0.87$\pm$0.15 (1) & -1.19$\pm$0.07 ~(5) \\ 
Fnx-mem0633 & -0.92$\pm$0.02 (44) & -0.52$\pm$0.10 (7) &                    & -1.26$\pm$0.11 (1) & -0.97$\pm$0.11 (2) & -1.43$\pm$0.07 ~(9) \\ 
Fnx-mem0634 & -0.26$\pm$0.07 (41) & -0.11$\pm$0.25 (3) &                    & -0.30$\pm$0.43 (1) & -0.75$\pm$0.43 (1) & -0.87$\pm$0.14 (10) \\ 
Fnx-mem0638 & -0.70$\pm$0.03 (51) & -0.26$\pm$0.11 (6) & -1.92$\pm$0.23 (1) & -0.78$\pm$0.23 (1) & -0.74$\pm$0.16 (2) & -1.13$\pm$0.08 ~(9) \\ 
Fnx-mem0647 & -1.48$\pm$0.02 (37) & -0.82$\pm$0.12 (5) &                    & -1.47$\pm$0.13 (1) &                    & -1.65$\pm$0.05 ~(8) \\ 
Fnx-mem0654 & -1.92$\pm$0.03 (33) & -1.33$\pm$0.10 (3) &                    & -1.38$\pm$0.18 (1) &                    & -1.84$\pm$0.07 ~(7) \\ 
Fnx-mem0664 & -2.31$\pm$0.02 (22) & -1.63$\pm$0.13 (2) &                    & -1.65$\pm$0.09 (1) &                    & -2.28$\pm$0.06 ~(2) \\ 
Fnx-mem0675 & -0.81$\pm$0.03 (55) & -0.49$\pm$0.09 (6) & -1.55$\pm$0.22 (1) & -1.02$\pm$0.22 (1) & -0.92$\pm$0.22 (1) & -1.11$\pm$0.09 ~(7) \\ 
Fnx-mem0678 & -0.67$\pm$0.03 (41) & -0.20$\pm$0.20 (1) &                    & -1.07$\pm$0.20 (1) &                    & -0.74$\pm$0.09 ~(5) \\ 
Fnx-mem0682 & -0.76$\pm$0.03 (50) & -0.46$\pm$0.09 (5) &                    & -0.96$\pm$0.20 (1) &                    & -1.05$\pm$0.07 ~(9) \\ 
Fnx-mem0704 & -2.55$\pm$0.02 (20) & -1.65$\pm$0.09 (3) & -1.05$\pm$0.09 (1) &                    & -1.37$\pm$0.09 (1) & -1.86$\pm$0.11 ~(3) \\ 
Fnx-mem0712 & -2.09$\pm$0.05 (23) & -1.48$\pm$0.13 (3) &                    & -1.65$\pm$0.22 (1) &                    & -1.62$\pm$0.31 ~(6) \\ 
Fnx-mem0714 & -1.79$\pm$0.07 (26) & -1.05$\pm$0.21 (3) &                    & -1.83$\pm$0.36 (1) &                    & -1.80$\pm$0.15 ~(6) \\ 
Fnx-mem0715 & -0.98$\pm$0.03 (49) & -0.65$\pm$0.12 (4) &                    & -1.09$\pm$0.23 (1) & -0.87$\pm$0.16 (2) & -1.30$\pm$0.12 ~(4) \\ 
Fnx-mem0717 & -0.30$\pm$0.04 (56) &  0.08$\pm$0.31 (1) & -1.59$\pm$0.22 (2) & -0.61$\pm$0.31 (1) & -0.44$\pm$0.22 (2) & -0.43$\pm$0.15 ~(4) \\ 
Fnx-mem0732 & -2.40$\pm$0.02 (31) & -1.57$\pm$0.08 (3) &                    & -1.78$\pm$0.13 (1) &                    & -1.97$\pm$0.07 ~(8) \\ 
Fnx-mem0747 & -1.39$\pm$0.02 (26) & -0.87$\pm$0.12 (3) &                    & -1.35$\pm$0.08 (1) & -1.32$\pm$0.08 (1) & -1.44$\pm$0.06 ~(8) \\ 
Fnx-mem0754 & -0.64$\pm$0.03 (43) & -0.18$\pm$0.11 (4) & -1.27$\pm$0.22 (1) & -0.81$\pm$0.22 (1) & -0.41$\pm$0.22 (1) & -1.22$\pm$0.07 ~(9) \\ 
Fnx-mem0779 & -2.05$\pm$0.05 (30) & -1.57$\pm$0.18 (2) &                    & -2.16$\pm$0.25 (1) &                    & -2.05$\pm$0.12 ~(4) \\ 
Fnx-rgb0507 & -0.77$\pm$0.04 (44) & -0.28$\pm$0.13 (5) &                    & -1.07$\pm$0.28 (1) &                    & -1.09$\pm$0.09 ~(9) \\ 
Fnx-rgb0509 & -1.15$\pm$0.03 (41) & -0.43$\pm$0.14 (3) & -1.88$\pm$0.22 (1) & -1.15$\pm$0.22 (1) &                    & -1.67$\pm$0.10 ~(9) \\ 
Fnx-rgb0522 & -1.87$\pm$0.04 (29) & -1.31$\pm$0.13 (3) &                    & -1.36$\pm$0.22 (1) &                    & -1.70$\pm$0.08 ~(8) \\ 
Fnx-rgb0539 & -0.60$\pm$0.03 (39) & -0.13$\pm$0.13 (2) &                    & -0.70$\pm$0.18 (1) &                    & -1.03$\pm$0.07 ~(6) \\ 
Fnx-rgb0553 & -0.48$\pm$0.03 (27) &                    &                    & -0.91$\pm$0.13 (1) &                    & -0.69$\pm$0.05 ~(7) \\ 
Fnx-rgb0556 & -0.54$\pm$0.03 (43) & -0.36$\pm$0.13 (2) & -1.53$\pm$0.19 (1) & -0.85$\pm$0.19 (1) & -0.49$\pm$0.19 (1) & -0.91$\pm$0.08 ~(5) \\ 
Fnx-rgb0561 & -0.97$\pm$0.03 (35) & -0.39$\pm$0.12 (4) & -1.39$\pm$0.16 (1) & -0.93$\pm$0.16 (1) & -0.46$\pm$0.16 (1) & -0.94$\pm$0.13 ~(6) \\ 
Fnx-rgb0574 & -0.97$\pm$0.02 (46) & -0.80$\pm$0.19 (4) & -1.22$\pm$0.10 (2) & -0.93$\pm$0.14 (1) & -0.90$\pm$0.14 (1) & -1.13$\pm$0.07 ~(9) \\ 
Fnx-rgb0590 & -1.51$\pm$0.03 (35) & -1.09$\pm$0.17 (4) &                    & -1.39$\pm$0.18 (1) & -1.26$\pm$0.18 (1) & -1.59$\pm$0.06 ~(9) \\ 
Fnx-rgb0596 & -2.68$\pm$0.04 (25) & -2.07$\pm$0.19 (1) &                    &                     &                   & -2.50$\pm$0.10 ~(4) \\ 
\newpage
     Star    &    [FeI/H]          &     [Sc/H]         &     [TiI/H]        &     [TiII/H]       &     [Cr/H]         &      [Ni/H]         \\ 
             &      dex            &       dex          &       dex          &       dex          &       dex          &        dex          \\ 
 \hline
Fnx-mem0514 & -0.70$\pm$0.04 (42) &                    & -1.04$\pm$0.10 (7) & -0.39$\pm$0.23 (1) & -1.16$\pm$0.23 (1) & -0.90$\pm$0.08~(9) \\ 
Fnx-mem0522 & -0.88$\pm$0.05 (39) &                    & -1.24$\pm$0.12 (7) & -0.90$\pm$0.31 (1) & -0.68$\pm$0.31 (1) & -0.98$\pm$0.10(10) \\ 
Fnx-mem0532 & -0.48$\pm$0.03 (29) &                    & -1.18$\pm$0.06 (8) & -0.46$\pm$0.16 (3) & -0.98$\pm$0.16 (1) & -0.79$\pm$0.05~(9) \\ 
Fnx-mem0539 & -0.70$\pm$0.03 (30) &                    & -1.13$\pm$0.07 (8) & -0.78$\pm$0.12 (3) & -0.88$\pm$0.14 (1) & -1.08$\pm$0.10~(8) \\ 
Fnx-mem0543 & -1.02$\pm$0.02 (39) & -1.24$\pm$0.10 (1) & -1.41$\pm$0.06 (7) & -0.84$\pm$0.22 (2) &                    & -1.16$\pm$0.08(12) \\ 
Fnx-mem0556 & -0.64$\pm$0.03 (42) &                    & -1.09$\pm$0.09 (6) & -0.54$\pm$0.16 (2) & -0.87$\pm$0.22 (1) & -0.88$\pm$0.12(10) \\ 
Fnx-mem0572 & -0.88$\pm$0.02 (44) & -0.80$\pm$0.16 (1) & -1.24$\pm$0.06 (7) & -0.93$\pm$0.22 (2) & -0.91$\pm$0.16 (1) & -1.03$\pm$0.05(10) \\ 
Fnx-mem0574 & -0.80$\pm$0.02 (40) &                    & -1.15$\pm$0.05 (8) & -1.15$\pm$0.48 (2) & -1.29$\pm$0.15 (1) & -1.17$\pm$0.06~(9) \\ 
Fnx-mem0584 & -2.17$\pm$0.04 (15) &                    &                    & -1.19$\pm$0.11 (2) & -1.86$\pm$0.16 (1) & -2.03$\pm$0.09~(3) \\ 
Fnx-mem0595 & -0.88$\pm$0.03 (30) &                    & -1.67$\pm$0.08 (5) & -0.87$\pm$0.13 (2) & -1.77$\pm$0.19 (1) & -1.38$\pm$0.10~(4) \\ 
Fnx-mem0598 & -0.85$\pm$0.02 (45) &                    & -1.39$\pm$0.06 (8) & -0.82$\pm$0.11 (2) & -1.61$\pm$0.15 (1) & -1.25$\pm$0.06~(7) \\ 
Fnx-mem0604 & -0.74$\pm$0.02 (32) &                    & -1.30$\pm$0.06 (9) & -0.88$\pm$0.11 (2) & -1.33$\pm$0.12 (1) & -1.23$\pm$0.06~(3) \\ 
Fnx-mem0606 & -0.62$\pm$0.03 (51) &                    & -1.56$\pm$0.12 (4) & -0.74$\pm$0.18 (2) & -0.84$\pm$0.23 (1) & -0.97$\pm$0.07(11) \\ 
Fnx-mem0607 & -0.96$\pm$0.02 (41) & -1.37$\pm$0.10 (1) & -1.44$\pm$0.04 (7) & -0.91$\pm$0.08 (3) & -1.47$\pm$0.10 (1) & -1.23$\pm$0.04~(9) \\ 
Fnx-mem0612 & -1.68$\pm$0.04 (25) &                    & -1.92$\pm$0.12 (3) & -1.52$\pm$0.38 (2) &                    & -1.54$\pm$0.12~(4) \\ 
Fnx-mem0613 & -0.44$\pm$0.03 (44) & -0.53$\pm$0.19 (1) & -0.79$\pm$0.12 (7) & -0.46$\pm$0.13 (2) & -0.38$\pm$0.19 (1) & -0.55$\pm$0.10(10) \\ 
Fnx-mem0626 & -1.00$\pm$0.03 (47) &                    & -1.27$\pm$0.07 (8) & -0.83$\pm$0.12 (3) & -1.31$\pm$0.21 (1) & -1.26$\pm$0.09~(6) \\ 
Fnx-mem0629 & -1.02$\pm$0.02 (27) & -0.88$\pm$0.11 (1) & -1.39$\pm$0.05 (6) & -0.68$\pm$0.08 (2) & -1.37$\pm$0.14 (1) & -1.08$\pm$0.04~(8) \\ 
Fnx-mem0631 & -0.82$\pm$0.02 (40) &                    & -1.30$\pm$0.06 (6) & -1.00$\pm$0.11 (2) &                    & -1.04$\pm$0.05~(9) \\ 
Fnx-mem0633 & -0.92$\pm$0.02 (44) &                    & -1.55$\pm$0.05 (5) & -0.85$\pm$0.11 (1) & -1.58$\pm$0.11 (1) & -1.32$\pm$0.04(10) \\ 
Fnx-mem0634 & -0.26$\pm$0.07 (41) & -0.71$\pm$0.43 (1) & -1.26$\pm$0.15 (8) & -0.24$\pm$0.30 (2) &                    & -0.49$\pm$0.14(10) \\ 
Fnx-mem0638 & -0.70$\pm$0.03 (51) &                    & -1.05$\pm$0.08 (9) & -0.52$\pm$0.23 (1) & -0.97$\pm$0.23 (1) & -1.03$\pm$0.09~(7) \\ 
Fnx-mem0647 & -1.48$\pm$0.02 (37) & -1.34$\pm$0.13 (1) & -1.97$\pm$0.13 (3) & -1.44$\pm$0.17 (3) & -1.76$\pm$0.13 (1) & -1.59$\pm$0.07(10) \\ 
Fnx-mem0654 & -1.92$\pm$0.03 (33) &                    & -1.89$\pm$0.28 (2) & -1.61$\pm$0.18 (1) &                    & -1.72$\pm$0.12(10) \\ 
Fnx-mem0664 & -2.31$\pm$0.02 (22) &                    &                    & -2.20$\pm$0.19 (2) & -2.55$\pm$0.09 (1) & -2.23$\pm$0.12~(5) \\ 
Fnx-mem0675 & -0.81$\pm$0.03 (55) & -0.94$\pm$0.22 (1) & -1.35$\pm$0.11 (4) & -1.12$\pm$0.16 (2) &                    & -1.07$\pm$0.07(10) \\ 
Fnx-mem0678 & -0.67$\pm$0.03 (41) &                    & -1.25$\pm$0.10 (8) & -0.72$\pm$0.19 (3) & -0.78$\pm$0.20 (1) & -0.91$\pm$0.08(11) \\ 
Fnx-mem0682 & -0.76$\pm$0.03 (50) & -1.08$\pm$0.20 (1) & -1.22$\pm$0.11 (8) & -1.02$\pm$0.34 (2) & -0.65$\pm$0.20 (1) & -1.10$\pm$0.07~(8) \\ 
Fnx-mem0704 & -2.55$\pm$0.02 (20) &                    &                    & -1.90$\pm$0.09 (1) &                    & -2.09$\pm$0.11~(2) \\ 
Fnx-mem0712 & -2.09$\pm$0.05 (23) &                    &                    & -1.40$\pm$0.21 (2) &                    & -2.11$\pm$0.22~(1) \\ 
Fnx-mem0714 & -1.79$\pm$0.07 (26) &                    & -1.76$\pm$0.25 (2) & -1.31$\pm$0.36 (1) & -2.46$\pm$0.36 (1) & -1.98$\pm$0.18~(4) \\ 
Fnx-mem0715 & -0.98$\pm$0.03 (49) &                    & -1.53$\pm$0.08 (8) & -0.62$\pm$0.25 (2) & -1.51$\pm$0.23 (1) & -1.06$\pm$0.06(14) \\ 
Fnx-mem0717 & -0.30$\pm$0.04 (56) &                    & -0.62$\pm$0.18 (7) & -0.97$\pm$0.31 (1) & -0.75$\pm$0.31 (1) & -0.75$\pm$0.12(10) \\ 
Fnx-mem0732 & -2.40$\pm$0.02 (31) &                    & -2.49$\pm$0.13 (1) & -1.75$\pm$0.13 (1) & -2.44$\pm$0.09 (2) & -2.37$\pm$0.06~(5) \\ 
Fnx-mem0747 & -1.39$\pm$0.02 (26) &                    & -1.65$\pm$0.12 (7) & -0.82$\pm$0.07 (2) & -1.74$\pm$0.06 (2) & -1.63$\pm$0.06~(6) \\ 
Fnx-mem0754 & -0.64$\pm$0.03 (43) &                    & -1.10$\pm$0.09 (7) & -0.90$\pm$0.37 (2) & -0.92$\pm$0.22 (1) & -1.21$\pm$0.11~(6) \\ 
Fnx-mem0779 & -2.05$\pm$0.05 (30) &                    &                    & -1.55$\pm$0.25 (1) & -2.23$\pm$0.25 (1) & -1.96$\pm$0.16~(4) \\ 
Fnx-rgb0507 & -0.77$\pm$0.04 (44) & -0.57$\pm$0.28 (1) & -1.53$\pm$0.11 (6) & -0.96$\pm$0.28 (1) & -0.49$\pm$0.28 (1) & -0.88$\pm$0.10~(8) \\ 
Fnx-rgb0509 & -1.15$\pm$0.03 (41) &                    & -1.67$\pm$0.08 (7) & -1.26$\pm$0.23 (2) & -1.71$\pm$0.22 (1) & -1.32$\pm$0.12~(7) \\ 
Fnx-rgb0522 & -1.87$\pm$0.04 (29) & -1.69$\pm$0.22 (1) & -2.09$\pm$0.38 (2) & -1.23$\pm$0.28 (3) & -1.94$\pm$0.16 (2) & -1.84$\pm$0.08~(8) \\ 
Fnx-rgb0539 & -0.60$\pm$0.03 (39) &                    & -0.69$\pm$0.09 (5) & -0.45$\pm$0.13 (2) &                    & -0.90$\pm$0.11~(7) \\ 
Fnx-rgb0553 & -0.48$\pm$0.03 (27) &                    & -0.76$\pm$0.09 (3) & -0.64$\pm$0.09 (2) & -0.33$\pm$0.15 (1) & -0.81$\pm$0.14~(7) \\ 
Fnx-rgb0556 & -0.54$\pm$0.03 (43) &                    & -1.17$\pm$0.07 (8) & -0.73$\pm$0.13 (2) & -1.02$\pm$0.19 (1) & -0.79$\pm$0.06(11) \\ 
Fnx-rgb0561 & -0.97$\pm$0.03 (35) & -0.81$\pm$0.16 (1) & -1.31$\pm$0.06 (8) & -1.28$\pm$0.16 (1) & -1.43$\pm$0.16 (1) & -1.08$\pm$0.09(10) \\ 
Fnx-rgb0574 & -0.97$\pm$0.02 (46) &                    & -1.17$\pm$0.06 (5) & -1.30$\pm$0.14 (1) &                    & -1.05$\pm$0.04(14) \\ 
Fnx-rgb0590 & -1.51$\pm$0.03 (35) &                    & -1.72$\pm$0.11 (5) & -1.28$\pm$0.20 (3) & -1.64$\pm$0.18 (1) & -1.43$\pm$0.13~(6) \\ 
Fnx-rgb0596 & -2.68$\pm$0.04 (25) &                    & -2.28$\pm$0.19 (1) & -2.63$\pm$0.19 (1) & -3.26$\pm$0.19 (1) & -2.95$\pm$0.20~(1) \\ 
\newpage
     Star    &    [FeI/H]          &      [Y/H]         &     [Ba/H]         &     [La/H]         &     [Nd/H]         &     [Eu/H]         \\ 
             &      dex            &        dex         &       dex          &       dex          &       dex          &       dex          \\ 
 \hline
Fnx-mem0514 & -0.70$\pm$0.04 (42) &                    & -0.31$\pm$0.16 (2) & -0.43$\pm$0.16 (2) & -0.21$\pm$0.16 (2) &                    \\ 
Fnx-mem0522 & -0.88$\pm$0.05 (39) &                    & -0.33$\pm$0.31 (1) & -0.67$\pm$0.31 (1) & -0.40$\pm$0.31 (1) & -0.50$\pm$0.31 (1) \\ 
Fnx-mem0532 & -0.48$\pm$0.03 (29) &                    & -0.17$\pm$0.11 (2) & -0.39$\pm$0.11 (2) & -0.49$\pm$0.16 (1) & -0.15$\pm$0.16 (1) \\ 
Fnx-mem0539 & -0.70$\pm$0.03 (30) &                    & -0.36$\pm$0.10 (2) &                    & -0.44$\pm$0.14 (1) & -0.24$\pm$0.14 (1) \\ 
Fnx-mem0543 & -1.02$\pm$0.02 (39) &                    & -0.65$\pm$0.10 (1) & -0.85$\pm$0.13 (2) &                    & -0.65$\pm$0.10 (1) \\ 
Fnx-mem0556 & -0.64$\pm$0.03 (42) & -0.76$\pm$0.22 (1) &  0.03$\pm$0.22 (1) & -0.39$\pm$0.22 (1) &  0.49$\pm$0.16 (2) &                    \\ 
Fnx-mem0572 & -0.88$\pm$0.02 (44) &                    & -0.51$\pm$0.11 (2) & -0.71$\pm$0.27 (2) & -0.60$\pm$0.16 (1) & -0.50$\pm$0.16 (1) \\ 
Fnx-mem0574 & -0.80$\pm$0.02 (40) &                    & -0.61$\pm$0.13 (2) & -0.88$\pm$0.15 (1) & -0.45$\pm$0.11 (2) & -0.66$\pm$0.15 (1) \\ 
Fnx-mem0584 & -2.17$\pm$0.04 (15) &                    &                    &                    & -0.38$\pm$0.16 (1) &                    \\ 
Fnx-mem0595 & -0.88$\pm$0.03 (30) & -0.71$\pm$0.19 (1) &                    &                    & -0.19$\pm$0.19 (1) &                    \\ 
Fnx-mem0598 & -0.85$\pm$0.02 (45) &                    & -0.65$\pm$0.15 (1) & -1.15$\pm$0.11 (2) & -0.76$\pm$0.11 (2) & -0.78$\pm$0.15 (1) \\ 
Fnx-mem0604 & -0.74$\pm$0.02 (32) &                    & -0.43$\pm$0.10 (1) & -0.56$\pm$0.10 (1) & -0.76$\pm$0.21 (2) & -0.65$\pm$0.10 (1) \\ 
Fnx-mem0606 & -0.62$\pm$0.03 (51) &                    & -0.24$\pm$0.23 (1) & -0.81$\pm$0.23 (1) & -0.45$\pm$0.16 (2) & -0.57$\pm$0.23 (1) \\ 
Fnx-mem0607 & -0.96$\pm$0.02 (41) &                    &                    & -0.66$\pm$0.07 (2) & -0.39$\pm$0.10 (1) & -0.46$\pm$0.10 (1) \\ 
Fnx-mem0612 & -1.68$\pm$0.04 (25) &                    & -1.97$\pm$0.21 (1) &                    &                    & -0.75$\pm$0.21 (1) \\ 
Fnx-mem0613 & -0.44$\pm$0.03 (44) &                    & -0.11$\pm$0.19 (1) & -0.27$\pm$0.13 (2) & -0.26$\pm$0.13 (2) &                    \\ 
Fnx-mem0626 & -1.00$\pm$0.03 (47) &                    & -0.44$\pm$0.15 (2) & -0.69$\pm$0.15 (2) & -0.29$\pm$0.21 (1) &                    \\ 
Fnx-mem0629 & -1.02$\pm$0.02 (27) & -0.92$\pm$0.11 (1) & -0.81$\pm$0.08 (2) & -0.66$\pm$0.33 (2) &                    & -0.57$\pm$0.11 (1) \\ 
Fnx-mem0631 & -0.82$\pm$0.02 (40) &                    & -0.27$\pm$0.15 (1) & -0.67$\pm$0.11 (2) & -0.51$\pm$0.25 (2) & -0.26$\pm$0.15 (1) \\ 
Fnx-mem0633 & -0.92$\pm$0.02 (44) &                    & -0.93$\pm$0.11 (1) & -1.12$\pm$0.08 (2) & -0.73$\pm$0.11 (1) & -1.02$\pm$0.11 (1) \\ 
Fnx-mem0634 & -0.26$\pm$0.07 (41) &                    &                    & -0.47$\pm$0.30 (2) &                    & -0.08$\pm$0.43 (1) \\ 
Fnx-mem0638 & -0.70$\pm$0.03 (51) &                    & -0.26$\pm$0.23 (1) & -0.53$\pm$0.16 (2) & -0.43$\pm$0.23 (1) & -0.60$\pm$0.23 (1) \\ 
Fnx-mem0647 & -1.48$\pm$0.02 (37) &                    & -1.27$\pm$0.13 (1) &                    &                    & -1.02$\pm$0.13 (1) \\ 
Fnx-mem0654 & -1.92$\pm$0.03 (33) &                    &                    &                    &                    & -1.12$\pm$0.18 (1) \\ 
Fnx-mem0664 & -2.31$\pm$0.02 (22) &                    & -2.13$\pm$0.09 (1) &                    &                    &                    \\ 
Fnx-mem0675 & -0.81$\pm$0.03 (55) & -0.98$\pm$0.22 (1) & -0.68$\pm$0.22 (1) &                    & -0.16$\pm$0.22 (1) & -0.46$\pm$0.22 (1) \\ 
Fnx-mem0678 & -0.67$\pm$0.03 (41) &                    &                    & -0.55$\pm$0.14 (2) & -0.30$\pm$0.19 (2) & -0.56$\pm$0.20 (1) \\ 
Fnx-mem0682 & -0.76$\pm$0.03 (50) &                    & -0.57$\pm$0.20 (1) & -1.03$\pm$0.14 (2) &                    & -0.59$\pm$0.20 (1) \\ 
Fnx-mem0704 & -2.55$\pm$0.02 (20) &                    & -2.15$\pm$0.09 (1) &                    &                    & -1.45$\pm$0.09 (1) \\ 
Fnx-mem0712 & -2.09$\pm$0.05 (23) &                    & -2.17$\pm$0.22 (1) &                    &                    &                    \\ 
Fnx-mem0714 & -1.79$\pm$0.07 (26) &                    & -1.54$\pm$0.28 (2) &                    &                    &                    \\ 
Fnx-mem0715 & -0.98$\pm$0.03 (49) &                    &                    & -0.82$\pm$0.33 (2) & -0.44$\pm$0.23 (1) & -0.61$\pm$0.23 (1) \\ 
Fnx-mem0717 & -0.30$\pm$0.04 (56) &                    &                    & -0.36$\pm$0.31 (1) &  0.62$\pm$0.31 (1) &                    \\ 
Fnx-mem0732 & -2.40$\pm$0.02 (31) &                    & -2.02$\pm$0.13 (1) &                    &                    &                    \\ 
Fnx-mem0747 & -1.39$\pm$0.02 (26) &                    & -1.38$\pm$0.10 (1) & -1.31$\pm$0.08 (1) &                    & -0.71$\pm$0.08 (1) \\ 
Fnx-mem0754 & -0.64$\pm$0.03 (43) &                    & -0.32$\pm$0.16 (2) & -0.66$\pm$0.16 (2) & -0.34$\pm$0.22 (1) & -0.43$\pm$0.22 (1) \\ 
Fnx-mem0779 & -2.05$\pm$0.05 (30) &                    & -1.97$\pm$0.25 (1) &                    &                    &                    \\ 
Fnx-rgb0507 & -0.77$\pm$0.04 (44) &                    & -0.31$\pm$0.28 (1) & -0.57$\pm$0.28 (1) &                    & -0.29$\pm$0.28 (1) \\ 
Fnx-rgb0509 & -1.15$\pm$0.03 (41) &                    & -0.67$\pm$0.22 (1) & -1.44$\pm$0.16 (2) &                    & -0.91$\pm$0.22 (1) \\ 
Fnx-rgb0522 & -1.87$\pm$0.04 (29) &                    & -2.22$\pm$0.22 (1) &                    &                    & -1.11$\pm$0.22 (1) \\ 
Fnx-rgb0539 & -0.60$\pm$0.03 (39) &                    & -0.17$\pm$0.18 (2) & -0.28$\pm$0.18 (1) &                    & -0.06$\pm$0.38 (1) \\ 
Fnx-rgb0553 & -0.48$\pm$0.03 (27) & -0.76$\pm$0.13 (1) & -0.30$\pm$0.13 (1) &                    &                    & -0.25$\pm$0.13 (1) \\ 
Fnx-rgb0556 & -0.54$\pm$0.03 (43) &                    & -0.29$\pm$0.19 (1) & -0.55$\pm$0.21 (2) & -0.22$\pm$0.19 (1) & -0.13$\pm$0.19 (1) \\ 
Fnx-rgb0561 & -0.97$\pm$0.03 (35) &                    & -0.30$\pm$0.11 (2) & -0.34$\pm$0.16 (1) & -0.19$\pm$0.16 (1) & -0.52$\pm$0.16 (1) \\ 
Fnx-rgb0574 & -0.97$\pm$0.02 (46) &                    & -0.37$\pm$0.14 (1) & -0.62$\pm$0.14 (1) & -0.76$\pm$0.14 (1) & -0.55$\pm$0.14 (1) \\ 
Fnx-rgb0590 & -1.51$\pm$0.03 (35) &                    & -1.22$\pm$0.18 (1) & -0.71$\pm$0.14 (2) &                    & -0.84$\pm$0.18 (1) \\ 
Fnx-rgb0596 & -2.68$\pm$0.04 (25) &                    & -3.53$\pm$0.18 (2) &                    & -0.98$\pm$0.19 (1) &                    \\ 
\end{longtable}
\end{longtab}

\newpage

\onllongtab{
\begin{longtab}
\scriptsize
\LTcapwidth=\textwidth
\begin{longtable}{cccrrrrrrrrrrrrr}
\caption{List of all the lines, their atomic parameters and their measured equivalent widths (EWs in m\AA), measured by DAOSPEC for our Fornax members.}\label{EW}\\
\hline\hline
\endfirsthead
\caption{continued.}\\
\hline\hline
\endhead
\hline
\endfoot
$\lambda(\AA)$&Elem&$\chi_{ex}$&log~gf& m0514 & m0522 & m0532 & m0539 & m0543 & m0556 & m0572 & m0574 & m0584 & m0595 & m0598 & m0604 \\ 
\hline
 6141.73 & Ba2 & 0.700 & -0.077 &       &       &       &       & 239.0 &       &       &       &       &       &       &       \\ 
 6496.91 & Ba2 & 0.600 & -0.380 & 212.0 &       &       & 230.0 &       &       &       &       &       &       &       &       \\ 
 6122.23 & Ca1 & 1.890 & -0.320 &       &       &       & 233.0 &       &       &       &       &       &       &       &       \\ 
 6156.03 & Ca1 & 2.520 & -2.390 &  48.0 &       &       &       &  26.0 &       &  28.0 &  41.0 &       &       &       &       \\ 
 6161.30 & Ca1 & 2.520 & -1.270 & 136.0 & 143.0 & 108.0 & 129.0 & 133.0 & 160.0 & 143.0 & 122.0 &  26.0 & 158.0 & 137.0 & 141.0 \\ 
 6162.17 & Ca1 & 1.900 & -0.320 &       &       &       &       &       &       &       &       & 139.0 &       &       &       \\
 6166.44 & Ca1 & 2.520 & -1.140 & 124.0 & 151.0 & 121.0 &       & 103.0 & 119.0 & 104.0 & 142.0 &       & 135.0 & 135.0 & 143.0 \\
\end{longtable}
\end{longtab}
} 

\section{Figures}

   \begin{figure*}[htp!]
   \centering
   \includegraphics[angle=0,width=0.8\textwidth]{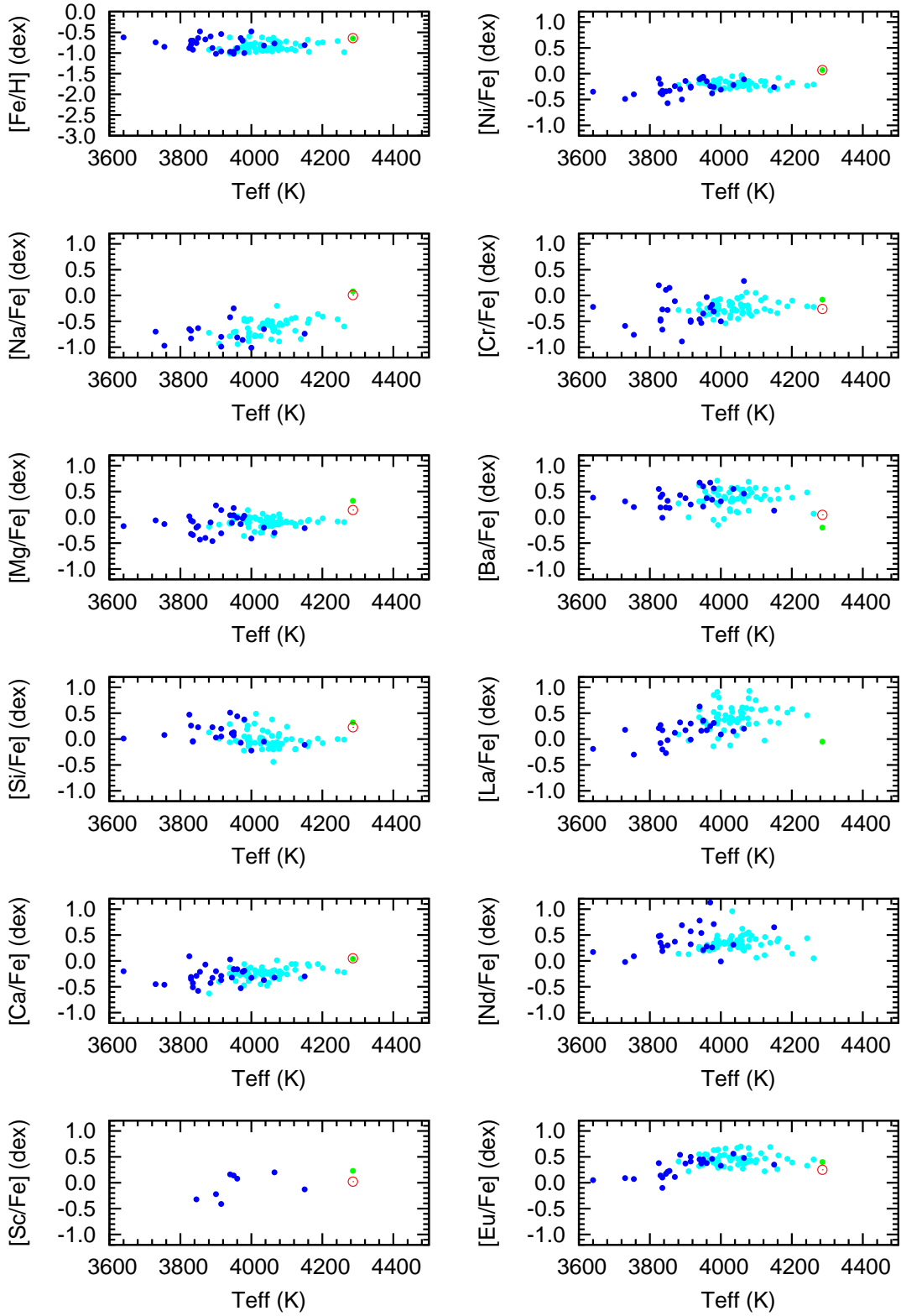}
      \caption{Abundances ratios vs. T$_{\rm eff}$ for RGB stars with --1.0$<$[Fe/H]$<$--0.5 dex in our sample, in blue; and in the sample of \cite{Let2010}, in cyan. We also show the values determined for Arcturus by \cite{vdS2013} in green; and by ourselves, using the same method as for our Fornax sample, in red.}
         \label{abvsTeff}
   \end{figure*}

   \begin{figure*}[htp!]
   \centering
   \includegraphics[angle=0,width=\textwidth]{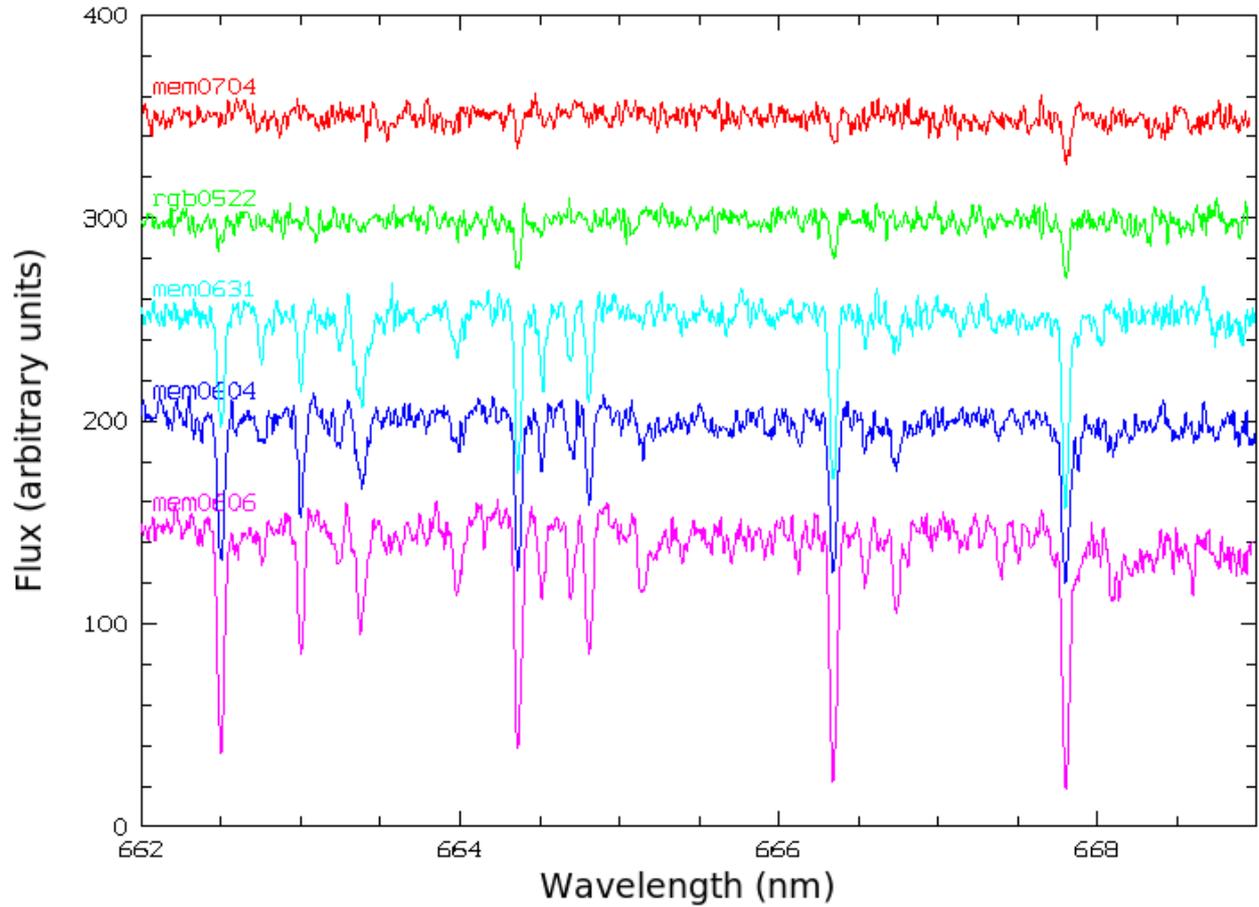}
      \caption{Typical sample spectra for stars with different T$_{\rm eff}$ (mem0704:~4330~K, rgb0522:~4220~K, mem0631:~4035~K, mem0604:~3730~K, mem0606:~3640~K) are shown to illustrate how molecular features change as a function of T$_{\rm eff}$. The spectra have been offset at arbitrary fluxes. The three coolest stars in this sample have similar, relatively high metallicities (--0.82, --0.74 and --0.62 dex) while the two hottest stars are metal-poor (--2.55 and -1.87 dex).}
         \label{5spectra}
   \end{figure*}

\end{appendix}


\begin{thebibliography}{}
   \bibitem[Aaronson \& Mould(1980)]{AM1980} Aaronson, M., Mould, J., 1980,
	ApJ 240, 804
   \bibitem[Aaronson \& Mould(1985)]{AM1985} Aaronson, M., Mould, J., 1985, 
	ApJ 290, 191
   \bibitem[Alonso et al.(1999)]{Alonso1999} Alonso, A., Arribas, S., Mart\' inez-Roger, C. 1999,
	A\&AS 140, 261
   \bibitem[Amorisco \& Evans(2012c)]{Am2012c} Amorisco, N. C., Evans, N. W., 2012,
	ApJ 756, 2
   \bibitem[Azzopardi et al.(1999)]{Azz1999} Azzopardi, M., Muratorio, G., Breysacher, J., Westerlund, B. E., 1999, 
	in IAU Symposium, Vol. 192, The Stellar Content of Local Group Galaxies, ed. P. Whitelock \& R. Cannon, 144
   \bibitem[Battaglia et al.(2006)]{Batt2006} Battaglia, G., Tolstoy, E., Helmi, A., Irwin, M. J., Letarte, B., Jablonka, P., Hill, V., Venn, K. A., Shetrone, M. D., Arimoto, N., Primas, F., Kaufer, A., Fran\c cois, P., Szeifert, T., Abel, T., Sadakane, K., 2006,
	A\&A 459, 423
   \bibitem[Battaglia et al.(2008a)]{Batta2008a} Battaglia, G., Irwin, M., Tolstoy, E., Hill, V., Helmi, A., Letarte, B., Jablonka, P., 2008,
	MNRAS 383, 183
   \bibitem[Battaglia \& Starkenburg(2012)]{Batt2012} Battaglia, G., Starkenburg, E., 2012,
	A\&A 539, A123
   \bibitem[Bensby et al.(2005)]{Ben2005} Bensby, T., Feltzing, S., Lundstr\"{o}m, I., Ilyin, I., 2005.
	A\&A 433, 185
   \bibitem[Bergemann \& Cescutti(2010)]{Berg2010} Bergemann, M., Cescutti, G., 2010,
	A\&A 522, A9
   \bibitem[Bergemann(2011)]{Berg2011} Bergemann, M., 2011,
	MNRAS 413, 2184
   \bibitem[Bersier(2000)]{Bersier2000} Bersier,  D., 2000,
	ApJ 534, 23
   \bibitem[Bersier \& Wood(2002)]{BW2002} Bersier, D., Wood, P. R., 2002,
	AJ 123, 840
   \bibitem[Blanco-Cuaresma et al.(2014)]{Blanco2014} Blanco-Cuaresma, S., Soubiran, C., Jofre, P., Heiter, U., 2014,
	A\&A 566, A98
   \bibitem[Buonanno et al.(1985)]{Buo1985} Buonanno, R., Corsi, C. E., Fusi Pecci, F. Hardy, E., Zinn, R., 1985,
	A\&A 152, 65
   \bibitem[Buonanno et al.(1999)]{Buo1999} Buonanno, R., Corsi, C. E., Castellani, M., Marconi, G., Fusi Pecci, F., Zinn, R. 1999, 
	AJ, 118, 1671
   \bibitem[Busso et al.(1999)]{Bus1999} Busso, M., Gallino, R., Wasserburg, G. J., 1999,
	ARA\&A 37, 239
   \bibitem[Busso et al.(2001)]{Bus2001} Busso, M., Gallino, R., Lambert, D. L., Travaglio, C., Smith, V. V., 2001,
	ApJ 557, 802
   \bibitem[Cayrel et al.(2004)]{Cayrel2004} Cayrel, R., Depagne, E., Spite, M., Hill, V., Spite, F., Fran\c cois, P., Plez, B., Beers, T., Primas, F., Andersen, J., Barbuy, B., Bonifacio, P., Molaro, P., Nordstr\"{o}m, B., 2004,
	A\&A 416, 1117	
   \bibitem[Coleman et al.(2004)]{Col2004} Coleman, M., Da Costa, G. S., Bland-Hawthorn, J., Martínez-Delgado, D., Freeman, K. C., Malin, D., 2004,
	AJ 127, 832
   \bibitem[Coleman et al.(2005b)]{Col2005b} Coleman, M. G., Da Costa, G. S., Bland-Hawthorn, J., Freeman, K. C., 2005b,
	AJ 129, 1443
   \bibitem[Coleman \& de Jong(2008)]{Col2008} Coleman, M. G., de Jong, J. T. A., 2008,
	ApJ 685, 933
   \bibitem[de Boer et al.(2012a)]{deBoer2012a} de Boer, T. J. L., Tolstoy, E., Hill, V., Saha, A., Olsen, K., Starkenburg, E., Lemasle, B., Irwin, M. J., Battaglia, G., 2012, 
	A\&A 539, A103
   \bibitem[de Boer et al.(2012b)]{deBoer2012b} de Boer, T. J. L., Tolstoy, E., Hill, V., Saha, A., Olszewski, E. W., Mateo, M., Starkenburg, E., Battaglia, G., Walker, M. G., 2012,
	A\&A 544, A73
   \bibitem[de Boer et al.(2013)]{deBoer2013} de Boer, T. J. L., Tolstoy, E., Saha, A., Olszewski, E. W., 2013,
	A\&A 551, A103
   \bibitem[del Pino et al.(2013)]{delPino2013} del Pino, A., Hidalgo, S. L., Aparicio, A., Gallart, C., Carrera, R., Monelli, M., Buonanno,  R., Marconi, G., 2013,
	MNRAS 433, 1505
   \bibitem[De Rijcke et al.(2004)]{deRijcke2004} De Rijcke, S., Dejonghe, H., Zeilinger, W. W., Hau, G. K. T., 2004, 
	A\&A, 426, 53 
   \bibitem[Dotter et al.(2008)]{Dotter2008} Dotter, A., Chaboyer, B., Jevremovi\'c, D., Kostov, V., Baron, E., Ferguson, J. W., 2008, 
	ApJS 178, 89
   \bibitem[Ellison et al.(2010)]{Ell2010} Ellison, S. L., Patton, D. R., Simard, L., McConnachie, A. W., Baldry, I. K., Mendel, J. T., 2010,
	MNRAS 407, 1514
   \bibitem[Fran\c cois et al.(2007)]{Fran2007} Fran\c cois, P., Depagne, E., Hill, V., Spite, M., Spite, F., Plez, B., Beers, T. C., Andersen, J., James, G., Barbuy, B., Cayrel, R., Bonifacio, P., Molaro, P., Nordstr\"{o}öm, B., Primas, F., 2007,
	A\&A 476, 935
   \bibitem[Frebel(2010)]{Frebel2010} Frebel, A., 2010, 
	AN 331, 474
   \bibitem[Freiburghaus et al.(1999)]{Frei1999} Freiburghaus, C., Rosswog, S., Thielemann, F.-K.. 1999,
	ApJ 525, 121
   \bibitem[Gallart et al.(2005)]{Gall2005} Gallart, C., Aparicio, A., Zinn, R., Buonanno, R., Hardy, E., Marconi, G., 2005, 
	in IAU Colloq. 198: Near-fields cosmology with dwarf elliptical galaxies, ed. H.~Jerjen \& B.~Binggeli, 25--29
   \bibitem[Gilmore \& Wyse(1991)]{Gil1991} Gilmore, G., Wyse, R. F. G., 1991,
	ApJ 367, L55-L58
   \bibitem[Greco et al.(2005)]{Greco2005} Greco, C., Clementini, G., Held, E. V., Poretti, E., Catelan, M., Dell'Arciprete, L., Gullieuszik, M., Maio, M., Rizzi, L., Smith, H. A., Pritzl, B. J., Rest, A., De Lee, N., 2005, 
	astro-ph/0507244
   \bibitem[Gullieuszik et al.(2007)]{Gull2007} Gullieuszik, M., Held, E. V., Rizzi, L., Saviane, I., Momani, Y., Ortolani, S., 2007,
	A\&A 467, 1025
   \bibitem[Gustafsson et al.(2008)]{Gus2008} Gustafsson, B., Edvardsson, B., Eriksson, K., J\o{}rgensen, U. G., Nordlund, \AA., Plez, B., 2008,
	A\&A 486, 951
   \bibitem[Harbeck et al.(2001)]{Har2001} Harbeck, D., Grebel, E. K., Holtzman, J., Guhathakurta, P., Brandner, W., Geisler, D., Sarajedini, A., Dolphin, A., Hurley-Keller, D., Mateo, M., 2001, 
	AJ 122, 3092
   \bibitem[Hendricks et al.(2014)]{Hen2014} Hendricks, B., Koch, A., Lanfranchi, G. A., Boeche, C., Walker, M., Johnson, C. I., Penarrubia, J., Gilmore, G., 2014,
	arXiv1402.6328
   \bibitem[Hodge(1961a)]{Hodge1961a} Hodge, P.W., 1961a,
	AJ 66, 83
   \bibitem[Irwin \& Hatzidimitriou(1995)]{Irwin1995} Irwin, M., Hatzidimitriou, D., 1995, 
	MNRAS 277, 1354
   \bibitem[Iwamoto et al.(1999)]{Iwa1999} Iwamoto, K., Brachwitz, F., Nomoto, K., Kishimoto, N., Umeda, H., Hix, W. R., Thielemann, F.-K., 1999,
 	ApJ SS 125, 439
   \bibitem[Jofre et al.(2013)]{Jofre2013} Jofre, P., Heiter, U., Soubiran, C., Blanco-Cuaresma, S., Pancino, E., Bergemann, M., Cantat-Gaudin, T., Gonzalez Hernandez, J. I., Hill, V., Lardo, C., de Laverny, P., Lind, K., Magrini, L., Masseron, T., Montes, D., Mucciarelli, A., Nordlander, T., Recio-Blanco, A., Sobeck, J., Sordo, R., Sousa, S. G., Tabernero, H., Vallenari, A., Van Eck, S., Worley, C. C., 2013, 
	arXiv 1309.1099
   \bibitem[Johnson(2002)]{Johnson2002} Johnson, J. A., 2002,
	ApJSS 139, 219
   \bibitem[Kirby et~al.(2010)]{Kirby2010} Kirby, E.~N., Guhathakurta, P., Simon, J.~D., Geha, M. C., Rockosi, C. M., Sneden, C., Cohen, J. G., Sohn, S. T., Majewski, S. R., Siegel, M., 2010,
	ApJ SS 191, 352
   \bibitem[Kirby et~al.(2011)]{Kirby2011} Kirby, E.~N., Cohen, J.~G., Smith, G.~H., Majewski, S. R., Sohn, S. T., Guhathakurta, P., (2011),
	ApJ 727, 79
   \bibitem[Kirby et~al.(2011b)]{Kirby2011b} Kirby, E. N., Martin, C. L., Finlator, K., 2011,
	ApJ 742, 25
   \bibitem[Kordopatis et al.(2013)]{Kordo2013} Kordopatis, G., Hill, V., Irwin, M., Gilmore, G., Wyse, R. F. G., Tolstoy, E., de Laverny, P., Recio-Blanco, A., Battaglia, G., Starkenburg, E., 2013,
	A\&A 555, A12
   \bibitem[Kroupa et al.(2013)]{Krou2013} Kroupa, P., Weidner, C., Pflamm-Altenburg, J., Thies, I., Dabringhausen, J., Marks, M., Maschberger, T., 2013,
	in {\it Planets, Stars and Stellar Systems}, ed. T. D. Oswalt \& G. Gilmore (Dordrecht: Springer), 115
   \bibitem[Lanfranchi et al.(2003)]{Lan2003} Lanfranchi, G. A., Matteucci, F., 2003,
	MNRAS 345, 71
   \bibitem[Lanfranchi et al.(2008)]{Lan2008} Lanfranchi, G. A., Matteucci, F., Cescutti, G., 2008,
	A\&A 481, 635
   \bibitem[Larsen et al.(2012)]{Lars2012} Larsen, S. S., Brodie, J. P., Strader, J., 2012,
	A\&A 546, A53
   \bibitem[Lemasle et al.(2012)]{Lem2012} Lemasle, B., Hill, V., Tolstoy, E., Venn, K. A., Shetrone, M. D., Irwin, M. J., de Boer, T. J. L., Starkenburg, E., Salvadori, S., 2012,
	A\&A 538, A100
   \bibitem[Letarte et al.(2006)]{Let2006} Letarte, B., Hill, V., Jablonka, P., Tolstoy, E., Fran\c cois, P., Meylan, G., 2006,
	A\&A 453, 547
   \bibitem[Letarte et al.(2010)]{Let2010} Letarte, B., Hill, V., Tolstoy, E., Jablonka, P., Shetrone, M., Venn, K. A., Spite, M., Irwin, M. J., Battaglia, G., Helmi, A., Primas, F., Fran\c cois, P., Kaufer, A., Szeifert, T., Arimoto, N., Sadakane, K., 2010,
	A\&A 523, A17
   \bibitem[Li et al.(2013a)]{Li2013} Li, H., Cui, W., Zhang, B., 2013a,
	ApJ 775, 12
   \bibitem[Li et al.(2013b)]{Li2013b} Li, H., Shen, X., Liang, S., Cui, W., Zhang, B., 2013b,
	PASP 125, 143
   \bibitem[Lind et al.(2011)]{Lind2011} Lind, K., Asplund, M., Barklem, P. S., Belyaev, A. K., 2011,
	A\&A 528, A103
    \bibitem[Magain(1984)]{Mag1984} Magain, P., 1984,
       A\&A 134, 189
    \bibitem[Mateo(1998)]{Mat1998r} Mateo, M., 1998,
	ARA\&A 36, 435
    \bibitem[Matteucci \& Brocato(1990)]{Matt1990} Matteucci, F., Brocato, E., 1990,
         ApJ 365, 539
    \bibitem[Matteucci(2003)]{Matt2003} Matteucci, F., 2003,
	Ap\&SS 284, 539
   \bibitem[Mayer et al.(2006)]{Mayer2006} Mayer, L., Mastropietro, C., Wadsley, J., Stadel, J., Moore, B., 2006,
	MNRAS 369, 1021
    \bibitem[McWilliam et al.(1995)]{McWil1995} McWilliam, A., Preston, G. W., Sneden, C., Searle, L., 1995,
	AJ 109, 2757
    \bibitem[McWilliam \& Smecker-Hane(2005b)]{MS2005b} McWilliam, A., Smecker-Hane, T. A., 2005,
	in ''Cosmic Abundances as Records of Stellar Evolution and Nucleosynthesis in honor of David L. Lambert'', ed. T. G. Barnes III \& F. N. Bash, ASPC 336, 221
    \bibitem[McWilliam et al.(2013)]{McWil2013} McWilliam, A., Wallerstein, G., Mottini, M., 2013,
        ApJ 778, 149
    \bibitem[Minor(2013)]{Minor2013} Minor, Q. E., 2013,
	ApJ 779, 116
    \bibitem[Nichols et al.(2014)]{Nichols2014} Nichols, M., Revaz, Y., Jablonka, P., 2014,
	Accepted in A\&A, arXiv1402.4480
    \bibitem[Nissen \& Schuster(1997)]{Niss1997} Nissen, P. E., Schuster, W. J., 1997,
	 A\&A 326, 751
    \bibitem[Nissen \& Schuster(2010)]{Niss2010} Nissen, P. E., Schuster, W. J., 2010,
	A\&A 511, 10
    \bibitem[Schuster et al.(2012)]{Schu2012} Schuster, W. J., Moreno, E., Nissen, P. E., Pichardo, B., 2012,
	A\&A 538, A21
    \bibitem[Oey(2011)]{Oey2011} Oey, M. S., 2011,
	ApJL 739, 46 
    \bibitem[Olszewski et al.(2006)]{Ol2006} Olszewski, E. W., Mateo, M., Harris, J., Walker, M. G., Coleman, M. G., Da Costa, G. S., 2006,
	AJ 131, 912
    \bibitem[Pasetto et al.(2011)]{Pas2011} Pasetto, S., Grebel, E. K., Berczik, P., Chiosi, C., Spurzem, R., 2011,
        A\&A 525, A99
    \bibitem[Pasquini et al.(2002)]{Pas2002} Pasquini, L., Avila, G., Blecha, A., Cacciari, C., Cayatte, V., Colless, M., Damiani, F., de Propris, R., Dekker, H., di Marcantonio, P., Farrell, T., Gillingham, P., Guinouard, I., Hammer, F., Kaufer, A., Hill, V., Marteaud, M., Modigliani, A., Mulas, G., North, P., Popovic, D., Rossetti, E., Royer, F., Santin, P., Schmutzer, R., Simond, G., Vola, P., Waller, L., \& Zoccali, M., 2002,
	The Messenger, 110, 1
   \bibitem[Pe\~{n}arrubia et al.(2008)]{Pena2008} Pe\~{n}arrubia, J., McConnachie, A. W., Navarro, J. F., 2008,
	ApJ 672, 904
    \bibitem[Pietrinferni et al.(2004)]{Piet2004} Pietrinferni, A., Cassisi, S., Salaris, M., Castelli, F., 2004, 
	ApJ 612, 168
    \bibitem[Pietrzy\'nski et al.(2003)]{Pietr2003} Pietrzy\'nski, G., Gieren W., Udalski A., 2003, 
	AJ 125, 2494
    \bibitem[Pietrzy\'nski et al.(2009)]{Pietr2009} Pietrzy\'nski, G., G\'orski, M., Gieren, W., Ivanov, V. D., Bresolin, F., Kudritzki, R-P., 2009, 
	AJ 138, 459
    \bibitem[Pomp\'eia et al.(2008)]{Pomp2008} Pomp\' eia, L., Hill, V., Spite, M., Cole, A., Primas, F., Romaniello, M., Pasquini, L., Cioni, M.-R., Smecker Hane, T., 2008,
	A\&A 480, 379
    \bibitem[Pont et al.(2004)]{Pont2004} Pont, F., Zinn, R., Gallart, C., Hardy, E., Winnick, R., 2004,
	AJ 127, 840
    \bibitem[Ram\` irez \& Mel\' endez (2005)]{Rami2005} Ram\` irez, I., Mel\' endez, J., 2005,
	ApJ 626, 465
    \bibitem[Revaz et al.(2009)]{Revaz2009} Revaz, Y., Jablonka, P., Sawala, T., Hill, V., Letarte, B., Irwin, M., Battaglia, G., Helmi, A., Shetrone, M. D., Tolstoy, E., Venn, K. A., 2009,
	A\&A 501, 189
    \bibitem[Revaz \& Jablonka(2012)]{Revaz2012} Revaz, Y., Jablonka, P., 2012,
	A\&A 538, A82
    \bibitem[Rizzi et al.(2007)]{Rizzi2007} Rizzi, L., Held, E. V., Saviane, I., Tully, R. B., Gullieuszik, M., 2007, 
	MNRAS 380, 1255
    \bibitem[Romano \& Starkenburg(2013)]{Romano2013} Romano, D., Starkenburg, E., 2013,
	MNRAS 434, 471
    \bibitem[Sakari et al.(2011)]{Saka2011} Sakari, C. M., Venn, K. A., Irwin, M., Aoki, W., Arimoto, N., Dotter, A., 2011,
	ApJ 740, 106	
    \bibitem[Saviane et al.(2000)]{Savi2000} Saviane, I., Held, E. V., Bertelli, G., 2000, 
	A\&A 355, 56
    \bibitem[Sawala et al.(2011)]{Sawa2011} Sawala, T., Guo, Q., Scannapieco, C., Jenkins, A., White, S., 2011,
	MNRAS 413, 659
    \bibitem[Sbordone et al.(2007)]{Sbo2007} Sbordone, L., Bonifacio, P., Buonanno, R., Marconi, G., Monaco, L., Zaggia, S., 2007,
	A\&A 465, 815
    \bibitem[Schlegel et al.(1998)]{Schle1998} Schlegel, D. J., Finkbeiner, D. P., Davis, M., 1998,
	ApJ 500, 525
    \bibitem[Scudder et al.(2012)]{Scu2012} Scudder, J. M., Ellison, S. L., Torrey, P., Patton, D. R., Mendel, J. T., 2012,
	MNRAS 426, 549
    \bibitem[Shapley(1938)]{Sha1938} Shapley, H., 1938,
	Nature, 142, 715
    \bibitem[Shetrone et al.(2003)]{She2003} Shetrone, M. D., Venn, K. A., Tolstoy, E., Primas, F., Hill, V., Kaufer, A., 2003, 
	AJ 125, 684
    \bibitem[Simmerer et al.(2004)]{Simm2004} Simmerer, J., Sneden, C., Cowan, J. J., Collier, J., Woolf, V. M., Lawler, J. E., 2004, 
	ApJ 617, 1091
    \bibitem[Smecker-Hane \& McWilliam(2002)]{SM2002} 	Smecker-Hane, T. A., McWilliam, A., 2002, 
	arXiv:astro-ph/0205411
    \bibitem[Sneden et al.(2008)]{Sne2008} Sneden, C., Cowan, J. J., Gallino, R., 2008, 
	ARA\&A 46, 241
    \bibitem[Spite(1967)]{Spite1967} Spite, M., 1967,
	AnAp 30, 211
    \bibitem[Starkenburg et al.(2010)]{Stark2010} Starkenburg, E., Hill, V., Tolstoy, E., Gonz\'{a}lez Hern\'{a}ndez, J. I., Irwin, M., Helmi, A., Battaglia, G., Jablonka, P., Tafelmeyer, M., Shetrone, M., Venn, K., de Boer, T., 2010,
	A\&A 513, A34
    \bibitem[Starkenburg et al.(2013)]{Stark2013} Starkenburg, E., Helmi, A., De Lucia, G., Li, Y.-S., Navarro, J. F., Font, A. S., Frenk, C. S., Springel, V., Vera-Ciro, C. A., White, S. D. M., 2013, MNRAS 429, 725
    \bibitem[Stetson et al.(1998)]{Stet1998} Stetson, P. B., Hesser, J. E., Smecker-Hane, T. A., 1998,
	PASP 110, 533
    \bibitem[Stetson \& Pancino(2008)]{StetPan2008} Stetson, P. B., Pancino, E., 2008,
	PASP 120, 1332
    \bibitem[Tafelmeyer et~al.(2010)]{Taf2010} Tafelmeyer, M., Jablonka, P., Hill, V., Shetrone, M., Tolstoy, E., Irwin, M. J., Battaglia, G., Helmi, A., Starkenburg, E., Venn, K. A., Abel, T., Fran\c cois, P., Kaufer, A., North, P., Primas, F., Szeifert, T., 2010,
	A\&A 524, A58
    \bibitem[Timmes et al.(2003)]{Tim2003} Timmes, F. X., Brown, E. F., Truran, J. W., 2003,
	ApJ 590, 83
    \bibitem[Tinsley(1979)]{Tin1979} Tinsley, B. M., 1979,
	ApJ 229,1046
    \bibitem[Tolstoy et al.(2001)]{Tolstoy2001} Tolstoy, E., Irwin, M.~J., Cole, A.~A., Pasquini, L., Gilmozzi, R., Gallagher, J. S., 2001, 
	MNRAS 327, 918
    \bibitem[Tolstoy et al.(2003)]{Tolstoy2003} Tolstoy, E., Venn, K. A., Shetrone, M., Primas, F., Hill, V., Kaufer, A., Szeifert, T., 2003,
       AJ 125, 707
    \bibitem[Tolstoy et al.(2006)]{Tolstoy2006} Tolstoy, E., Hill, V., Irwin, M., Helmi, A., Battaglia, G., Letarte, B., Venn, K., Jablonka, P., Shetrone, M., Arimoto, N., Abel, T., Primas, F., Kaufer, A., Szeifert, T., Fran\c cois, P., Sadakane, K., 2006,
	Msngr 123, 33
    \bibitem[Tolstoy et al.(2009)]{Tolstoy2009} Tolstoy, E., Hill, V., Tosi, M., 2009,
	ARA\&A 47, 371
    \bibitem[Travaglio et al.(2005)]{Tra2005} Travaglio, C., Hillebrandt, W., Reinecke, M., 2005,
	A\&A 443, 1007
    \bibitem[Tsujimoto et al.(1995)]{Tsuji1995} Tsujimoto, T., Nomoto, K., Yoshii, Y., Hashimoto, M., Yanagida, S., Thielemann, F.-K., 1995,
	MNRAS 277, 945
    \bibitem[Tsujimoto(2011)]{Tsuji2011} Tsujimoto, T., 2011,
	ApJ 778, 149
    \bibitem[Van der Swaelmen et al.(2013)]{vdS2013} Van der Swaelmen, M., Hill, V., Primas, F., Cole, A. A., 2013,
	A\&A 560, A44
    \bibitem[Venn et al.(2004)]{Venn2004} Venn, K. A., Irwin, M., Shetrone, M. D., Tout, C. A., Hill, V., Tolstoy, E., 2004,
	AJ 128, 1177
    \bibitem[Venn et al.(2012)]{Venn2012} Venn, K. A., Shetrone, M. D., Irwin, M. J., Hill, V., Jablonka, P., Tolstoy, E., Lemasle, B., Divell, M., Starkenburg, E., Letarte, B., Baldner, C., Battaglia, G., Helmi, A., Kaufer, A., Primas, F., 2012,
	ApJ 751, 102
    \bibitem[Walker et al.(2009)]{Walk2009} Walker, M. G., Mateo, M., Olszewski, E. W., Sen, B., Woodroofe, M., 2009,
	AJ 137, 3109
    \bibitem[Woosley \& Weaver(1995)]{WooWea1995} Woosley, S. E., Weaver, T. A., 1995,
	ApJS 101, 181
    \bibitem[Yozin \& Bekki(2012)]{Yo2012} Yozin, C., Bekki, K., 2012,
	ApJ 756, 18
\end{thebibliography}
\end{document}